\documentclass[prc,aps,preprint,tightenlines,nofootinbib,superscriptaddress]{revtex4-1}

\usepackage{subcaption}
\usepackage{appendix}
\usepackage{graphicx}
\usepackage{slashed}
\usepackage{amsmath}
\usepackage{hyperref}
\hypersetup{colorlinks=true,citecolor=blue,linkcolor=blue,urlcolor=blue}

\begin{document}
\title{Probing superfast quarks in nuclei through dijet production at the LHC}
\author{Adam J. Freese}
\author{Misak M. Sargsian}
\affiliation{Florida International University, Miami, FL 33199}
\author{Mark I. Strikman}
\affiliation{Pennsylvania State University, University Park, PA 16802}
\date{\today}

\begin{abstract}
We investigate dijet production from proton-nucleus collisions at the Large Hadron Collider (LHC) as a means
for observing superfast quarks in nuclei with Bjorken $x>1$.
Kinematically, superfast quarks can be identified through directly measurable jet kinematics.
Dynamically, their description requires understanding several elusive properties of nuclear QCD,
such as nuclear forces at very short distances,
as well as medium modification of parton distributions in nuclei.
In the present work, we develop a model for nuclear parton distributions at large $x$
in which the nuclear dynamics at short distance scales are described by two- and three-nucleon short range
correlations (SRCs). 
Nuclear modifications are accounted for using the color screening model,
and an improved description of the EMC effect is reached by using a
structure function parametrization that includes higher-twist contributions.
We apply QCD evolution at the leading order to obtain nuclear parton distributions in the kinematic regime
of the LHC,
and based on the obtained distributions calculate the cross section for dijet production.
We find that the rates of the dijet production in $pA$ collisions
at kinematics accessible by ATLAS and CMS are sufficient not only to observe 
superfast quarks but also to get information about the 
practically unexplored three-nucleon SRCs in nuclei.
Additionally, the LHC can extend our knowledge of the EMC effect to large $Q^2$
where higher-twist effects are negligible.
\end{abstract}

\maketitle
 
\section{Introduction}
\label{sec:intro}

The dynamics of quantum chromodynamics (QCD) in the nuclear medium is one of the most interesting areas of
modern nuclear physics.
Many aspects of it are currently being investigated,
including the formation of collective quark-gluon states such as
quark-gluon plasma,
the shadowing of the small $x$ parton densities in nuclei,
the hadronization of quarks and gluons within the nucleus,
and the medium modification of partonic distribution functions in nuclei.

Another interesting aspect of nuclear QCD is the possibility for quarks to carry
a light cone momentum fraction higher than that of a free nucleon at rest.
Deep inelastic scattering (DIS) from such a parton will correspond to Bjorken $x > 1$
(where $x$ is normalized to run from $0$ to the nuclear mass number $A$),
and we will hereafter refer to quarks with such a light cone momentum fraction as superfast quarks.

Due to the short-range nature of strong interactions,
detecting a superfast quark in a nucleus requires probing the nucleus at extremely short distance scales.
The characteristic space-time distances in nuclei become shorter with an increase in $Q^2$
for fixed $x$ owing to a property of QCD evolution,
namely that a probed parton at high $Q^2$ will have come from a parent quark with a higher light cone momentum fraction.
Thus, the theoretical expectation is that superfast quarks at large $Q^2$ will allow one to probe unprecedented
small space-time distances in nuclei,
on the order of $1/xm_N$\cite{Frankfurt:1988nt,Sargsian:2002wc}.

Our current understanding of the dynamics of nuclei at short distances is extremely limited.
Due to the short range nature of the strong interaction,
one expects that they will be dominated by multi-nucleon short range correlations
(SRCs)\cite{Frankfurt:1981mk}, which may include non-nucleonic degrees of freedom
(such as $\Delta \Delta$ and $NN^*$ components),
followed by the transition from baryonic to quark-gluon degrees of freedom.

There has been considerable progress recently made in studies of two-nucleon SRCs in inclusive
and semi-inclusive nuclear processes, which have been dominated by quasi-elastic (QE) scattering
of a high-energy probe (either an electron or a proton) off of a nucleon in the
SRC\cite{Frankfurt:1993sp,Aclander:1999fd,Tang:2002ww,Egiyan:2003vg,Egiyan:2005hs,Piasetzky:2006ai,
Shneor:2007tu,Subedi:2008zz,Frankfurt:2008zv,Fomin:2011ng,Arrington:2011xs,Arrington:2012ax,Korover:2014dma}.
These studies unambiguously established the existence of two-nucleon SRCs and measured the
probability of a nucleon existing in one for the given nuclei.
Due to the quasi-elastic nature of the scattering processes, however,
it is much more difficult to reach kinematics dominated by three-nucleon SRCs\cite{Egiyan:2003vg}.
Moreover, probing the transition to quark and gluon degrees of freedom has so far remained elusive.

The possibility of probing superfast quarks in hard nuclear processes provides a new venue in studies of SRCs.
From the theoretical point of view, as was mentioned above, the QCD evolution of quark distributions gives an
additional handle on the ability to probe exceedingly small space-time distances through an increase in $Q^2$
(at the scale of Ioffe time $1/m_Nx_B$).
From the experimental point of view, deep inelastic processes do not restrict the phase space of the final states
(as is the case for quasielastic processes),
and hence one can use closure to express the cross section through the parton distribution functions of the nucleus. 
Measurement of nuclear parton distributions at $x\ge 1$
allows one to explore different aspects of nuclear short-range phenomena.

One of the conventional methods for probing superfast quarks in nuclei is deep inelastic scattering (DIS)
from nuclei at Bjorken $x >1$.
A number of attempts have been made over the years.
The attempts to measure quarks at $x\gtrsim1$ were undertaken at CERN using a muon beam\cite{Benvenuti:1994bb}
and at FNAL using a neutrino beam\cite{Vakili:1999qt}, with mutually contradictory results.
Measurements using electron beams were taken for $x\sim1$ at SLAC\cite{Bosted:1992fy}
with $Q^2\leq10$~GeV$^2$ and at Jefferson Lab for $x>1$ and $Q^2=7$~GeV$^2$ \cite{Fomin:2010ei}.
However, this $x$ range gets a significant contribution from higher-twist quasieleastic scattering
up to fairly large $Q^2$ ($\sim15$~GeV$^2$).
For example, the nucleus/deuteron cross section ratio is reduced for $x=1$ and $Q^2=10$~GeV$^2$
by a factor of $2$ to $2.5$ due to QE contributions\cite{Frankfurt:1993sp}.
The only way to avoid this is to probe larger values of $Q^2$ $\ge 30$~GeV$^2$,
for which the quasieleastic contribution will be a small correction\cite{Sargsian:2002wc}.
Such experiments are currently included in the physics program for the 12 GeV upgrade of
Jefferson Lab\cite{Dudek:2012vr},
and the first experimental data will be available within the next few years\cite{Arrington:pr2006}.

In this work, we propose a new approach for probing superfast quarks by considering dijet production
in proton-nucleus collisions at LHC kinematics.
This approach is based on the possibility of relating the light cone momentum fractions of the initial partons 
to the measured kinematics of dijets;
by selecting transverse jet momenta and pseudo-rapidities, one can
isolate scattering off the superfast quarks within the nucleus.
We develop a theoretical framework for calculating this reaction,
which requires addressing several issues,
such as modeling the high-momentum (short range) properties of the nuclear wave function,
and calculating the medium modification of parton distributions within the nucleus and evolving this modification
to the large $Q^2$ values relevant to the LHC.
Within the framework we develop, we calculate the absolute cross section for the reaction and study its sensitivity
to two- and three-nucleon SRCs in nuclei.

The article is organized as follows:
In Sec.~\ref{sec:formal} we review the formalism of the dijet production reaction,
including its kinematics and the cross section formula.
The cross section for dijet production depends on the nuclear parton distribution functions (PDFs),
which are discussed in Sec.~\ref{sec:pdf}.
Sec.~\ref{sec:pdf} also discusses short range correlations and medium modifications,
and how they factor into and affect the nuclear PDFs.
This section contains a description of the structure of the elusive three-nucleon SRCs,
and provides a good fit of EMC effect data using the color screening model of medium modifications.
Sec.~\ref{sec:subprocess} discusses the hard subprocesses that contribute to dijet production
and justifies the use of the leading order of QCD in calculating them.
In Sec.~\ref{sec:num}, we present numerical estimates for the cross section,
and find that at a characteristic $pA$ luminosity reached at the LHC, rates are high enough to measure $x \ge 1$.
Additionally, we find high rates for kinematics corresponding to the EMC effect,
which at LHC energies allows the EMC effect to be studied with negligible higher-twist contributions.
In this section we also discuss potential issues related to the accuracy of jet reconstruction.
Conclusions and outlook are given in Sec.~\ref{sec:theend}.
The Appendix gives detailed derivations of the factorization formula and the SRC parts of the nuclear light
cone fraction distributions.

%%%%%%%%%%%%%%%%%%%%%%%%%%%%%%%%%%%%%%%%%%%%%%%%%%%%%%%%%%%%%%%%%%%%%%%%%%%%%%%%%%%%%%%%%%%%%%%%%%%%%%%%%%%%%%%%%%%%%%%%
%  Basic formalism
%%%%%%%%%%%%%%%%%%%%%%%%%%%%%%%%%%%%%%%%%%%%%%%%%%%%%%%%%%%%%%%%%%%%%%%%%%%%%%%%%%%%%%%%%%%%%%%%%%%%%%%%%%%%%%%%%%%%%%%%

\section{Basic Formalism}
\label{sec:formal}

\begin{figure}
  \centering
  \includegraphics[scale=0.5]{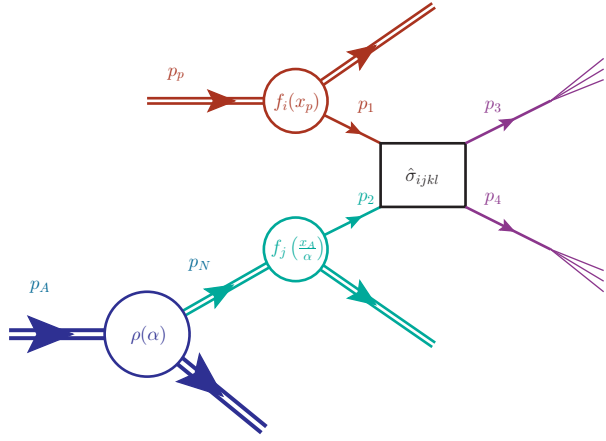}
  \caption{(Color online) Diagram of reaction}
  \label{fig:pAjj}
\end{figure}

The reaction we consider is the production of a dijet from a proton-nucleus collision,
\begin{equation}
  p + A \rightarrow \mathrm{dijet}+X
  \label{eqn:reaction}
  ,
\end{equation}
where the kinematics of the two outgoing jets are fully determined.
The reaction is treated at the leading order (LO) in perturbative QCD (pQCD),
meaning
that the jets are produced by a two-body parton-parton scattering process,
in which one parton originates in the proton and the other in the nucleus.
In our derivation, we consider the nuclear parton to have originated within a bound nucleon.
This scenario is presented in Fig.~\ref{fig:pAjj},
where we also define the kinematic variables of the reaction (\ref{eqn:reaction}).

%%%%%%%%%%%%%%%%%%%%%%%%%%%%%%%%%%%%%%%%%%%%%%%%%%%%%%%%%%%%%%%%%%%%%%%%%%%%%%%%%%%%%%%%%%%%%%%%%%%%%%%%%%%%%%%%%%%%%%%%
%  Jet kinematics
%%%%%%%%%%%%%%%%%%%%%%%%%%%%%%%%%%%%%%%%%%%%%%%%%%%%%%%%%%%%%%%%%%%%%%%%%%%%%%%%%%%%%%%%%%%%%%%%%%%%%%%%%%%%%%%%%%%%%%%%

\subsection{Jet Kinematics}
\label{sec:formal:kine}

We consider a reference frame where the incoming proton moves in the $+z$ direction and the heavy nucleus,
with charge number $Z$ and mass number $A$, moves in the $-z$ direction.
The four-momenta are described using light cone coordinates, namely
\begin{equation}
  p^\mu \equiv (p^+,p^-,\mathbf{p}_T)
  \label{eqn:lcp}
  ,
\end{equation}
where $p^\pm = E \pm p_z$ and $\mathbf{p}_T$ is the two-component transverse momentum.
Using the on-mass shell condition and the fact that the energies of the proton and nucleus greatly exceed their masses,
in the collider reference frame one has
\begin{align}
  p_p^\mu &= \left(p_p^+,\frac{m_p^2}{p_p^+},\mathbf{0}_T\right) = (2E_0,0,\mathbf{0}_T)
    = \left(\sqrt{\frac{As_{NN}^{\mathrm{avg.}}}{Z}},0,\mathbf{0}_T\right) \label{eqn:lc:pp} \\
  p_A^\mu &= \left(\frac{M_A^2}{p_A^-},p_A^-,\mathbf{0}_T\right) = (0,2ZE_0,\mathbf{0}_T)
    = \left(0,\sqrt{AZs_{NN}^{\mathrm{avg.}}},\mathbf{0}_T\right)\label{eqn:lc:pA}
  ,
\end{align}
where $E_0$ is the beam energy per proton in the same reference frame,
and $s_{NN}^{\mathrm{avg.}} = 4\frac{Z}{A}E_0^2$ is the square of the average center-of-mass energy per nucleon.
As an example, a lead-proton collision with a beam energy of $4$~TeV per proton would have an average center-of-mass
energy per nucleon of $\sqrt{s_{NN}^{\mathrm{avg.}}}\approx5.02$~TeV.
However, since the motion of the bound nucleon inside the nucleus will in general be variable,
the actual center-of-mass energy per nucleon $\sqrt{s_{NN}}$ is not a fixed parameter,
though it will be equal to $\sqrt{s_{NN}^{\mathrm{avg.}}}$ in the limit where the nucleons
do not interact and all move forward with equal momenta.

At leading order, the collision results in an interaction between two partons, one each from the proton and the nucleus.
Their respective four-momenta are labeled $p_1$ and $p_2$.
We use a collinear approximation, in which the initial partons are treated as having zero transverse momentum.
Moreover, they are treated as massless and on-shell, so
\begin{align*}
  p_1 &= \left(p_1^+,0;\mathbf{0}\right) \\
  p_2 &= \left(0,p_2^-;\mathbf{0}\right)
  .
\end{align*}
The light cone momentum fractions are defined for each parton as
\begin{align}
  x_p &=   \frac{p_1^+}{p_p^+} = \sqrt\frac{Z}{A}\frac{p_1^+}{\sqrt{s_{NN}^{\mathrm{avg.}}}} \label{eqn:xp} \\
  x_A &= A \frac{p_2^-}{p_A^-} = \sqrt\frac{A}{Z}\frac{p_2^-}{\sqrt{s_{NN}^{\mathrm{avg.}}}} \label{eqn:xA}
  .
\end{align}
Note that $x_A$ is scaled by a factor of $A$.
The rationale behind this is that the parton from the nucleus is found within one of its nucleons,
and in a limiting case where the nucleons are all non-interacting and carry equal momenta,
their light cone momentum is equal to $p_{N,\mathrm{avg}}^- = \frac{p_A^-}{A}$, meaning $x_A \leq 1$ in this case.

In reality, however, the bound nucleons do interact and it is possible that $p_2^- > p_{N,\mathrm{avg}}^-$.
In this situation, the nuclear parton originates from a nucleon which has a larger-than-average momentum
($p_N^- > p_{N,\mathrm{avg}}^-$).
This indicates that finding an exceedingly large $x_A > 1$ will identify a bound nucleon with momentum
significantly larger than average.

The parton momentum fractions cannot be directly measured.
However, they can be related to the kinematic parameters of the jets.
The jets from the proton and the nucleus are respectively ascribed four-momenta $p_3$ and $p_4$.
At leading order, they come from the fragmentation of two partons (with the same momenta),
which are treated as massless and on-shell.
From energy-momentum conservation, it follows that
\begin{equation}
  p_1 + p_2 = p_3 + p_4
  \label{eqn:conservation}
  ,
\end{equation}
and due to the assumed collinear approximation,
$\mathbf{p}_{3T} = -\mathbf{p}_{4T} \equiv \mathbf{p}_T$.
Using this relation, and neglecting the masses of the produced jets, we obtain:
\begin{equation}
  p_3^+p_3^- = p_4^+p_4^- = p_T^2
  .
\end{equation}
To proceed, we define the rapidity $\eta$ as
\begin{equation}
  \eta = \frac{1}{2}\log\left(\frac{p^+}{p^-}\right)
  ,
\end{equation}
and use this, with the massless limit (in which $p^+p^- = p_T^2$) to obtain
\begin{align}
  p^\pm &= p_T e^{\pm\eta}
  .
\end{align}
Applied to the jets of reaction (\ref{eqn:reaction}), this results in
$p_3^\pm = p_T e^{\pm\eta_3}$ and $p_4^\pm = p_T e^{\pm\eta_4}$.
Using these relations, and energy momentum conservation, {\sl viz.}~Eq.~(\ref{eqn:conservation}),
the light cone momenta of the initial partons can be expressed through jet kinematics in the following form:
\begin{align*}
  p_1^+ &= p_3^+ + p_4^+ = p_T \left(e^{ \eta_3} + e^{ \eta_4}\right) \\
  p_2^- &= p_3^- + p_4^- = p_T \left(e^{-\eta_3} + e^{-\eta_4}\right)
  .
\end{align*}
These relations can be used to express the momentum fractions $x_p$ and $x_A$ in terms of jet observables,
namely\footnote{Note that for the forward kinematics we are interested in, the measurement of the emission angle and energy
  of the forward jet already provides accurate determination of $x$ for large-$x$ partons.
  See the discussion in Sec.~\ref{sec:jetres}.
}
\begin{align}
  x_p &= \sqrt\frac{Z}{A}\frac{p_T}{\sqrt{s_{NN}^{\mathrm{avg.}}}}
    \left(e^{ \eta_3} + e^{ \eta_4}\right) \label{eqn:xp:eta} \\
  x_A &= \sqrt\frac{A}{Z}\frac{p_T}{\sqrt{s_{NN}^{\mathrm{avg.}}}}
    \left(e^{-\eta_3} + e^{-\eta_4}\right) \label{eqn:xA:eta}
  .
\end{align}

\begin{figure}
  \centering
  \includegraphics[scale=0.5]{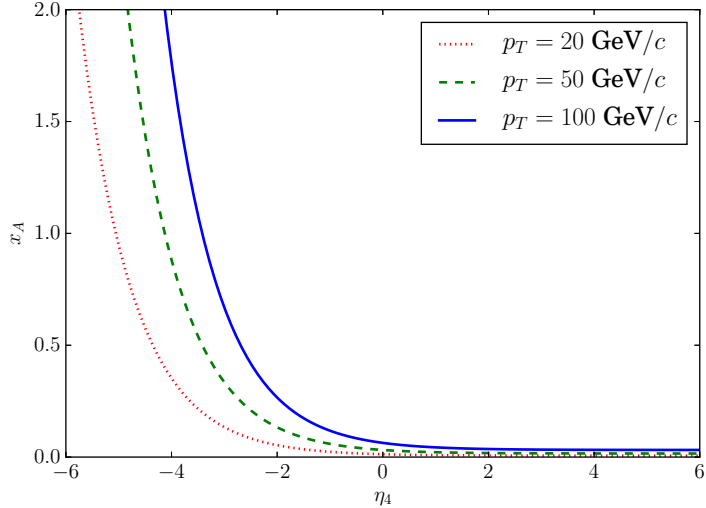}
  \caption{Dependence of $x_A$ on rapidity $\eta_4$ of parton from nucleus,
    for several transverse jet momenta. $\eta_3=0$
    and $\sqrt{s_{NN}^{\mathrm{avg.}}}=5.02$~TeV in this plot.}
  \label{fig:kine:xA}
\end{figure}

The main question that concerns us is whether
partons with $x_A > 1$ can be seen in proton-nucleus collisions at the LHC.
Eq.~(\ref{eqn:xA:eta}) suggests that we should look for three conditions: large $p_T$,
small (or somewhat negative) $\eta_3$, and small (or highly negative) $\eta_4$.
Negative $\eta_3$ will require that the parton originating from the proton reverse its direction before
fragmenting into a jet, which is highly unlikely owing to the large momentum transfer necessary to effect this.
Thus, for a given $p_T$, the most plausible scenario is to keep $\eta_3$ small ($\sim 0$),
and to look for jets with high rapidity in the nucleus beam direction.
Such a situation is presented in Fig.~\ref{fig:kine:xA},
where the dependence of $x_A$ on $\eta_4$ is given at $\eta_3=0$ for different values of $p_T$.
More practically, as will be discussed in Sec.~\ref{sec:num},
one can integrate over an $\eta_3$ range centered around $\eta_3=0$ in order to increase the cross section.

In addition to the momentum fractions, we introduce a characteristic ``hardness'' scale $Q^2$,
defined through the invariant momentum transfer:
\begin{align*}
  Q^2 &\equiv -t = -(p_1-p_3)^2 = 2(p_1 \cdot p_3) = p_1^+p_3^-
    = p_T^2 \left(1 + e^{-y_3 + y_4}\right) \approx p_T^2
  .
\end{align*}
This scale is used as both a renormalization and factorization scale.

%%%%%%%%%%%%%%%%%%%%%%%%%%%%%%%%%%%%%%%%%%%%%%%%%%%%%%%%%%%%%%%%%%%%%%%%%%%%%%%%%%%%%%%%%%%%%%%%%%%%%%%%%%%%%%%%%%%%%%%%
%  Dijet cross section
%%%%%%%%%%%%%%%%%%%%%%%%%%%%%%%%%%%%%%%%%%%%%%%%%%%%%%%%%%%%%%%%%%%%%%%%%%%%%%%%%%%%%%%%%%%%%%%%%%%%%%%%%%%%%%%%%%%%%%%%

\subsection{Dijet cross section}
\label{sec:formal:crx}

The factorization formula for the hadronic cross section
can be presented in terms of the partonic cross section as follows:
\begin{equation}
  \sigma_{pA} = \sum_{ij} \int_0^1 dx_p \int_0^A dx_A
    f_{i/p}(x_p,Q^2) f_{j/A}(x_A,Q^2) \sigma_{\mathrm{hard}}
  \label{eqn:factorize}
  ,
\end{equation}
where $f_{i/p}(x_p,Q^2)$ and $f_{j/A}(x_A,Q^2)$ are parton distribution functions (PDFs) for the proton and nucleus,
respectively.
This is similar in form to the standard factorization formula for $pp$ collisions\cite{Collins:2011zzd,Ellis:1991qj},
and in fact reduces to it in the
free nucleon limit. 

Since, at leading order, the momenta of the outgoing partons coincide with the jet momenta,
we can integrate out the transverse momentum of one of the jets and express differentials
through the rapidities of the outgoing jets.
For this purpose, we present the differential form of Eq.~(\ref{eqn:factorize}) as
\begin{align}
  d\sigma_{pA}
    &= \sum_{ijkl} f_{i/p}(x_p,Q^2) f_{j/A}(x_A,Q^2)
    \frac{1}{4(p_1 \cdot p_2)}
    \frac{\overline{\left|\mathcal{M}_{ij\rightarrow kl}\right|^2}}{1+\delta_{kl}}
    (2\pi)^4 
    2
    \delta^{(1)}(p_1^+-p_3^+-p_4^+)
  \notag \\ & \qquad \times
    \delta^{(1)}(p_2^--p_3^--p_4^-)
    \delta^{(2)}(\mathbf{p}_{3T}+\mathbf{p}_{4T})
    \frac{d^3\mathbf{p}_3}{2E_3(2\pi)^3}
    \frac{d^3\mathbf{p}_4}{2E_4(2\pi)^3}
    dx_p dx_A
  \label{eqn:dsig:step1}
  ,
\end{align}
where we have expressed four-momentum conservation through the light cone momenta.
Here, the indices $i$, $j$, $k$, and $l$ indicate parton types,
and $\mathcal{M}_{ij\rightarrow kl}$ is the invariant Feynman amplitude for the hard partonic scattering process.

Using the definitions of the light cone momenta fractions in Eqs.~(\ref{eqn:xp},\ref{eqn:xA}),
we can express the delta functions for $p^\pm$ conservation as:
\begin{align}
  2 \delta^{(1)}(p_1^+-p_3^+-p_4^+) \delta^{(1)}(p_2^--p_3^--p_4^-)
  = \frac{2 A}{p_p^+ p_A^-}
    \delta^{(1)}\left(x_p -  \frac{p_3^+ + p_4^+}{p_p^+}\right)
    \delta^{(1)}\left(x_A - A\frac{p_3^- + p_4^-}{p_A^-}\right)
  \label{eqn:dsig:step2}
  .
\end{align}
Inserting this relation into Eq.~(\ref{eqn:dsig:step1}),
one can integrate out $\mathbf{p}_{4T}$, $x_p$ and $x_A$, resulting in
\begin{align*}
  d\sigma_{pA} &= \sum_{ijkl}
    \frac{A}{16\pi}
    \frac{1}{p_p^+ p_A^-}
    \frac{1}{2(p_1 \cdot p_2)} 
    f_{i/p}(x_p,Q^2) f_{j/A}(x_A,Q^2)
    \frac{\overline{\left|\mathcal{M}_{ij\rightarrow kl}\right|^2}}{1+\delta_{kl}}
    \frac{dp_{3z}}{E_3} \frac{dp_{4z}}{E_4} dp_T^2
  .
\end{align*}
The $p_z$ elements can be rewritten as rapidities, since $d\eta = \frac{dp_z}{E}$.
In addition, as $2(p_1 \cdot p_2) = \frac{x_p x_A}{A} p_p^+ p_A^-$, we have
\begin{align*}
  d\sigma_{pA} &= \sum_{ijkl}
    \frac{1}{16\pi}
    \left(\frac{A}{p_p^+ p_A^-}\right)^2
    \frac{f_{i/p}(x_p,Q^2)}{x_p}
    \frac{f_{j/A}(x_A,Q^2)}{x_A}
    \frac{\overline{\left|\mathcal{M}_{ij\rightarrow kl}\right|^2}}{1+\delta_{kl}}
    d\eta_3 d\eta_4 dp_T^2
  .
\end{align*}
Since the square of the average center-of-mass energy per nucleon is
\begin{equation}
  s_{NN}^{\mathrm{avg.}} = \frac{p_p^+ p_A^-}{A}
  ,
\end{equation}
the differential cross section can be written in the standard form:
\begin{equation}
  \frac{d^3\sigma}{d\eta_3 d\eta_4 dp_T^2} =
  \sum_{ijkl}
    \frac{1}{16\pi(s_{NN}^{\mathrm{avg.}})^2}
    \frac{f_{i/p}(x_p,Q^2)}{x_p} \frac{f_{j/A}(x_A,Q^2)}{x_A}
    \frac{\overline{\left|\mathcal{M}_{ij\rightarrow kl}\right|^2}}{1+\delta_{kl}}
  \label{eqn:diffcrx}
  .
\end{equation}

%%%%%%%%%%%%%%%%%%%%%%%%%%%%%%%%%%%%%%%%%%%%%%%%%%%%%%%%%%%%%%%%%%%%%%%%%%%%%%%%%%%%%%%%%%%%%%%%%%%%%%%%%%%%%%%%%%%%%%%%
%  Nuclear PDF
%%%%%%%%%%%%%%%%%%%%%%%%%%%%%%%%%%%%%%%%%%%%%%%%%%%%%%%%%%%%%%%%%%%%%%%%%%%%%%%%%%%%%%%%%%%%%%%%%%%%%%%%%%%%%%%%%%%%%%%%

\section{Nuclear PDFs}
\label{sec:pdf}

The main theoretical issue to be worked out is the nuclear PDFs $f_{j/A}(x_A,Q^2)$.
Currently, there is a lack of strong experimental constraints on nuclear PDFs.
Phenomenological parametrizations exist for several nuclei,
and are based on experimental measurements of the nuclear structure function $F_2^{(A)}(x,Q^2)$
in deep inelastic scattering (DIS) experiments in a restricted range of $x$ and $Q^2$.
({\sl cf. e.g.}
Refs.~\cite{Eskola:1998df,deFlorian:2003qf,Hirai:2007sx,Eskola:2009uj,deFlorian:2011fp,Honkanen:2013goa}.)
Most treatments of the nuclear PDF parametrize the ratio between the nucleus and the nucleon,
which only makes sense for treating $x_A$ up to $1$.
Ref.~\cite{deFlorian:2003qf} 
%is an exception to this, and
instead relates the nuclear and nucleonic
PDFs though a Mellin convolution, and parametrizes the function that is convoluted with the nucleonic PDF.
This approach has the possibility of accounting for $x_A > 1$,
but owing to a lack of experimental constraint, Ref.~\cite{deFlorian:2003qf} did not treat this region.

In order to describe $x_A > 1$ in the absence of robust empirical parametrizations of the nuclear PDFs,
we must account for them in terms of the nucleonic PDF theoretically.
To this end, we must look at the theoretical relationship between the nuclear and nucleonic parton distributions.

Treating the nucleus strictly as a collection of unmodified nucleons
with a light cone distribution normalized by the baryon sum rule
leads to a contradiction with experimental data---in particular
for the ratio of nuclear and deuteron DIS cross sections.
(See Sec.~\ref{sec:pdf:emc} for details.)
However, the introduction of {\sl a priori}~unknown bound nucleon PDFs $f_{j/N}^{(b)}$
allows for the nuclear PDF to be expressed as a convolution between these PDFs and
the nuclear light cone fraction distribution\cite{Frankfurt:1988nt,Arneodo:1992wf,Geesaman:1995yd}.
In the Bjorken limit,
where the photon virtuality $Q^2$ and energy $q_0$ both go to infinity
at a fixed $x_A = \frac{AQ^2}{2M_Aq_0}$,
the convolution formula takes the following form:
\begin{equation}
  f_{i/A}(x_A,Q^2) = \sum_N \int_x^A \frac{d\alpha}{\alpha} \int d^2\mathbf{p}_T f_{N/A}(\alpha,\mathbf{p}_T)
    f^{(b)}_{i/N}\left(\frac{x_A}{\alpha},\alpha,\mathbf{p}_T,Q^2\right)
  \label{eqn:pdf:convolution}
  ,
\end{equation}
which can be derived as an impulse approximation (see Appendix~\ref{appendix:factorize} for details).
Here, $f_{N/A}(\alpha,\mathbf{p}_T)$ is the light cone fraction distribution
of a nucleon $N$ in the nucleus $A$, described in Sec.~\ref{sec:pdf:lcd};
$\alpha$ is a scaled light cone momentum fraction given in Eq.~(\ref{eqn:alpha}) of the same section,
and $\mathbf{p}_T$ is the transverse momentum of the nucleon.
The $(b)$ in the superscript of $f^{(b)}_{i/N}\left(\frac{x_A}{\alpha},\alpha,\mathbf{p}_T,Q^2\right)$
signifies that this is an bound nucleonic PDF,
which differs from the free nucleon PDF owing to modifications from the nuclear medium. 
In particular, it is a function of $\alpha$ and $\mathbf{p}_T$ 
in addition to $x_N = \frac{x_A}{\alpha}$ and the factorization scale $Q^2$ because of medium modifications.

Eq.~(\ref{eqn:pdf:convolution}) indicates that to construct nuclear PDFs, one needs to address two theoretical issues:
the nuclear light cone fraction distribution $f_{N/A}(\alpha,\mathbf{p}_T)$,
and possible medium modification effects on the bound nucleon PDFs.

%%%%%%%%%%%%%%%%%%%%%%%%%%%%%%%%%%%%%%%%%%%%%%%%%%%%%%%%%%%%%%%%%%%%%%%%%%%%%%%%%%%%%%%%%%%%%%%%%%%%%%%%%%%%%%%%%%%%%%%%
%  Light cone density matrix
%%%%%%%%%%%%%%%%%%%%%%%%%%%%%%%%%%%%%%%%%%%%%%%%%%%%%%%%%%%%%%%%%%%%%%%%%%%%%%%%%%%%%%%%%%%%%%%%%%%%%%%%%%%%%%%%%%%%%%%%

\subsection{Light cone distribution and SRCs}
\label{sec:pdf:lcd}

We formally define the light cone fraction distribution in terms of the nuclear wave function as:
\begin{align}
  f_{N/A}(\alpha,\mathbf{p}_T)
    =
    \int
  &
%    \prod_{j=1}^A \frac{d\alpha d^2\mathbf{p}_T}{\alpha}
    \prod_{j=1}^A \frac{d\alpha_j d^2\mathbf{p}_{j,T}}{\alpha_j}
    \psi^\dagger \left(\alpha_1,\mathbf{p}_{T1}, \ldots, \alpha_A, \mathbf{p}_{TA} \right)
    \psi \left(\alpha_1,\mathbf{p}_{T1}, \ldots, \alpha_A, \mathbf{p}_{TA} \right)
  \notag \\ & \times
    \delta^{(1)}\left(A-\sum_{i=1}^A\alpha_i\right)
    \delta^{(2)}\left(\sum_{i=1}^A\mathbf{p}_{iT}\right)
    \left\{
      \sum_{i=1}^A
      \delta^{(1)}(\alpha-\alpha_i)
      \delta^{(2)}(\mathbf{p}_T-\mathbf{p}_{iT})
    \right\}
                \label{eqn:fNA}
\end{align}
where the (scaled) light cone momentum fraction $\alpha$ given by
\begin{equation}
  \alpha = A\frac{p_N^+}{p_A^+}
  \label{eqn:alpha}
  ,
\end{equation}
and $\mathbf{p}_T$ is the transverse momentum of the nucleon.
The light cone fraction distribution is normalized to obey two sum rules,
namely, the baryon number and momentum sum rules:
\begin{align}
  \int_0^A d\alpha \int d^2\mathbf{p}_T f_{N/A}(\alpha,\mathbf{p}_T) &= 1
    \label{eqn:sumrule:baryon} \\
  \int_0^A d\alpha \int d^2\mathbf{p}_T \alpha f_{N/A}(\alpha,\mathbf{p}_T) &= 1
    \label{eqn:sumrule:momentum}
  .
\end{align}
The light cone distribution thus defined can be related to the nuclear
light cone density matrix $\rho_{N/A}(\alpha,\mathbf{p}_T)$
({\sl cf.~e.g.}~\cite{Frankfurt:1977vc,Frankfurt:1981mk,Frankfurt:1988nt})
in the following way:
\begin{equation}
  f_{N/A}(\alpha,\mathbf{p}_T) = \frac{1}{\alpha}\rho_{N/A}(\alpha,\mathbf{p}_T)
  .
\end{equation}
We employ the light cone fraction distribution $f_{N/A}(\alpha,\mathbf{p}_T)$, however, 
because it represents a direct analogy to PDFs,
although for the distribution of nucleons in nuclei rather than of partons in hadrons.

For our calculations, we expand the light cone fraction distribution
as a sum of contributions from the nuclear mean field and multi-nucleon short range correlations (SRCs).
The decomposition takes the form:
\begin{equation}
  f_{N/A}(\alpha,\mathbf{p}_T) = f_{N/A}^{(MF)}(\alpha,\mathbf{p}_T)
    + \sum_{j=2}^A f_{N/A}^{(j)}(\alpha,\mathbf{p}_T)
  \label{eqn:rhoA:decomp}
  ,
\end{equation}
where $f_{N/A}^{(MF)}(\alpha,\mathbf{p}_T)$ is the mean field part,
and $f_{N/A}^{(j)}(\alpha,\mathbf{p}_T)$ is the distribution of $j$-nucleon SRCs.

%%%%%%%%%%%%%%%%%%%%%%%%%%%%%%%%%%%%%%%%%%%%%%%%%%%%%%%%%%%%%%%%%%%%%%%%%%%%%%%%%%%%%%%%%%%%%%%%%%%%%%%%%%%%%%%%%%%%%%%%
%  Mean field
%%%%%%%%%%%%%%%%%%%%%%%%%%%%%%%%%%%%%%%%%%%%%%%%%%%%%%%%%%%%%%%%%%%%%%%%%%%%%%%%%%%%%%%%%%%%%%%%%%%%%%%%%%%%%%%%%%%%%%%%

\subsubsection{Mean field distribution}

The mean field part of the light cone fraction distribution describes how the nucleons in the nucleus would be distributed
if they were only acted upon the mean field generated by the $(A-1)$ other nucleons.
The mean field distribution can be related to the wave function of the nucleus, which is calculated in the non-relativistic
limit since
the relevant momenta are smaller than the typical Fermi momentum of heavy nuclei.
The simplest mean field model considers the heavy nucleus as a degenerate Fermi gas,
and the typical Fermi momentum is around $250$ MeV$/c$\cite{Moniz:1971mt}.
More sophisticated 
momentum distributions (calculated based on, for example, Hartree-Fock approximations)
still fall off sharply above the Fermi momentum,
so the leading order relativistic corrections
are at most on the order of the mean field momentum distribution strength above $k_F$,
which $<1\%$ in magnitude.

The mean field distribution is related to the non-relativistic momentum-space wave function using
the sum rules of Eqs.~(\ref{eqn:sumrule:baryon},\ref{eqn:sumrule:momentum}).
If short range correlations are neglected, the mean field distribution obeys
Eqs.~(\ref{eqn:sumrule:baryon},\ref{eqn:sumrule:momentum}) by itself.
The momentum-space wave function $\Psi_{MF}^{(N)}(p)$ (which carries an index of $(N)$ for isospin since the
wave function will in general be different for protons and neutrons) is likewise normalized to unity, so we equate
\begin{equation*}
  \left|\Psi_{MF}^{(N)}(p)\right|^2 d^3\mathbf{p} = f_{N/A}^{(MF)}(\alpha,\mathbf{p}_T) d\alpha d^2\mathbf{p}_T
  .
\end{equation*}
In the non-relativistic limit, we can write, in the nuclear center of mass frame,
\begin{align*}
  \alpha &= A\frac{E + p_z}{m_A} \approx A\frac{m_N + p_z}{m_A} \\
  p_z &\approx \frac{m_A}{A}\alpha - m_N \\
  dp_z & \approx \frac{m_A}{A} d\alpha
  .
\end{align*}
Consequently, we can identify
\begin{equation}
  f_{N/A}^{(MF)}(\alpha,\mathbf{p}_T) = \frac{m_A}{A} \left| \Psi_{MF}^{(N)}(p) \right|^2
  .
\end{equation}

One modification must be made when accounting for SRCs.
It is not the mean field part of the light cone distribution,
but the light cone distribution taken as a whole that is normalized to satisfy the sum rules
of Eqs.~(\ref{eqn:sumrule:baryon},\ref{eqn:sumrule:momentum}).
Therefore, $f_{N/A}^{(MF)}(\alpha,\mathbf{p}_T)$ must be scaled down by a factor of $a_1^{(N)}(A)$,
which is defined as
\begin{equation}
  a_1^{(N)}(A) = 1 - \sum_{j=2}^A \int d\alpha d^2\mathbf{p}_T f_{N/A}^{(j)}(\alpha,\mathbf{p}_T)
  \label{eqn:a1}
  ,
\end{equation}
{\sl i.e.} it subtracts off the probability that the nucleon is in a short range correlation.
For instance, if a nucleon is in a short range correlation $25\%$ of the time, then $a_1^{(N)}(A)$ will be $0.75$.
In general, however, $a_1^{(N)}$ will be different for protons and neutrons.

For numerical estimates, we will use the momentum distributions calculated in
Ref.~\cite{Zverev:1986mv} for the mean field.

%%%%%%%%%%%%%%%%%%%%%%%%%%%%%%%%%%%%%%%%%%%%%%%%%%%%%%%%%%%%%%%%%%%%%%%%%%%%%%%%%%%%%%%%%%%%%%%%%%%%%%%%%%%%%%%%%%%%%%%%
%  2N SRCs
%%%%%%%%%%%%%%%%%%%%%%%%%%%%%%%%%%%%%%%%%%%%%%%%%%%%%%%%%%%%%%%%%%%%%%%%%%%%%%%%%%%%%%%%%%%%%%%%%%%%%%%%%%%%%%%%%%%%%%%%

\subsubsection{Two-nucleon correlations}
\label{sec:pdf:lcd:2N}

Around $25\%$ of the time, a nucleon in a heavy nucleus is in a two-nucleon
short range correlation\cite{Frankfurt:1993sp}.
A short range correlation occurs when two nucleons are separated by
a distance on the order of $1$~fm and consequently have a relative momentum larger than $k_F$.
The analysis of recent experiments indicates\cite{Frankfurt:2008zv}
that in the momentum range from $k_F$ to $\sim 650$~MeV$/c$,
the nucleus is well-described by two-nucleon SRCs.
Due to the short distance between the nucleons,
the dynamics of the nucleons are primarily influenced by their mutual interaction
rather than by the mean field.
The most immediate consequences of two-nucleon SRCs are a large tail in the momentum distribution above the Fermi momentum,
and large excitation energies of the residual system produced by the removal of the fast nucleons,
which are not reproduced by mean field models\cite{Arrington:2011xs}.

Within the last decade, there has been considerable experimental and theoretical effort put into studying two-nucleon
SRCs and their properties.
There are a fair number of triple-coincidence experiments demonstrating their
existence\cite{Shneor:2007tu,Subedi:2008zz,Korover:2014dma}
and demonstrating that two-nucleon SRCs primarily form between a proton and a
neutron\cite{Piasetzky:2006ai,Shneor:2007tu,Subedi:2008zz}.
The fact that most two-nucleon SRCs are predominantly $pn$ pairs has an important implication for momentum distributions
(and thus for the 2N light cone fraction distribution):
for neutron rich nuclei,
a given proton is more likely to be in a short range correlation
than a given neutron\cite{McGauley:2011qc,Feldmeier:2011qy,Sargsian:2012sm,Hen:2014nza}.
This means that $f_{p/A}^{(2)}(\alpha,\mathbf{p}_T) > f_{n/A}^{(2)}(\alpha,\mathbf{p}_T)$, 
and consequently (for the sum rules Eqs.~(\ref{eqn:sumrule:baryon},\ref{eqn:sumrule:momentum}) to be satisfied),
$a_1^{(p)} < a_1^{(n)}$.

Another important aspect of two-nucleon SRCs is the universal
high-momentum tail that they introduce to the momentum distribution of
nuclei\cite{Frankfurt:1988nt,Feldmeier:2011qy,Alvioli:2013qyz,Wiringa:2013ala}.
Most significantly, because of the dominance of spin-one, isosinglet SRCs,
the high-momentum tail behaves to a good approximation
like a scaled version of the high-momentum tail in the deuteron momentum distribution.

Based on the arguments discussed above, a model of the two-nucleon SRC contribution to the nuclear
light cone fraction distribution should incorporate two main properties:
first,
the universality of the form of the high momentum tail and its proportionality to the deuteron momentum distribution,
and second, 
different high-momentum sharing between protons and neutrons in asymmetric nuclei.
To account for universality, we use the light cone approximation of
Refs.~\cite{Frankfurt:1977vc,Frankfurt:1981mk,Frankfurt:1988nt},
which uses the requirement of rotational invariance to relate the light cone deuteron wave function to
the non-relativistic wave function by using the internal $pn$ light cone momentum
\begin{equation}
  k = \sqrt{\frac{m^2(\alpha-1)^2 + p_T^2}{\alpha(2-\alpha)}}
  \label{eqn:k}
  ,
\end{equation}
which is obtained by the definitions $\mathbf{k}_T = \mathbf{p}_T$ and
\begin{equation}
  \alpha = 1 + \frac{k_z}{\sqrt{m^2+k^2}}
  ,
\end{equation}
and which also uses high momentum scaling to relate $f_{N/d}(\alpha,\mathbf{p}_T)$ to $f_{N/A}^{(2)}(\alpha,\mathbf{p}_T)$.
To account for isopsin-singlet dominance, we use the model of Ref.~\cite{Sargsian:2012sm,Sargsian:2013gea},
in which the high-momentum part of the nuclear momentum distribution is inversely proportional
to the relative fraction of protons or neutrons.
The model additionally includes a scaling factor $a_2(A)$, which is extracted
from SRC studies in inclusive $eA$ processes\cite{Frankfurt:1993sp,Egiyan:2003vg,Egiyan:2005hs}.

Accounting for these two effects leads to the two-nucleon SRC light cone fraction distribution:
\begin{equation}
  f_{N/A}^{(2)}(\alpha,\mathbf{p}_T)
    = \frac{a_2(A)}{2\chi_N} \frac{\overline{\left|\psi_{d}(k)\right|^2}}{\alpha(2-\alpha)} \Theta(k-k_F)
  \label{eqn:rho2}
  ,
\end{equation}
where the extra factor of $\alpha(2-\alpha)$ is from the phase space of the struck and spectator nucleons.
The details of the derivation
of Eq.~(\ref{eqn:rho2}),
can be found in Appendix \ref{appendix:lcd:2N}.
Here,
$\psi_{d}(k)$ is the relativistic, light-cone deuteron wave function,
related to the non-relativistic wave function by
\begin{equation}
  \left|\psi_d(k)\right|^2 = \sqrt{m^2+k^2}\left|\psi_{NR}(k)\right|^2
  \label{eqn:wf:lc:nr}
  ,
\end{equation}
and $\chi_N$ is the relative abundance of nucleons of type $N$ in the nucleus, {\sl i.e.}
$\chi_p = \frac{Z}{A}$ and $\chi_n = \frac{A-Z}{A}$.
The factor of $\chi_N$ is present in Eq.~(\ref{eqn:rho2}) results in the equality
\begin{equation*}
  \sum_p f_{p/A}^{(2)}(\alpha,\mathbf{p}_T) = \sum_n f_{n/A}^{(2)}(\alpha,\mathbf{p}_T)
  .
\end{equation*}
This relation is confirmed by quantum Monte-Carlo calculations for light nuclei up to $A=12$,
and for medium to heavy nuclei it was confirmed using a correlated basis
calculation of the nuclear momentum distributions in a non-relativistic approach\cite{Vanhalst:2014cqa}.
Such a relation is also in agreement with the recent analysis of momentum sharing in heavy nuclei\cite{Hen:2014nza}.

In the present model we neglect by center of mass motion of the SRC\footnote{
  Inclusion of the center-of-mass motion leads to practically the same light cone fraction distribution,
  but with a somewhat smaller $a_2$.
  Since we are interested in quantities sensitive to the light cone fraction distribution,
  the difference in the recoil between two models due to center of mass motion
        has no impact on our analysis.}.
As a result, the two-nucleon SRC contribution to the nuclear light cone distribution
sets in as soon as $k > k_F$,
which is signified by the step function $\Theta(k-k_F)$ in Eq.~(\ref{eqn:rho2}).

The remaining factor in Eq.~(\ref{eqn:rho2}), $a_2(A)$,
is a scaling factor that describes how large the high-momentum tail of the nucleus is
relative to the high-momentum tail of the deuteron.
It is determined experimentally by examining ratios of
quasielasitc inclusive cross sections of $A(e,e^\prime X)$ and $^2$H$(e,e^\prime X)$ reactions:
\begin{align*}
  a_2(A) &= \frac{2 \sigma_{eA}(x,Q^2)}{A \sigma_{ed}(x,Q^2)}
    \qquad \mathrm{at} ~~ x \gtrsim 1.4~~\mathrm{and}~~Q^2 \gtrsim 1.5~\mathrm{GeV}^2
  ,
\end{align*}
where a roughly flat plateau in the ratio of the quasi-elastic cross sections is
observed\cite{Frankfurt:1993sp,Egiyan:2003vg,Egiyan:2005hs,Fomin:2011ng,Arrington:2012ax}.
The values of $a_2(A)$ for various nuclei are well-constrained by
experiment\cite{Frankfurt:1993sp,Egiyan:2005hs,Fomin:2011ng}
and it is equal to roughly $5.6$ for heavy nuclei such as iron-56.
In this work, we will also use $a_2(^{208}\mathrm{Pb})=5.6$, as no experimental data for this quantity exist for lead.

For numerical estimates, we use the Paris potential for parameterizing the non-relativistic
deuteron wave function\cite{Lacombe:1980dr} in Eq.~(\ref{eqn:wf:lc:nr}).

%%%%%%%%%%%%%%%%%%%%%%%%%%%%%%%%%%%%%%%%%%%%%%%%%%%%%%%%%%%%%%%%%%%%%%%%%%%%%%%%%%%%%%%%%%%%%%%%%%%%%%%%%%%%%%%%%%%%%%%%
%  3N SRCs
%%%%%%%%%%%%%%%%%%%%%%%%%%%%%%%%%%%%%%%%%%%%%%%%%%%%%%%%%%%%%%%%%%%%%%%%%%%%%%%%%%%%%%%%%%%%%%%%%%%%%%%%%%%%%%%%%%%%%%%%

\subsubsection{Three-nucleon correlations}
\label{sec:pdf:lcd:3N}

When the light cone momentum fraction $\alpha > 2$ (and probably already for $\alpha \sim 1.6$ or $1.7$),
one expects the nuclear light cone distribution to be dominated by three-nucleon short range correlations.
We consider a three-nucleon SRC to occur when the center of mass momentum of the 3N system is small,
and when the struck nucleon,
as well as the other nucleons in the SRC,
have large ($>k_F$) internal momentum.
Such a configuration can naturally generate a light cone momentum fraction exceeding 2.

There are two distinct mechanisms that could generate three-nucleon correlations.
The first is a sequence of short-range two-nucleon interactions,
and the second an irreducible three-nucleon interaction.
The second contributes a much larger removal energy part to
the nuclear spectral function than the first mechanism\cite{Sargsian:2005ru}.
Since the light cone momentum distribution is an integral over removal energies,
one expects it to be dominated by the first mechanism,
which occurs at lower values of removal energies than the irreducible 3N interaction.
Therefore, neglecting the irreducible 3N interactions,
we develop a model of three-nucleon SRCs where the struck nucleon obtains its momentum from a sequence
of two-nucleon short range interactions.
This model is based on the collinear approximation of Ref.~\cite{Frankfurt:1981mk},
where the three nucleons in the SRC are moving collinearly prior to the interactions that generate the SRC.

In the present work, we develop the model further using the recently observed dominance of $pn$ SRCs.
This allows us to express the three-nucleon SRC as occurring through a sequence of $pn$ interactions,
and thus the expressing 3N SRC part of the nuclear light cone fraction distribution
as a convolution of two 2N SRC light cone momentum distributions.
A full derivation of the three-nucleon SRC distribution is given in Appendix~\ref{appendix:lcd:3N},
with the final result\footnote{
  This is a large-$A$ approximation, which is expected to have $\frac{1}{A}$ corrections
  owing to surface effects, isospin asymmetry, and combinatorics of selecting $pn$ pairs.
}:
\begin{align}
  f_{N/A}^{(3)}(\alpha,\mathbf{p}_T) =
    \left\{a_2(A)\right\}^2
    \frac{1}{\alpha}
    \int & \frac{d\alpha_3d^2\mathbf{p}_{3T}}{\alpha_3(3-\alpha-\alpha_3)}
      \left\{\frac{3-\alpha_3}{2(2-\alpha_3)}\right\}^2
  \notag \\ &
      ~\overline{\left|{\psi_{d}(k_{12})}\right|^2}~
      \Theta(k_{12}-k_F)
      ~\overline{\left|{\psi_{d}(k_{23})}\right|^2}~
      \Theta(k_{23}-k_F)
  \label{eqn:rho3}
  .
\end{align}
Here, $k_{12}$ and $k_{13}$ are relative light cone momenta between pairs of nucleons in the three-nucleon SRC.
Their functional forms are given by Eqs.~(\ref{eqn:k12},\ref{eqn:k23}).
$\alpha_3$ and $\mathbf{p}_{3T}$ are, respectively, the light cone momentum fraction and the transverse momentum
of one of the spectator nucleons in the three-nucleon SRC.
Since the three-nucleon SRC is generated through a sequence of two short-range $pn$ interactions,
it can only occur if the conditions under which the individual $pn$ interactions would occur are satisfied.
For this reason, the threshold for three-nucleon SRCs to appear is that the relative light cone momenta
of the pairs should each satisfy the threshold condition for which short range two-nucleon
interactions occur, namely they should both be above the Fermi momentum $k_F$.
It is worth noting that $k_{12}$ and $k_{23}$ are both functions of the light cone momentum fraction $\alpha_3$
of a spectator nucleon, which is integrated over. ({\sl cf.}~Appendix~\ref{appendix:lcd:3N} for details.)
Because of this, the discontinuity of the $\Theta(~)$ function
is smeared out and $f_{N/A}^{(3)}(\alpha,\mathbf{p}_T)$ is itself a smooth function.

The factor $\left\{a_2(A)\right\}^2$ appearing in Eq.~(\ref{eqn:rho3}) is a consequence of the fact that three-nucleon
SRCs arise from a sequence of short-range $pn$ interactions.
In such a scenario, for heavy nuclei,
the probability of having a three-nucleon correlation should be proportional to the square of the probability of
having a two-nucleon correlation.

Eq.~(\ref{eqn:rho3}) may also be written in terms of a scaling factor $a_3(A)$,
similar to $a_2(A)$ the scaling factor for two-nucleon SRCs.
If it is assumed that three-nucleon SRCs are universal in their behavior,
the parameter $a_3(A)$ would be possible to extract in a similar way to $a_2(A)$ in the case of 2N SRCs.
In particular, as is the case for $a_2(A)$, it should be possible to extract $a_3(A)$ through the ratio
of quasielastic cross sections for $eA$ and $e^3\mathrm{He}$ processes at high $x > 2$ and $Q^2$, {\sl viz.}
\begin{equation}
  a_3(A) = \frac{3 \sigma_{eA}(x,Q^2)}{A \sigma_{e^3\mathrm{He}}(x,Q^2)} \qquad : \qquad x \gtrsim 2.4
  \label{eqn:a3}
  ,
\end{equation}
provided that a 3N SRC plateau is actually observed in this ratio.
In principle, the observation of $a_3(A)$ may provide an experimental test of the three-nucleon SRC model
presented in this work.
By definition, $a_3(^3\mathrm{He})=1$, and for Eq.~(\ref{eqn:rho3}) and Eq.~(\ref{eqn:a3}) to be consistent would
require that $a_3(A) \propto \left(\frac{a_2(A)}{a_2(^3\mathrm{He})}\right)^2$.
Consequently, we have
\begin{align}
  f_{N/A}^{(3)}(\alpha,\mathbf{p}_T) =
    \left\{a_2(^3\mathrm{He})\right\}^2
    \frac{a_3(A)}{\alpha}
    \int & \frac{d\alpha_3d^2\mathbf{p}_{3T}}{\alpha_3(3-\alpha-\alpha_3)}
      \left\{\frac{3-\alpha_3}{2(2-\alpha_3)}\right\}^2
  \notag \\ &
      ~\overline{\left|{\psi_{d}(k_{12})}\right|^2}~
      \Theta(k_{12}-k_F)
      ~\overline{\left|{\psi_{d}(k_{23})}\right|^2}~
      \Theta(k_{23}-k_F)
  .
\end{align}

The experimental status of a three-nucleon scaling plateau is, however, ambiguous.
While observation of an $x>2$ plateau was reported in Ref.~\cite{Egiyan:2005hs}
and in the analysis of SLAC data\cite{Frankfurt:1988nt},
a later experiment at similar kinematics and better resolution did not observe the plateau\cite{Fomin:2011ng}.
The discrepancy may be due to a resolution issue, as argued in Ref.~\cite{Higinbotham:2014xna}.
On the theoretical side, one still needs to investigate whether the kinematics covered by these experiments
correspond to high enough nucleon momentum that universality should be expected to set in. 

\begin{figure}
  \centering
  \begin{subfigure}[t]{.45\textwidth}
    \centering
    \includegraphics[width=\textwidth]{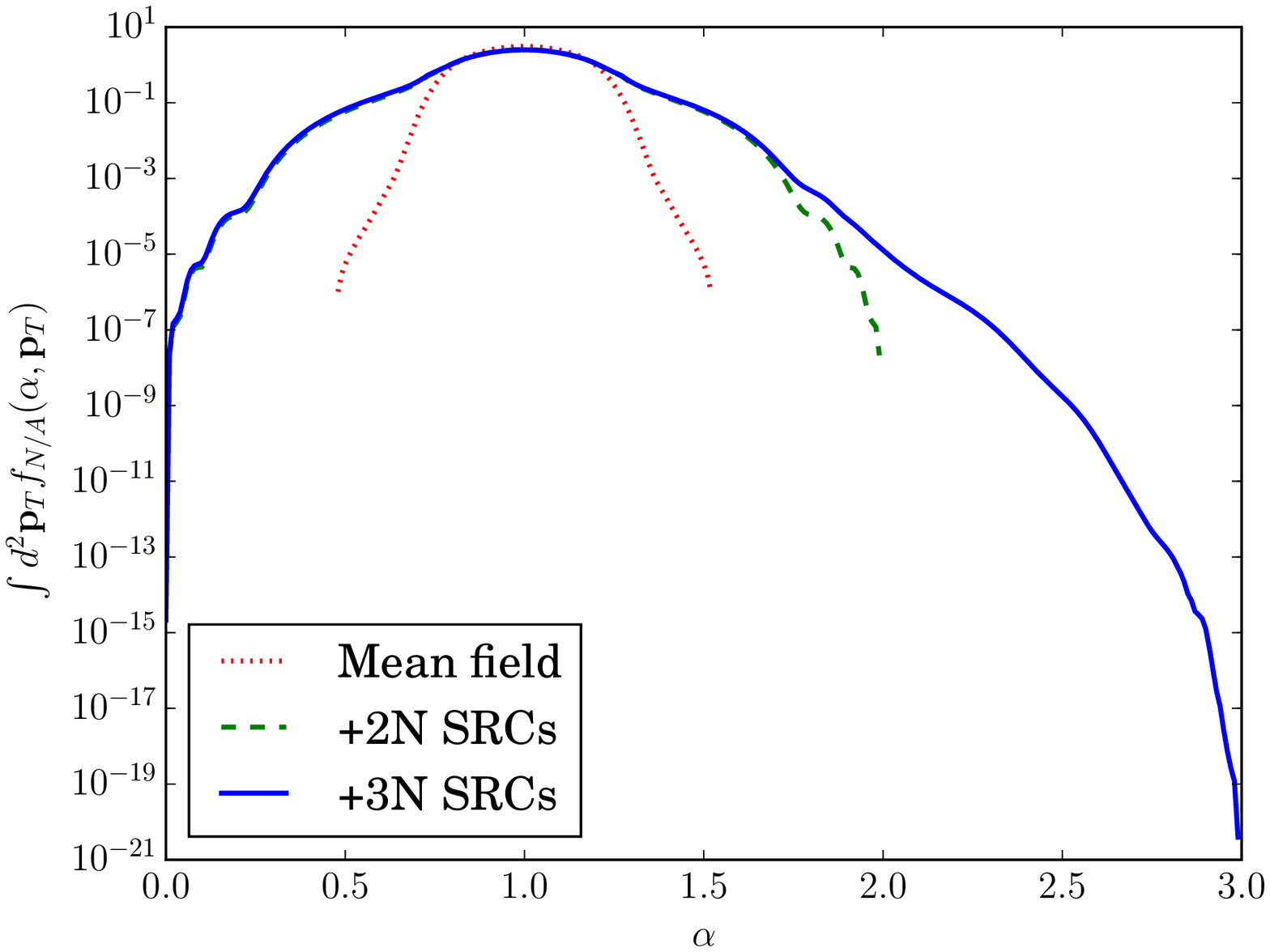}
    \caption{~}
    \label{fig:rho:all}
  \end{subfigure}
  \begin{subfigure}[t]{.45\textwidth}
    \centering
    \includegraphics[width=\textwidth]{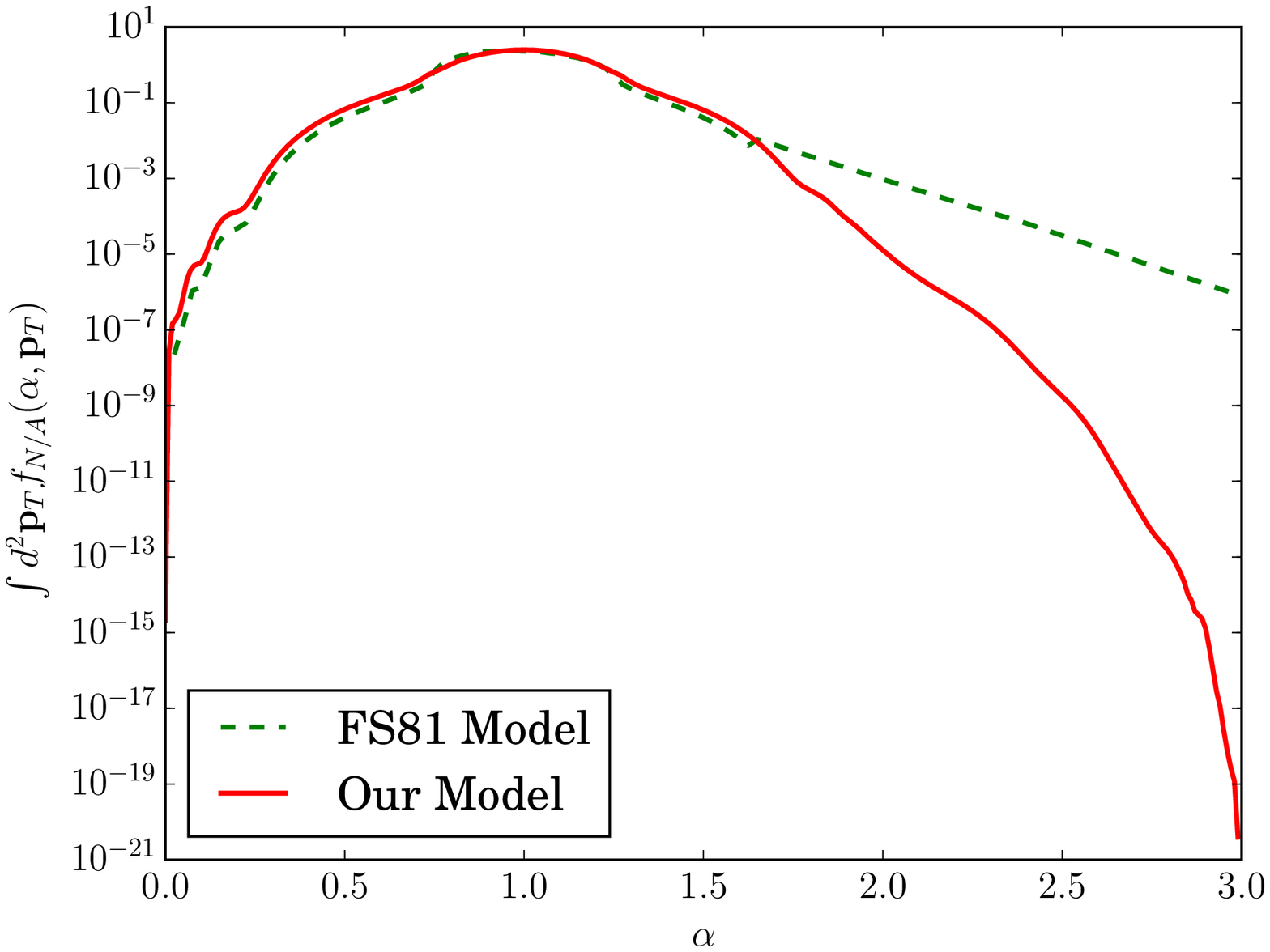}
    \caption{~}
    \label{fig:rho:fs}
  \end{subfigure}
  \caption{(Color online.) The $\alpha$ dependence of the nuclear light cone distribution for $^{208}$Pb.
    (a) Compares the mean field by itself to the light cone distribution with 2N and 3N SRCs.
    (b) Compares our model to the model in Chapter 5 of Ref.~\cite{Frankfurt:1981mk}.}
  \label{fig:rho}
\end{figure}

In lieu of experimental data for $a_3(A)$, we calculate $f_{N/A}^{(3)}(\alpha,\mathbf{p}_T)$ in this work using
Eq.~(\ref{eqn:rho3}), which requires knowing only the experimentally known quantity $a_2(A)$. 
In Fig.~\ref{fig:rho}, we present a calculation of the nuclear light cone distribution for $^{208}$Pb.
In Fig.~\ref{fig:rho:all}, we compare the light cone distribution using Eq.~(\ref{eqn:rhoA:decomp}),
considering only the mean field, and two-nucleon and three-nucleon correlations in addition.
As our calculations show, 3N SRCs dominate starting only at $\alpha \gtrsim 1.7$.

In Fig.~\ref{fig:rho:fs}, we have compared our model for the nuclear light cone density to a model
in Ref.~\cite{Frankfurt:1981mk}---in particular, in its Eq.~(5.11)---which we call the FS81 model.
In this model, the relative weight of $j\ge3$ SRCs was fixed based on a fit of
$p+A\rightarrow p+X$ data with 400~GeV protons\cite{Bayukov:1979vf}.
Since this model contains $j \ge 4$-nucleon SRCs,
it is expected to exceed our model, as can be seen in Fig.~\ref{fig:rho:fs}.

%%%%%%%%%%%%%%%%%%%%%%%%%%%%%%%%%%%%%%%%%%%%%%%%%%%%%%%%%%%%%%%%%%%%%%%%%%%%%%%%%%%%%%%%%%%%%%%%%%%%%%%%%%%%%%%%%%%%%%%%
%  EMC effect
%%%%%%%%%%%%%%%%%%%%%%%%%%%%%%%%%%%%%%%%%%%%%%%%%%%%%%%%%%%%%%%%%%%%%%%%%%%%%%%%%%%%%%%%%%%%%%%%%%%%%%%%%%%%%%%%%%%%%%%%

\subsection{Medium modifications}
\label{sec:pdf:emc}

Our next goal is to calculate PDFs for a bound nucleon,
$f^{(b)}_{i/N}\left(\frac{x}{\alpha},\alpha,\mathbf{p}_T,Q^2\right)$, 
which enter into Eq.~(\ref{eqn:pdf:convolution}).
In doing so we should take into account that the nuclear medium is strongly interacting,
and each nucleon has a high probability (20-30$\%$)
of being in a short-range correlation where the nucleons themselves overlap.
As a result, it should be expected that the parton distributions for nucleons immersed
in the nuclear medium should be modified in some way.

Deviations of the nuclear PDFs from expectations based on describing the nucleus as a system of unmodified nucleons
was first observed by the 
European Muon Collaboration\cite{Aubert:1983xm} in measurements of the ratio of cross sections
for deeply inelastic muon scattering for nuclear and deuteron targets,
{\sl i.e.} in
\begin{equation}
  \mathcal{R}_{\mathrm{EMC}}(x,Q^2) = \frac{2}{A} \frac{\sigma_{eA}}{\sigma_{ed}}
    \approx \frac{2}{A} \frac{F_2^{(A)}(x,Q^2)}{F_2^{(d)}(x,Q^2)}
  \label{eqn:REMC}
  .
\end{equation}
The ratio was mearued in DIS kinematics which
allowed the cross sections to be related to
the structure function $F_2(x,Q^2)$.
It was originally expected that, except for smearing from Fermi motion at high $x$, the structure functions
should just be the structure functions for free nucleons, and thus the ratio should be 1.
However, there is a dip below 1 in the range $0.3 < x <0.7$ that cannot be reproduced
by nucleonic motion alone\cite{Miller:2001tg,Smith:2002ci}.
This phenomenon is commonly referred to as the EMC effect.

This dip in the ratio $\mathcal{R}_{\mathrm{EMC}}$ is ubiquitous throughout nuclei,
and roughly saturates for large $A$.
Additionally, it is proportional to the local density of the nucleus considered\cite{Seely:2009gt},
suggesting that it is an effect of the nuclear medium.
Since the structure function $F_2(x,Q^2)$ is directly related to the parton distribution functions, through
\begin{equation}
  F_2(x,Q^2) = \sum_i e_i^2 x f_{i}(x,Q^2)
  \label{eqn:F2pdf}
  ,
\end{equation}
the EMC effect is commonly interpreted to be due to a modification of the PDF of a bound nucleon in the nuclear medium.
Even though there is a consensus about this,
the existing theoretical models based on modifications of the bound nucleon PDFs
differ due to different assumptions made about the nature of
nuclear medium effects
(for reviews of the EMC effect see~\cite{Norton:2003cb,Rith:2014tma}).
It was argued in Refs.~\cite{Frankfurt:1985cv,Kulagin:1994fz,Melnitchouk:1996vp}
that the EMC effect for the bound nucleon should be proportional
to the first approximation of its kinetic energy, or more precisely to the off-shellness of the bound nucleon.
This indicates that more modification should occur in the high momentum part of the nuclear wave function.

Since our goal is to study the effects of SRCs on the reaction (\ref{eqn:reaction}),
and since SRCs dominate the high momentum part of the nuclear wave function,
it will be necessary to account for the medium modifications that produce the EMC effect when studying this reaction.
In particular, the medium modification effects will be prominent in the nuclear parton distribution $f_{i/A}(x,Q^2)$
present in Eq.~(\ref{eqn:diffcrx}).
Numerical estimates of the EMC effect will be based on the
color screening model\cite{Frankfurt:1985cv,Frankfurt:1988nt},
which satisfactorily describes the phenomenology of the EMC effect observed in inclusive DIS reactions.
(See also Ref.~\cite{Frank:1995pv}.)

\begin{figure}
  \centering
  \begin{subfigure}[t]{.45\textwidth}
    \centering
    \includegraphics[width=\textwidth]{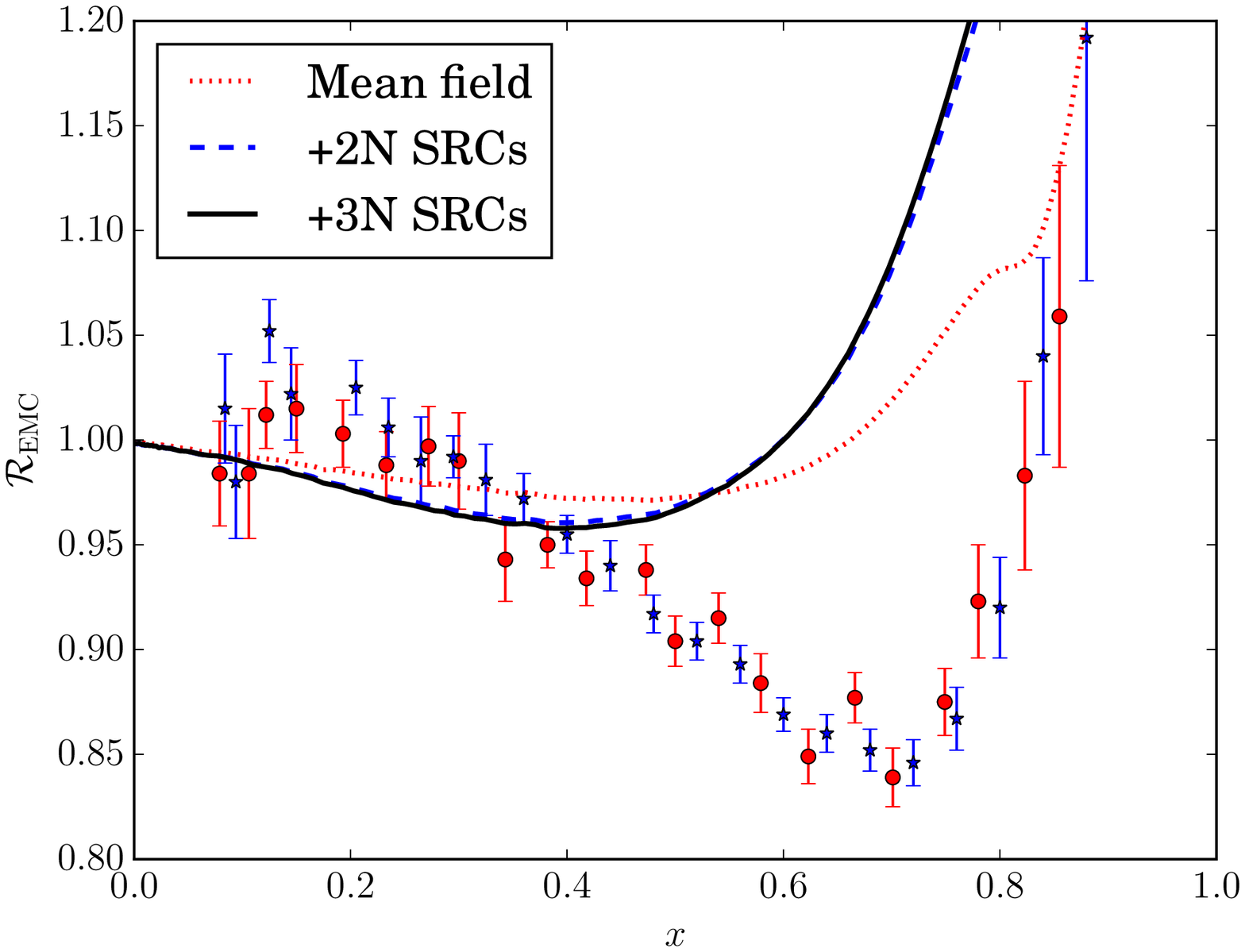}
    \caption{EMC ratio for $^{56}$Fe}
  \end{subfigure}
  \begin{subfigure}[t]{.45\textwidth}
    \centering
    \includegraphics[width=\textwidth]{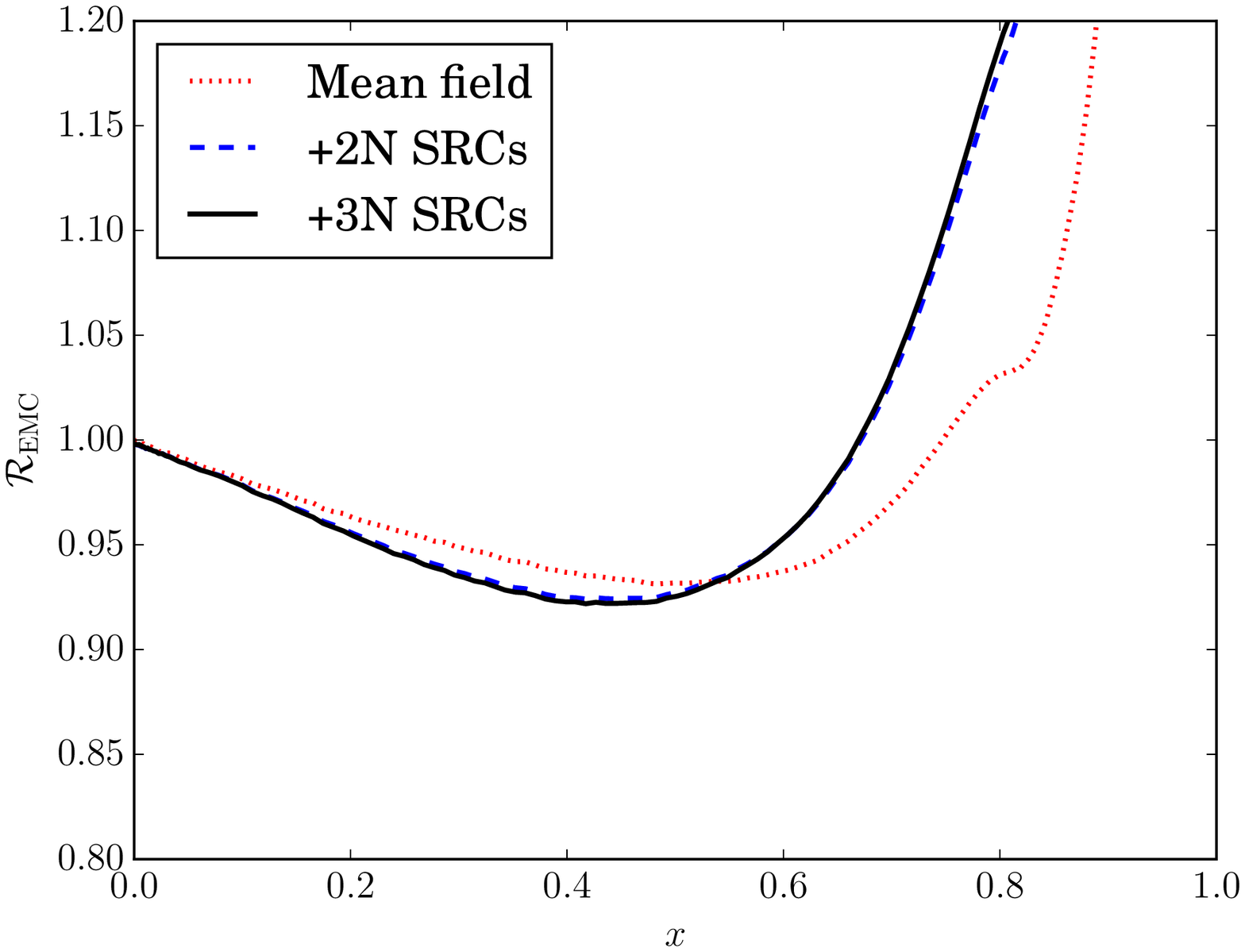}
    \caption{EMC ratio for $^{208}$Pb}
  \end{subfigure}
  \caption{
    (Color online.)
    EMC ratios at $Q^2=10$ GeV$^2$, without medium modifications.
    Data are from SLAC experiments:
    Stars (blue) are from \cite{Gomez:1993ri};
    circles (red) are from \cite{Arnold:1983mw}.
    The nucleon structure functions are computed with the Bodek-Ritchie
    parametrization\cite{Bodek:1979rx,Bodek:1980ar,Bodek:1981wr}.
  }
  \label{fig:EMC:unmod}
\end{figure}

Before proceeding with a specific medium modification model,
we present estimates of the ratio given in Eq.~(\ref{eqn:REMC}) in the absence of medium modifications
in order to assess the extent of the modifications that must occur.
Here, $F_2^{(A)}(x,Q^2)$ is calculated using
\begin{equation}
  F_2^{(A)}(x,Q^2) = \sum_N \int_x^A d\alpha \int d^2\mathbf{p}_T f_{N/A}(\alpha,\mathbf{p}_T)
    F^{(N,b)}_2\left(\frac{x}{\alpha},\alpha,\mathbf{p}_T,Q^2\right)
  \label{eqn:F2:convolution}
  ,
\end{equation}
which follows from Eq.~(\ref{eqn:F2pdf}) and the PDF convolution formula Eq.~(\ref{eqn:pdf:convolution}),
and a phenomenological parametrization for free nucleons is used to estimate
$F^{(N,b)}_2\left(\frac{x}{\alpha},\alpha,\mathbf{p}_T,Q^2\right)$\footnote{
  Hereafter, for $F^{N}_2\left(\frac{x}{\alpha},Q^2\right)$ of free nucleons we use phenomenological 
    parametrization of Refs.~\cite{Bodek:1979rx,Bodek:1980ar,Bodek:1981wr}.
}.
It should be noted that, in the absence of medium modifications,
the nucleonic structure function $F_2^{(N,\mathrm{free})}\left(\frac{x}{\alpha},Q^2\right)$
is not a function of $\alpha$ (except through $x_N=\frac{x}{\alpha}$) or $\mathbf{p}_T$.
The nuclear light cone distribution $f_{N/A}(\alpha,\mathbf{p}_T)$ in Eq.~(\ref{eqn:F2:convolution}) is calculated
using the formalism of Sec.~\ref{sec:pdf:lcd}.
As Fig.~\ref{fig:EMC:unmod} shows, there is a significant discrepancy between the calculation and the data
in the SRC-dominated region $x > 0.6$,
indicating that nucleons in SRCs are more strongly modified by the nuclear medium than nucleons in the mean field.
This is consistent with our expectation that modification should increase with nucleon virtuality,
since SRCs occur at significantly larger momenta.

\subsubsection{Correctly defining x}
\label{sec:xhat}

Up until now, data from DIS experiments have been plotted against the kinematic Bjorken scaling variable
$x_N = \frac{Q^2}{2m_p\nu}$, where $m_p$ is the mass of the proton.
However, the Bjorken scaling variable that enters into the dynamics of QCD,
including the convolution formulas of Eqs.~(\ref{eqn:pdf:convolution},\ref{eqn:F2:convolution}),
is instead $x_A = \frac{AQ^2}{2M_A\nu}$.
Thus, it is as arguments of $x_A$ and $x_d$ rather than of $x_N$ that the ratio of Eq.~(\ref{eqn:REMC})
should be presented in order to see dynamical QCD effects.
By contrast, presenting data as a function of $x_N$ instead of $x_{A/d}$ artificially shifts
the arguments of the nuclear and deuteron structure functions in Eq.~(\ref{eqn:REMC}) by different amounts,
since $x_A = \frac{Am_p}{M_A}x_N$ and $x_d = \frac{2m_p}{M_d}x_N$.
Since the deuteron is a loosely bound system with little binding energy,
the arguments for the nuclear and deuteron structure functions are shifted by substantially different amounts
when the nucleus has a high binding energy.

\begin{figure}
  \centering
  \begin{subfigure}[t]{.45\textwidth}
    \centering
    \includegraphics[width=\textwidth]{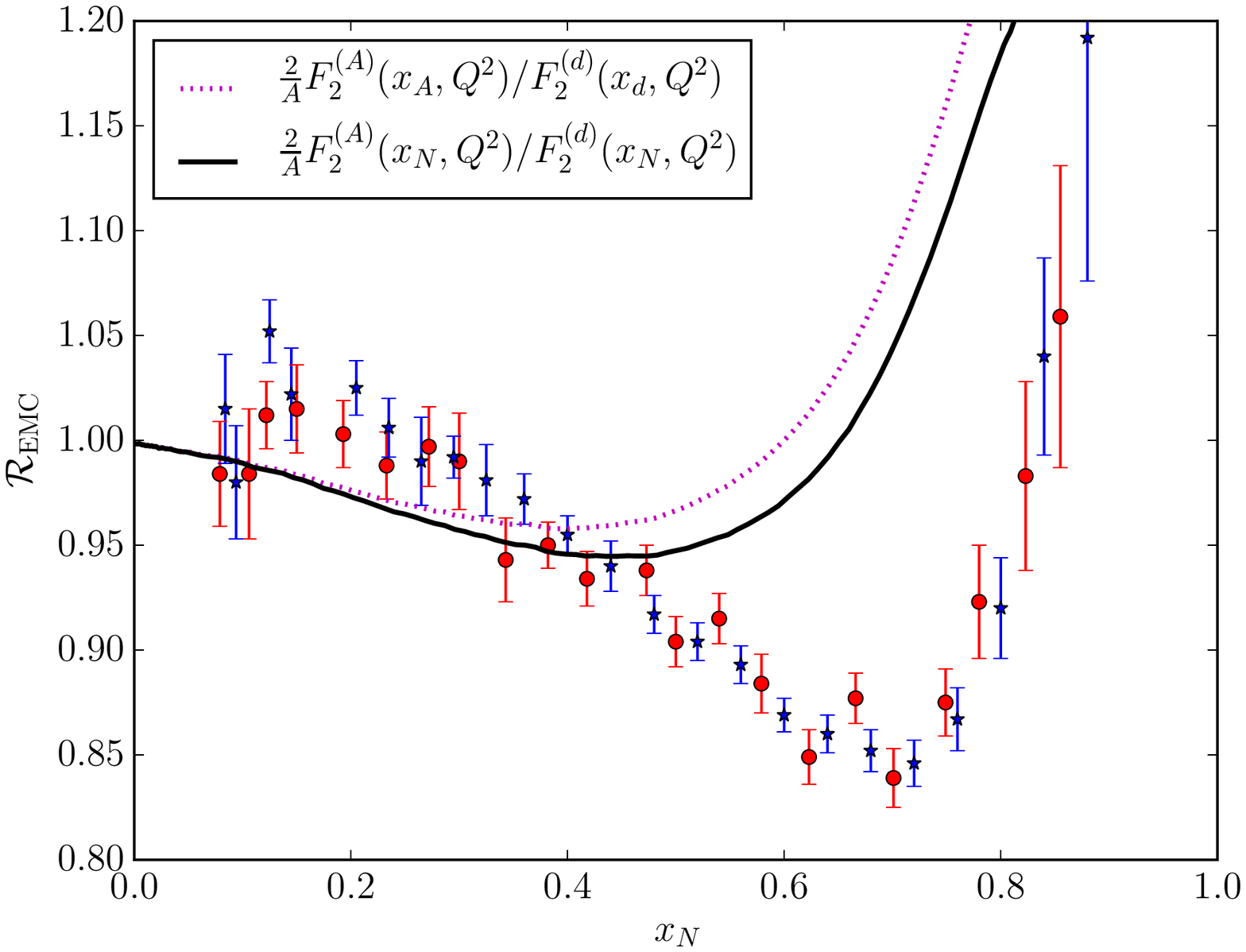}
    \caption{EMC ratio for $^{56}$Fe}
  \end{subfigure}
  \begin{subfigure}[t]{.45\textwidth}
    \centering
    \includegraphics[width=\textwidth]{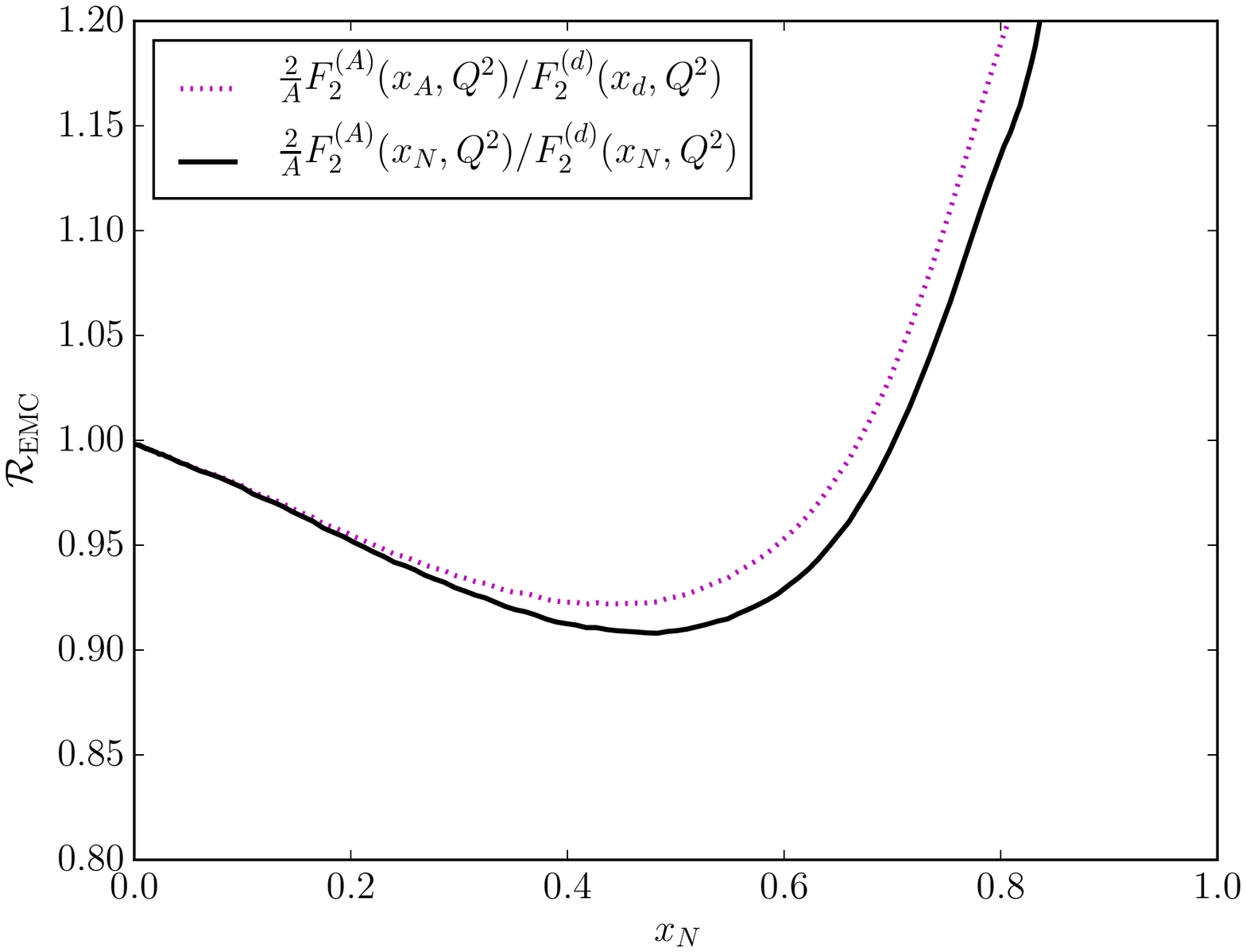}
    \caption{EMC ratio for $^{208}$Pb}
  \end{subfigure}
  \caption{
    (Color online.)
    EMC ratios at $Q^2=10$ GeV$^2$, without medium modifications.
    2N and 3N correlations are included in theory curves.
    The dotted (magenta) lines compare structure functions as
    functions of the dynamical QCD scaling variables $x_A$ and $x_d$,
    while solid (black) lines compare structure functions as functions of
    the kinematical variable $x_N$.
    Data and nucleonic structure function parametrizations are as in Fig.~\ref{fig:EMC:unmod}.
  }
  \label{fig:EMC:xshift}
\end{figure}

It has been observed\cite{Frankfurt:2012qs,Honkanen:2013goa} that presenting structure function ratios
as an argument of $x_N$ rather than $x_A$ and $x_d$ introduces an artificial dip in the ratio for $x_N\gtrsim0.5$.
In Fig.~\ref{fig:EMC:xshift}, we present theoretical calculations for
$\mathcal{R}_{\mathrm{EMC}}(x_A,x_d)$ (as a function of the dynamical QCD scaling variables $x_{A/d}$)
and $\mathcal{R}_{\mathrm{EMC}}(x_N)$ (as a function of the kinematical Bjorken scaling variable $x_N$)
for both $^{56}$Fe and $^{208}$Pb.
In these plots, the use of $x_N$ as a variable produces an artificial dip, which partially explains
the EMC effect up to $x_N\sim 0.5$.
However, there continues to be a discrepancy between theory and data, which requires medium modification to explain.

%%%%%%%%%%%%%%%%%%%%%%%%%%%%%%%%%%%%%%%%%%%%%%%%%%%%%%%%%%%%%%%%%%%%%%%%%%%%%%%%%%%%%%%%%%%%%%%%%%%%%%%%%%%%%%%%%%%%%%%%
%  Color screening
%%%%%%%%%%%%%%%%%%%%%%%%%%%%%%%%%%%%%%%%%%%%%%%%%%%%%%%%%%%%%%%%%%%%%%%%%%%%%%%%%%%%%%%%%%%%%%%%%%%%%%%%%%%%%%%%%%%%%%%%

\subsubsection{Color screening model of the EMC effect}

Any nucleon can be expected to spend some of its time in a point-like configuration (PLC),
in which its constituent quarks are compressed into a volume
that is much smaller than the average volume of the nucleon.
A PLC is largely invisible to the color force,
in analogy to neutrally charged atoms in a gas whose van der Waals forces become weaker if the atoms are compressed.
Due to the color neutrality of the nucleon,
any color interaction between nucleons owes to higher moments (dipole, quadrupole, {\sl etc.}),
which decrease with distance between the color-charged constituents.
Moreover, it can be shown by the renormalizability of QCD that meson exchange between nucleons,
one of which is in a PLC, is suppressed\cite{Frankfurt:1985cv}.

Since nucleons in an average-sized configuration (ASC) and a PLC will interact differently,
the probability that the nucleon can be found in either configuration should be modified
by the immersion of a nucleon in the nuclear medium.
In particular, PLCs are expected to be suppressed compared to ASCs since the bound nucleon will assume a configuration
that maximizes the binding energy and brings the nucleus to a lower-energy ground state.

The change in probability can be estimated using non-relativistic perturbation theory,
as has been done in Refs.~\cite{Frankfurt:1985cv,Frankfurt:1988nt}.
Since the PLC components in the nucleon PDFs are expected to dominate in the $x\ge 0.6$ region,
it was found that in this region the bound nucleon PDFs should be modified by a factor $\delta_A(k^2,x_N)$,
which depends on the nucleon momentum (or virtuality) as
\begin{align}
  \delta_A(k^2,x_N\gtrsim0.6) &= \frac{1}{(1+z)^2} \\
  z &= \frac{\frac{k^2}{m_p} + 2\epsilon_A}{\Delta E_A}
  \label{eqn:plc:z}
  ,
\end{align}
where $\epsilon_A$ is the nuclear biding energy and $\Delta E_A$ characterizes the 
nucleon excitation energy in the medium.
In analogy with electric charge screening, this is called the color screening model of the EMC effect.

We shall use the color screening model as an example for accounting for medium modifications 
in calculation of the dijet cross section.
For this, one should first fix the parameter $\Delta E_A$ as well as extrapolate the suppression factor 
$\delta$ to the whole range of $x$.
To do so we implement the color screening model in calculating the nuclear structure function $F^{(A)}_2(x,Q^2)$
and compare it with the available lepton DIS data.
Such an implementation is done by substituting 
\begin{equation}
  F^{(N,b)}_2\left(\frac{x}{\alpha},\alpha,\mathbf{p}_T,Q^2\right)
    = F^{(N)}_2\left(\frac{x}{\alpha},Q^2\right)\delta\left(k^2(\alpha,\mathbf{p}_T),\frac{x}{\alpha}\right)
  \label{eqn:F2:eff}
\end{equation}
in the convolution formula of Eq.~(\ref{eqn:F2:convolution}) and using the phenomenological parameterization
mentioned above\cite{Bodek:1979rx,Bodek:1980ar,Bodek:1981wr} for the free nucleon structure function $F^{(N)}_2(x,Q^2)$.
It is worth emphasizing that this phenomenological parametrization
also contains contributions from higher-twist effects,
so Eq.~(\ref{eqn:F2:eff}) implies that higher-twist effects are modified the same way as partonic distributions.
Justification for such an approximation follows from the fact that no strong $Q^2$ dependence
has been observed for the experimentally measured range
$2$~GeV$^2< Q^2 < 200$~GeV$^2$, though errors for large $Q^2$ are fairly large\cite{Norton:2003cb,Arrington:2003nt}.

In comparing the color screening predictions of Eq.~(\ref{eqn:F2:convolution}) with the data,
we assumed that no medium modification occurs at $x_N\lesssim0.45$,
and the region between $0.45 < x_N < 0.6$ is interpolated linearly\footnote{
  We neglect here a small effect of enhancement of the bound nucleon PDF at smaller $x$,
  which is implied by the baryon charge sum rule.}:
\begin{align}
  \delta_A(k^2,0.45<x_N<0.6) = 1 + \frac{x_N-0.45}{0.15}\left\{\delta_A(k^2,x_N\gtrsim0.6)-1\right\}
  .
\end{align}

\begin{figure}
  \centering
  \begin{subfigure}[t]{.45\textwidth}
    \centering
    \includegraphics[width=\textwidth]{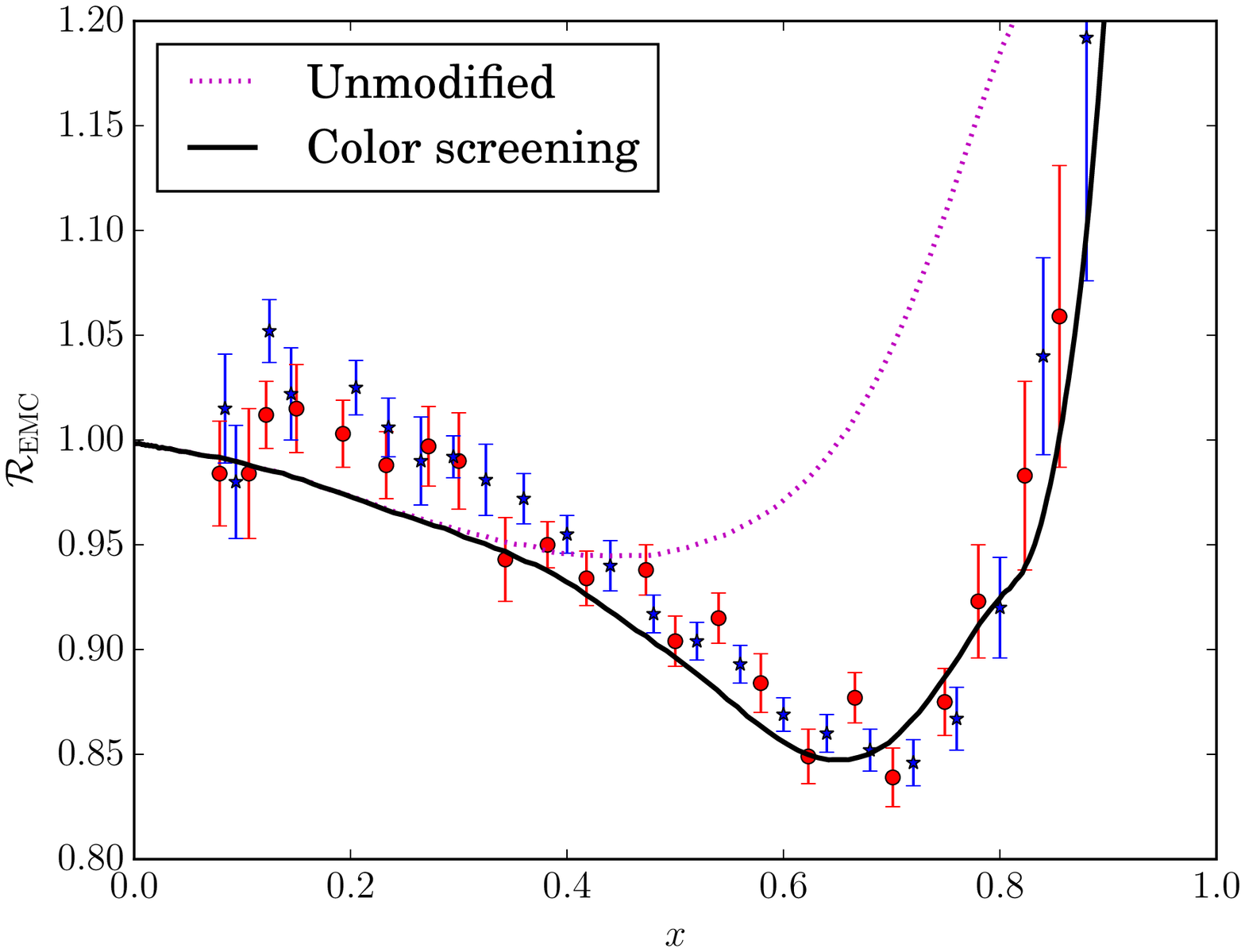}
    \caption{EMC ratio for $^{56}$Fe}
  \end{subfigure}
  \begin{subfigure}[t]{.45\textwidth}
    \centering
    \includegraphics[width=\textwidth]{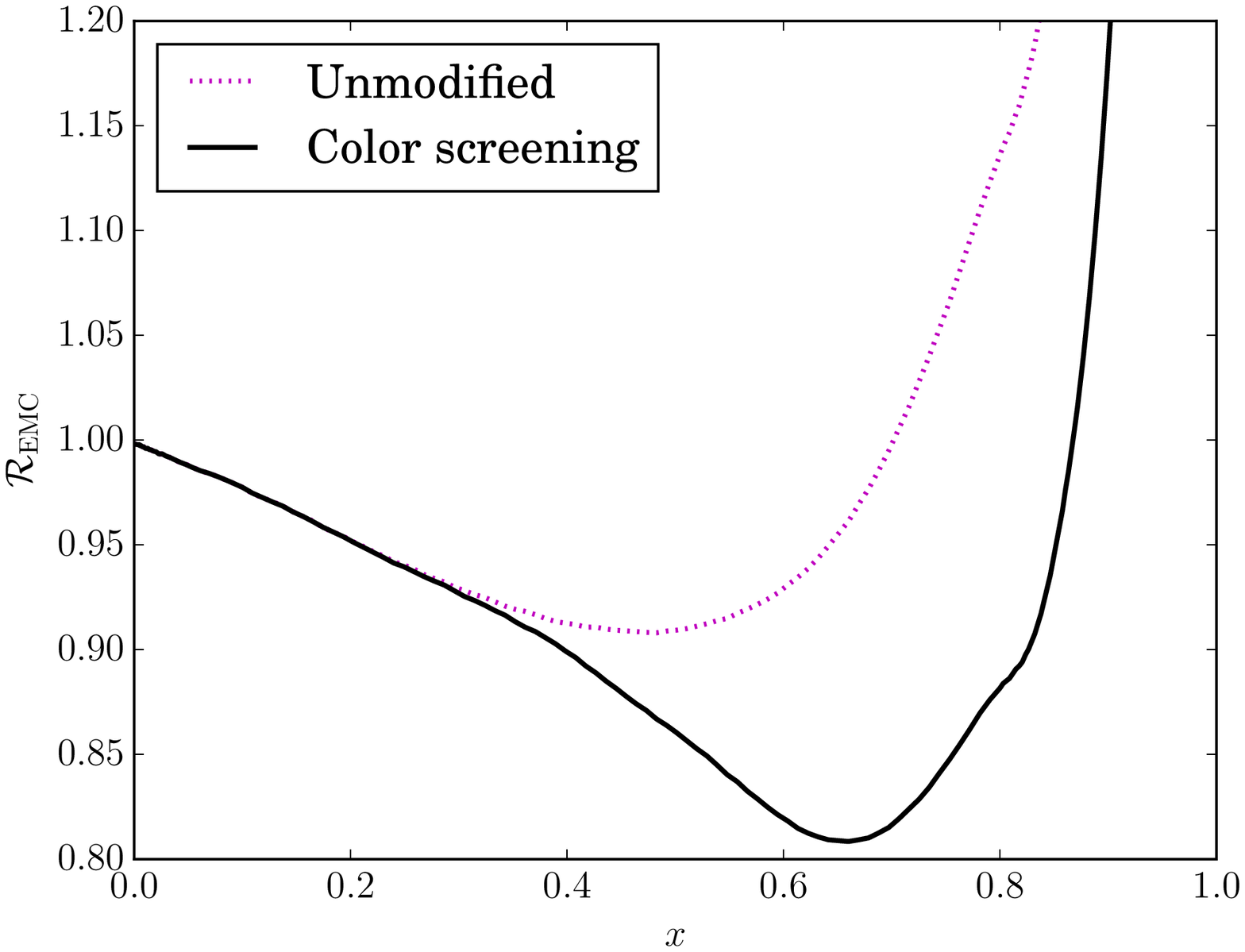}
    \caption{EMC ratio for $^{208}$Pb}
  \end{subfigure}
  \caption{
    (Color online.)
    EMC ratios with and without the color screening model of medium modifications.
    2N and 3N correlations are included.
    $Q^2=10$~GeV$^2$, and
    data and nucleonic structure function parametrizations are as in Fig.~\ref{fig:EMC:unmod}.
  }
  \label{fig:EMC:clrscn}
\end{figure}

Since the nucleon excitation energy $\Delta E_A$ is a dynamical parameter depending on
the spin-isopsin composition of interacting nucleons,
we fitted it separately for the mean field and SRC domains of nucleon momenta.
For a nucleon in the mean field of a heavy nucleus,
we expect the excitation energy $\Delta E_A$ to be in the range $300$-$500$~MeV, 
namely between the excitation energies of a $\Delta$ and an $N^*$ resonance.
The best fit to data appears to be $\Delta E_A \approx 500$~MeV.
For the deuteron, as well as for a nucleon in a 2N SRC, we expect the lowest excited state
to be a $\Delta\Delta$ configuration, giving $\Delta E_d \approx 600$ MeV.
We assumed the same $\Delta E = 600$~MeV excitation energy for a nucleon in a 3N SRC,
since such a correlation in our approach is generated through a sequence of 2N correlations.
We take $\epsilon_A$ to be the binding energy per nucleon of the nucleus under consideration when in the mean field
and neglect it for a nucleon in an SRC, since it is expected to be negligible compared to $k^2/m$ in an SRC.

The formula for $k$ in terms of $\alpha$ and $\mathbf{p}_T$ also depends on whether the nucleon is in the mean field
or a short range correlation.
The non-relativistic approximation is adequate for describing a nucleon in the mean field,
and for two-nucleon SRCs Eq.~(\ref{eqn:k}) is valid.
For three-nucleon SRCs, 
\begin{equation}
  k^2 = 2\frac{(\alpha-1)^2m^2+\mathbf{p}_T^2}{\alpha(3-\alpha)}
\end{equation}
is a reasonable approximation if the relative momentum between the spectator nucleons is small.

As can be seen in Fig.~\ref{fig:EMC:clrscn},
using the parameters described above results in reasonably good description of the EMC effect.
After fixing the $\delta$ factor of Eq.~(\ref{eqn:plc:z}) and its $x$-dependence,
the color screening model is now applied to quark PDFs rather than structure functions in the following form:
\begin{equation}
  f_{i/A}(x,Q^2) = \sum_{j=1}^3 \sum_N \int_x^A \frac{d\alpha}{\alpha} d^2\mathbf{p}_T
    f_{N/A}^{(j)}(\alpha,\mathbf{p}_T) f_{i/N}\left(\frac{x}{\alpha},Q^2\right)
    \delta_A^{(j)}\left(k^2(\alpha,\mathbf{p}_T),\frac{x}{\alpha}\right)
  \label{eqn:pdf:clrscn}
  .
\end{equation}
The sum over $j$ here indicates the sum over the mean field ($j=1$) and $j$-nucleon SRC ($j=2,3$) parts
of the light cone fraction distribution.
$\delta_A^{(j)}\left(k^2(\alpha,\mathbf{p}_T),\frac{x}{\alpha}\right)$
has an index $j$ here to indicate that the color screening model affects the mean field and SRCs differently.

\begin{figure}
  \centering
  \includegraphics[scale=0.5]{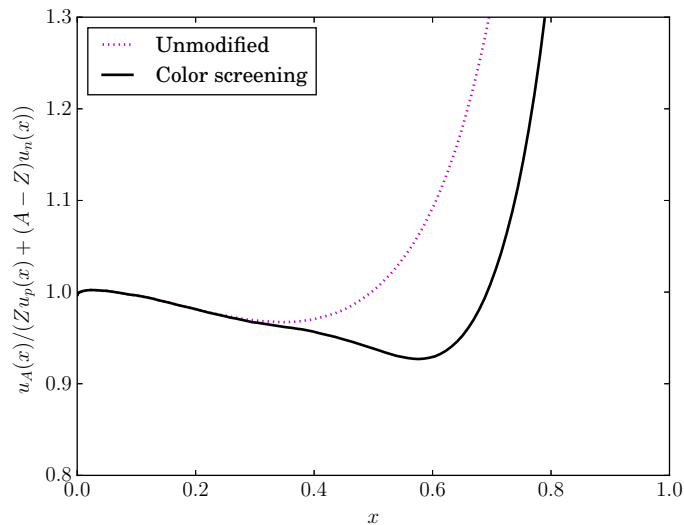}
  \caption{
    (Color online.)
    Isospin-weighted ratio of the $u$ quark distribution for $^{208}$Pb and the free nucleon,
    both with and without the color screening model.
    2N and 3N correlations are included in the $^{208}$Pb distribution.
    Curves are at $Q^2 = 10$~GeV$^2$.
    PDF parametrization is CT10\cite{Gao:2013xoa}.
  }
  \label{fig:pdf:clrscn}
\end{figure}

To estimate the medium modification for gluon distributions we notice that the medium modifications
occur predominantly at large $x$, where valence quarks dominate. 
Gluons in this region would originate as radiation from quarks,
and would therefore inherit medium modifications from the quarks.
For this reason, we apply Eq.~(\ref{eqn:pdf:clrscn}) to gluons as well as quarks and anti-quarks\footnote{
  Anti-quark and gluon contributions are, however, negligible in the region we are considering.
}.
The result of applying the color screening model to the nuclear PDF is demonstrated in Fig.~\ref{fig:pdf:clrscn}.

%%%%%%%%%%%%%%%%%%%%%%%%%%%%%%%%%%%%%%%%%%%%%%%%%%%%%%%%%%%%%%%%%%%%%%%%%%%%%%%%%%%%%%%%%%%%%%%%%%%%%%%%%%%%%%%%%%%%%%%%
%  DGLAP evolution
%%%%%%%%%%%%%%%%%%%%%%%%%%%%%%%%%%%%%%%%%%%%%%%%%%%%%%%%%%%%%%%%%%%%%%%%%%%%%%%%%%%%%%%%%%%%%%%%%%%%%%%%%%%%%%%%%%%%%%%%

\subsubsection{Evolution of medium modifications}

The EMC effect has so far been observed in a restricted range of $Q^2$,
around $Q^2\sim 2$-$200$~GeV$^2$, with the greatest modification at $x \gtrsim 0.6$.
The experimental errors, especially at large $Q^2$,
do not allow the identification of any appreciable $Q^2$ dependence of the EMC ratio.
However, from the point of view of QCD,
one should expect at least logarithmic $Q^2$ dependence,
owing to the evolution of high-$x$, low-$Q^2$ partons into lower $x$ with an increase of $Q^2$.
This will be the case with evolution to the high $Q^2 \sim 10^4$~GeV$^2$ relevant to the LHC.
In our estimates of medium modification in dijet production,
we predict how the EMC effect will manifest at $Q^2=10^4$~GeV$^2$
using DGLAP evolution\cite{Dokshitzer:1977sg,Gribov:1972ri,Altarelli:1977zs}.

It is necessary when considering medium modifications
to use DGLAP evolution to evolve the nuclear PDFs obtained at low $Q^2$ to high $Q^2$,
rather than to apply the medium-modified convolution formula Eq.~(\ref{eqn:pdf:clrscn}) to a high-$Q^2$ nucleonic PDF.
The factors $\delta_A^{(j)}(k^2,x_N)$ in the color screening model, for instance, contain $x$ dependence in the form
of a linear interpolation, preventing this term from being factored out of a DGLAP integral.
The $x$ dependence in this factor is particular to the model's purpose of explaining DIS cross section ratios
at low $Q^2$, so cannot {\sl a priori} be expected to have the same $x$ dependence at large $Q^2$.
Eq.~(\ref{eqn:pdf:clrscn}) cannot be directly applied at high $Q^2$ without evolving $\delta_A^{(j)}(k^2,x_N)$,
which has unknown $Q^2$ dependence.
However, by applying DGLAP evolution to the medium-modified nuclear PDF as a whole,
which is obtained from Eq.~(\ref{eqn:pdf:clrscn}) at low $Q^2$,
the correct $Q^2$ dependence of medium modification can be obtained.

\begin{figure}
  \centering
  \begin{subfigure}[t]{.45\textwidth}
    \centering
    \includegraphics[width=\textwidth]{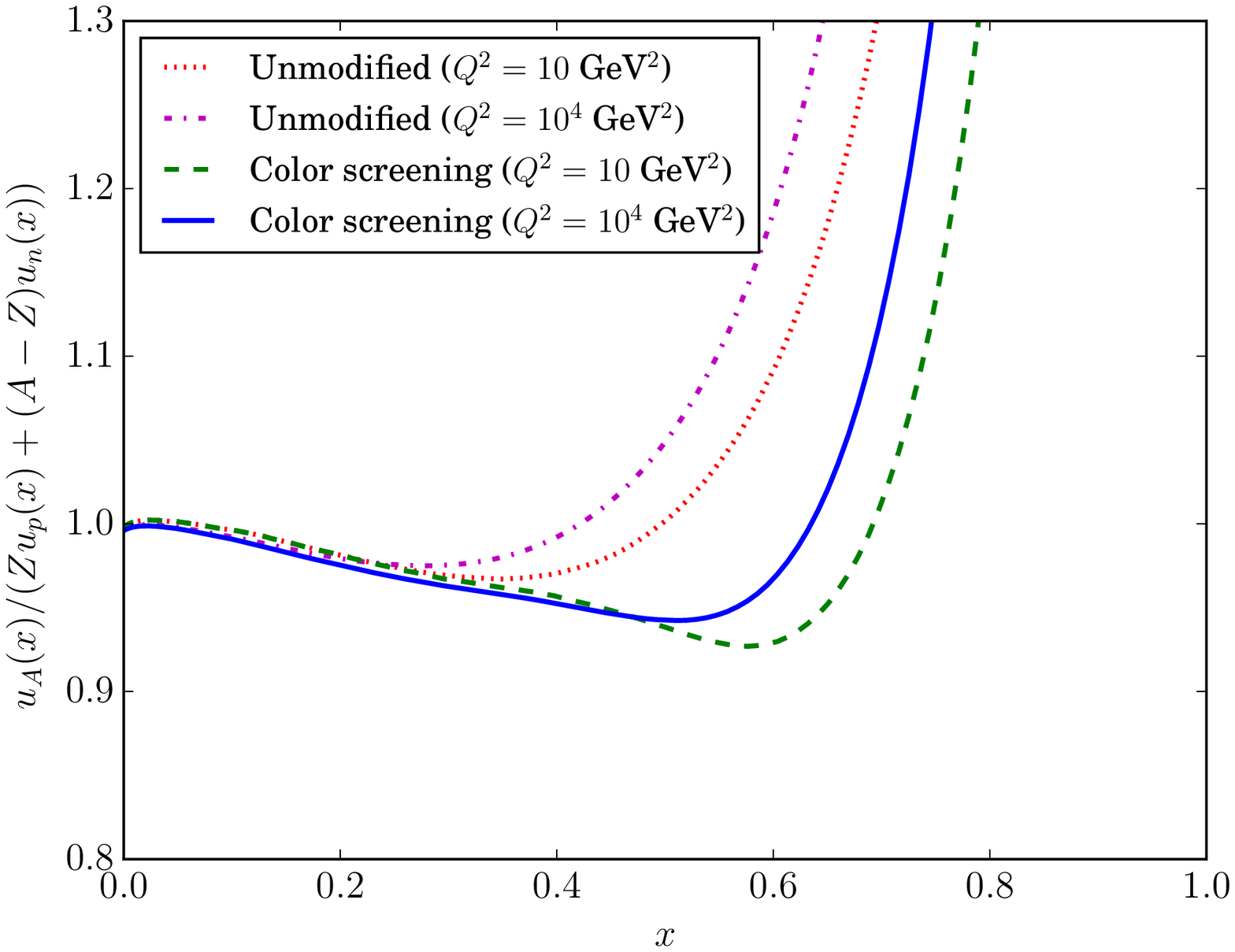}
    \caption{~}
    \label{fig:pdf:evolved:x}
  \end{subfigure}
  \begin{subfigure}[t]{.45\textwidth}
    \centering
    \includegraphics[width=\textwidth]{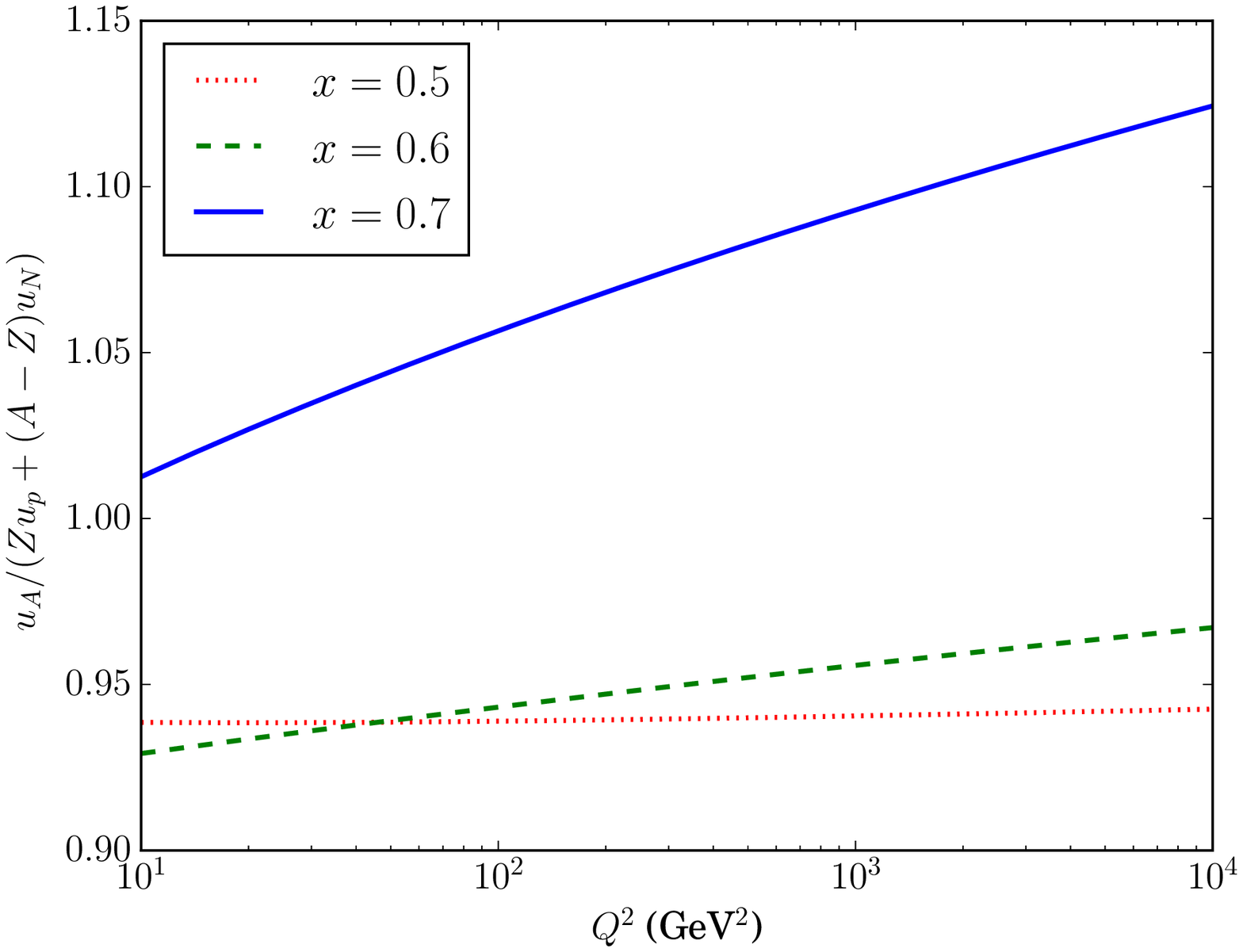}
    \caption{~}
    \label{fig:pdf:evolved:Q}
  \end{subfigure}
  \caption{
    (Color online.)
    Isospin-weighted ratio of $u$ quark distribution for $^{208}$Pb to that of the free nucleon.
    2N and 3N correlations are included in the $^{208}$Pb distribution.
    (a) As a function of $x$.
    Unevolved lines are at $Q^2=10$~GeV$^2$, and evolved at $Q^2=10000$~GeV$^2$.
    (b) As a function of $Q^2$, with the EMC effect accounted for by the color screening model.
    The $Q^2=10$~GeV$^2$ PDFs are computed using the CT10 parametrization\cite{Gao:2013xoa},
    as well as the convolution formula Eq.~(\ref{eqn:pdf:convolution}).
%    The evolved curves are computed from this using DGLAP evolution.
  }
  \label{fig:pdf:evolved}
\end{figure}

The standard DGLAP formula is modified to account for the possibility of $x>1$ inherent in dealing with nuclei;
in particular, we have
\begin{equation}
  \frac{\partial f_{i/A}(x,Q^2)}{\partial \log(Q^2)} = \frac{\alpha_s(Q^2)}{2\pi}
    \sum_{ij} \int_x^A \frac{dy}{y} P_{ij}\left(\frac{x}{y}\right) f_{j/A}(y,Q^2)
  \label{eqn:dglap}
  .
\end{equation}
Here, $P_{ij}$ are the Alterelli-Parisi splitting functions\cite{Altarelli:1977zs}.
In our calculation, we solved Eq.~(\ref{eqn:dglap}) numerically within the leading order approximation,
using the nuclear PDFs obtained in the previous sections as input.
(Next-to-leading order corrections to the evolution of medium modifications are expected to be negligible
at the high $x$ being considered here.)
The results of this calculation are shown in Fig.~\ref{fig:pdf:evolved},
which considers isospin-weighted ratios of the nuclear $u$ quark distribution
to the free nucleonic $u$ distribution, since the effects of evolution are most easily seen in ratios.
Fig.~\ref{fig:pdf:evolved:x} in particular
compares the effects of evolution on nuclear PDFs with and without medium modification accounted for.
Fig.~\ref{fig:pdf:evolved:Q} shows how the modified PDFs, as functions of $Q^2$,
evolve at several fixed values of $x$.
At $x\sim 0.5$, where the PDF ratio mostly deviates from $1$ due to medium modifications,
the ratio is roughly constant as $Q^2$ increases, suggesting little evolution of the medium modification itself.
On the other hand, the PDF ratio at $x\sim0.7$ is more strongly $Q^2$-dependent,
showing that at high $Q^2$, the enhancement of the nuclear PDF relative to the nucleonic PDF
due to Fermi motion and SRCs becomes more important at lower $x$---a
fact that can also be seen in Fig.~\ref{fig:pdf:evolved:x},
where the steep increase in the ratio occurs at a lower $x$ when $Q^2$ is increased.

The high-$x$ enhancement of nuclear PDFs moves to the left in Fig.~\ref{fig:pdf:evolved:x} because,
as a parton evolves, it radiates other partons which carry off part of its forward light cone momentum,
causing its light cone momentum fraction $x$ to decrease as $Q^2$ increases.
Accordingly, a quark with a somewhat moderate value of $x$ which is probed at the high $Q^2$ characteristic of the LHC
could have originated within a ``primordial,'' low-$Q^2$ quark with a higher momentum fraction $x$.
The extent to which lower-$x$, higher-$Q^2$ partons can be considered as originating within higher-$x$, lower-$Q^2$
partons is quantified by the QCD evolution trajectory, which is detailed in Ref.~\cite{Frankfurt:2011cs}.

The QCD evolution trajectory starts at a point $(x_0,Q^2_0)$ in $x$-$Q^2$ parameter space,
and describes the evolution of a particular parton flavor $i$.
Another point $(x,Q^2)$ lies on the evolution trajectory if half of the PDF
$f_{i/A}(x,Q^2)$ originated from evolution of the portion of the $Q^2=Q_0^2$ PDF at $x \geq x_0$, {\sl i.e.}~if,
given $\tilde{f}_{i/A}(x,Q^2;x_0,Q_0^2)$ defined by
\begin{align}
  \frac{\partial \tilde{f}_{i/A}(x,Q^2;x_0,Q^2_0)}{\partial \log(Q^2)} &= \frac{\alpha_s(Q^2)}{2\pi}
    \sum_{ij} \int_{x_0}^A \frac{dy}{y} P_{ij}\left(\frac{x}{y}\right) \tilde{f}_{j/A}(y,Q^2;x_0,Q^2_0)
  \label{eqn:dglap:x0}
  \\
  \tilde{f}_{i/A}(x,Q_0^2;x_0,Q^2_0) &= f_{i/A}(x,Q_0^2) \Theta(x-x_0)
  \label{eqn:dglap:x0:init}
  ,
\end{align}
then the $x$ for a given $Q^2$ on the trajectory is given by the requirement
\begin{equation}
  \tilde{f}_{i/A}(x,Q^2;x_0,Q_0^2) = \frac{1}{2}f(x,Q^2)
  \label{eqn:trajectory}
  .
\end{equation}
Note that Eq.~(\ref{eqn:dglap:x0:init}) defines an initial condition, which ensures that when its argument $x < x_0$,
the function $\tilde{f}_{i/A}(x,Q^2;x_0,Q_0^2)$ comes from the evolution of partons with $x \geq x_0$,
and thus that Eq.~(\ref{eqn:trajectory}) has the required interpretation.

\begin{figure}
  \centering
        \includegraphics[scale=0.45]{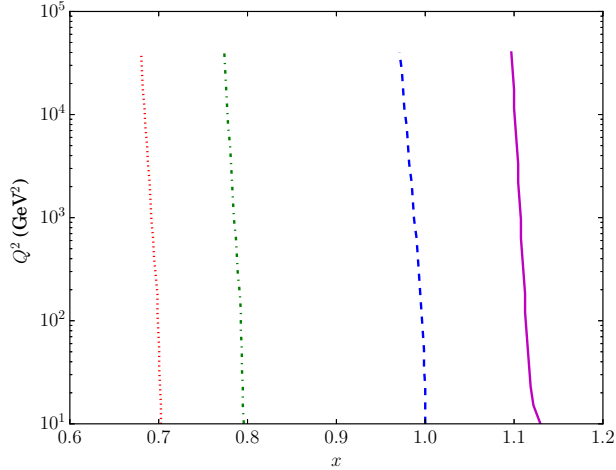}
  \caption{Evolution trajectories for $^{208}$Pb,
    from $Q^2 = 10$~GeV$^2$ to $Q^2 = 40000$~GeV$^2$.
    2N and 3N SRCs as well as color screening are considered.
    The $Q^2=10$~GeV$^2$ PDFs are computed using the CT10 parametrization\cite{Gao:2013xoa},
    and Eq.~(\ref{eqn:pdf:convolution}).}
  \label{fig:trajectory}
\end{figure}

With these considerations in mind, several QCD evolution trajectories for the nuclear $u$ quark distribution
in $^{208}$Pb are presented in Fig.~\ref{fig:trajectory}.
Consistent with the findings of Ref.~\cite{Frankfurt:2011cs},
the trajectories do not move very far to the left for large $x_0$,
so there is little hope of seeing primordially superfast ($x > 1$) quarks at moderate $x$ values;
even the trajectory which starts at $(x_0=1,Q^2_0=10$~GeV$^2)$
fails to reach $x=0.95$ by the time $Q^2 = 40000$~GeV$^2$,
corresponding to a $p_T = 200$~GeV$/c$ jet.

%%%%%%%%%%%%%%%%%%%%%%%%%%%%%%%%%%%%%%%%%%%%%%%%%%%%%%%%%%%%%%%%%%%%%%%%%%%%%%%%%%%%%%%%%%%%%%%%%%%%%%%%%%%%%%%%%%%%%%%%
%  Hard subprocesses
%%%%%%%%%%%%%%%%%%%%%%%%%%%%%%%%%%%%%%%%%%%%%%%%%%%%%%%%%%%%%%%%%%%%%%%%%%%%%%%%%%%%%%%%%%%%%%%%%%%%%%%%%%%%%%%%%%%%%%%%

\section{Hard subprocesses}
\label{sec:subprocess}

We have developed a model of nuclear PDFs at $x>0.1$ that accounts for both short range correlations
and effects from medium modification.
In addition to a model for PDFs, however, we also need to calculate the squares of the invariant matrix elements
$\overline{\left|\mathcal{M}_{ij}\right|^2}$
for the hard partonic subprocesses that contribute to Eq.~(\ref{eqn:diffcrx}).
We consider two partons, labeled $i$ and $j$, to exist in the initial state of the reaction---one from
each of the proton and the heavy nucleus participating in the reaction (\ref{eqn:reaction}).
At leading order, a dijet is associated with two partons in the final state, which we label $k$ and $l$.
Each parton can in principle be a quark, an anti-quark, or a gluon,
although what partonic subprocesses are possible is limited by quark flavor conservation.
A list of the possible two-parton to two-parton subprocesses
and their squared matrix elements is presented in Table~\ref{table:subprocess} ({\sl cf.}~\cite{Combridge:1977dm}).

\bgroup
\def\arraystretch{2}
\begin{table}
  \centering
  \begin{tabular}{| c | c |}
    \hline
    Subprocess & $\frac{\overline{\left|\mathcal{M}\right|^2}}{g_s^4}$   \\
    \hline    $q_j + q_k \rightarrow q_j + q_k$    
      &    $\frac{4}{9}\frac{s^2+u^2}{t^2}$   \\
    \hline    $q_j + q_j \rightarrow q_j + q_j$    
      &    $\frac{4}{9}\left(\frac{s^2+u^2}{t^2}+\frac{s^2+t^2}{u^2}\right) - \frac{8}{27}\frac{s^2}{ut}$    \\
    \hline    $q_j + \bar{q}_j \rightarrow q_k + \bar{q}_k$    
      &    $\frac{4}{9}\frac{t^2+u^2}{s^2}$    \\
    \hline    $q_j + \bar{q}_j \rightarrow q_j + \bar{q}_j$    
      &    $\frac{4}{9}\left(\frac{s^2+u^2}{t^2}+\frac{t^2+u^2}{s^2}\right) - \frac{8}{27}\frac{u^2}{st}$    \\
    \hline    $q_j + \bar{q}_j \rightarrow g + g$    
      &    $\frac{32}{27}\frac{u^2+t^2}{ut} - \frac{8}{3}\frac{u^2+t^2}{s^2}$    \\
    \hline    $g + g \rightarrow q_j + \bar{q}_j$    
      &    $\frac{1}{6}\frac{u^2+t^2}{ut} - \frac{3}{8}\frac{u^2+t^2}{s^2}$    \\
    \hline    $q_j + g \rightarrow q_j + g$    
      &    $-\frac{4}{9}\frac{u^2+s^2}{us} + \frac{8}{3}\frac{u^2+s^2}{t^2}$    \\
    \hline    $g + g \rightarrow g + g$    
      &    $\frac{9}{2}\left(3-\frac{ut}{s^2}-\frac{us}{t^2}-\frac{st}{u^2}\right)$
    \\
    \hline
  \end{tabular}
  \caption{
    Squared matrix element for hard partonic processes at leading order,
    summed over final spins and averaged over initial spins,
    and divided by $g_s^4$.
    $j\neq k$ in subprocesses where both indices are present.
    The table is from Ref.~\cite{Combridge:1977dm}.
  }
  \label{table:subprocess}
\end{table}

At next-to-leading order (NLO), a dijet can be associated either with two or with three partons in the final state,
provided two of the three partons have small enough differences in both their rapidities and their azimuthal angles,
so that $\Delta\phi^2 + \Delta\eta^2 < R^2$,
where $R$ is the cone radius used for defining jets.
Additionally, three-parton dijets are needed to cancel the infrared divergences in the NLO
contributions to the two-parton to two-parton matrix elements\cite{Ellis:1985er}.
Finally, identifying two partons as a single jet in the NLO contribution alters
the simple kinematical formula Eq.~(\ref{eqn:xA}) for determining $x_A$ through dijet kinematics.

These facts indicate that we should verify the adequacy of the leading order approximation.
To make this verification,
we perform a leading order calculation of the reaction (\ref{eqn:reaction}) for the case of $A=1$,
{\sl i.e.} for dijet production from proton-proton collisions, for which a fair amount of experimental data
in the kinematical regime of interest exist.
We compare the calculation within the leading order to experimental data from Ref.~\cite{Aad:2011fc}.
The data, however, are given in terms of a two-fold differential cross section $d^2\sigma/dm_{JJ}d|y^*|$,
where $m_{JJ}^2 = (p_3+p_4)^2$ and $y^*$ is the rapidity of an individual jet in the dijet center of mass frame.
The relationship of this two-fold differential cross section to the three-fold differential cross section
of Eq.~(\ref{eqn:diffcrx}) is given in Appendix~\ref{appendix:dsig2},
and a calculation of the leading-order calculation with the data is given in Fig.~\ref{fig:dsig2pp}.
As the comparison shows, we have a reasonably good agreement with the data that validates
the use of leading order for calculating the hard subprocesses that contribute to reaction (\ref{eqn:reaction}).

\begin{figure}
  \centering
  \includegraphics[scale=0.45]{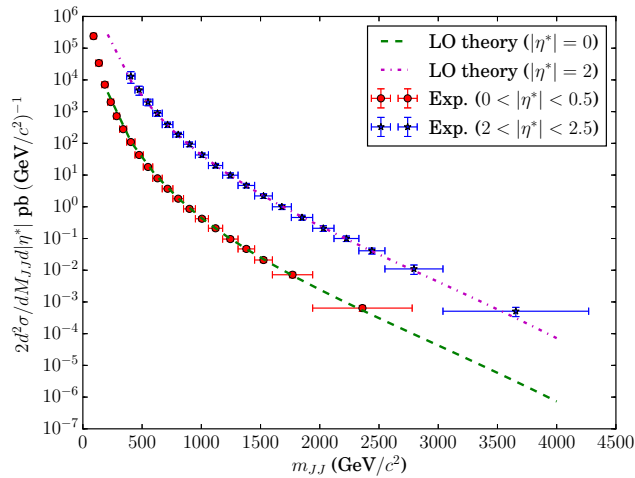}
  \caption{(Color online) Comparison of leading-order (LO) calculation of the two-fold differential
    cross section using Eq.~(\ref{eqn:dsig2}) to experimental data from Ref.~\cite{Aad:2011fc}.
    Proton PDF used is CT10\cite{Gao:2013xoa}
  }
  \label{fig:dsig2pp}
\end{figure}

%%%%%%%%%%%%%%%%%%%%%%%%%%%%%%%%%%%%%%%%%%%%%%%%%%%%%%%%%%%%%%%%%%%%%%%%%%%%%%%%%%%%%%%%%%%%%%%%%%%%%%%%%%%%%%%%%%%%%%%%
%  Numerical estimates
%%%%%%%%%%%%%%%%%%%%%%%%%%%%%%%%%%%%%%%%%%%%%%%%%%%%%%%%%%%%%%%%%%%%%%%%%%%%%%%%%%%%%%%%%%%%%%%%%%%%%%%%%%%%%%%%%%%%%%%%

\section{Numerical estimates}
\label{sec:num}

In the previous sections, we have developed all the components necessary to calculate the three-fold differential
cross section of Eq.~(\ref{eqn:diffcrx}) for the dijet production reaction (\ref{eqn:reaction}).
It is our intention to use them to determine the sensitivity of the dijet production reaction
(\ref{eqn:reaction}) to the presence of short range correlations in the nucleus,
as well as to medium modification effects.
Additionally, we will look at nuclear partons with $x_A > 1$ in particular.
Since $x_A$ can be related to measurable jet kinematics via Eq.~(\ref{eqn:xA}),
it is necessary to elaborate on the kinematics we will consider.
Once the kinematics considered are clarified,
we will present numerical estimates of the three-fold differential cross section of Eq.~(\ref{eqn:diffcrx})
at these kinematics,
and additionally we will estimate the partially and fully integrated cross sections to find out if
events corresponding to $x_A > 1$ are frequent enough to be measurable at the LHC.

\subsection{Kinematics considered}
\label{sec:num:kine}

Since one of our goals is to probe $x_A > 1$, we consider kinematics that maximize $x_A$\footnote{
  We are also interested in the EMC effect region, since current DIS data are either at low $Q^2$
  where higher-twist effects are significant, or have poor accuracy for $x \gtrsim 0.65$.
  The kinematics of interest here will also cover such moderate $x$ values.
}.
According to Eq.~(\ref{eqn:xA}), three variables are related to $x_A$.
The jet rapidities $\eta_3$ and $\eta_4$ should be as small or highly negative
as possible in order to maximize $x_A$.
Additionally, the transverse jet momentum $p_T$ should be as large as possible.
On the other hand, non-central jets are more difficult to detect and identify,
and the jet cross section for large $p_T$ is known to drop rapidly.
It is necessary to find a balance between kinematics that will increase $x_A$
and those that will produce an appreciable yield.

Since central jets are easier to detect, and since we are interested in superfast partons in the nucleus, 
we consider dijets with one jet produced in the central rapidity region where LHC detectors have the best resolution
and the second jet moving in the direction of the nucleus beam.
In particular, the rapidity ranges considered are $-2.5 < \eta_3 < 2.5$ and $3 < -\eta_4 < 5$.
We also consider a $p_T$ range from $40$~GeV$/c$ to $200$~GeV$/c$.

\subsection{Three-fold differential cross section}

Eq.~(\ref{eqn:diffcrx}) can be used to obtain numerical estimates of the three-fold differential cross section
$d^3\sigma/d\eta_3 d\eta_4 dp_T^2$ for the dijet production reaction (\ref{eqn:reaction}).
Numerical estimates are presented in Fig.~\ref{fig:crx:three},
where estimates with and without short range correlations are presented for comparison,
and likewise estimates with and without medium modification are presented.
As can be seen in Figs.~\ref{fig:crx:unmod:log},
the presence of SRCs increases the cross section for superfast ($x_A>1$) quarks considerably.
Fig.~\ref{fig:crx:unmod:ratio} demonstrates the relative importance of three-nucleon SRCs in particular
at large $x_A$.
On the other hand, medium modification decreases the cross section,
especially at large $x$ where highly-modified SRCs dominate the cross section,
as can be seen in Figs.~\ref{fig:crx:clrscn:log} and \ref{fig:crx:clrscn:ratio}.

\begin{figure}
  \centering
  \begin{subfigure}[t]{.45\textwidth}
    \includegraphics[width=\textwidth]{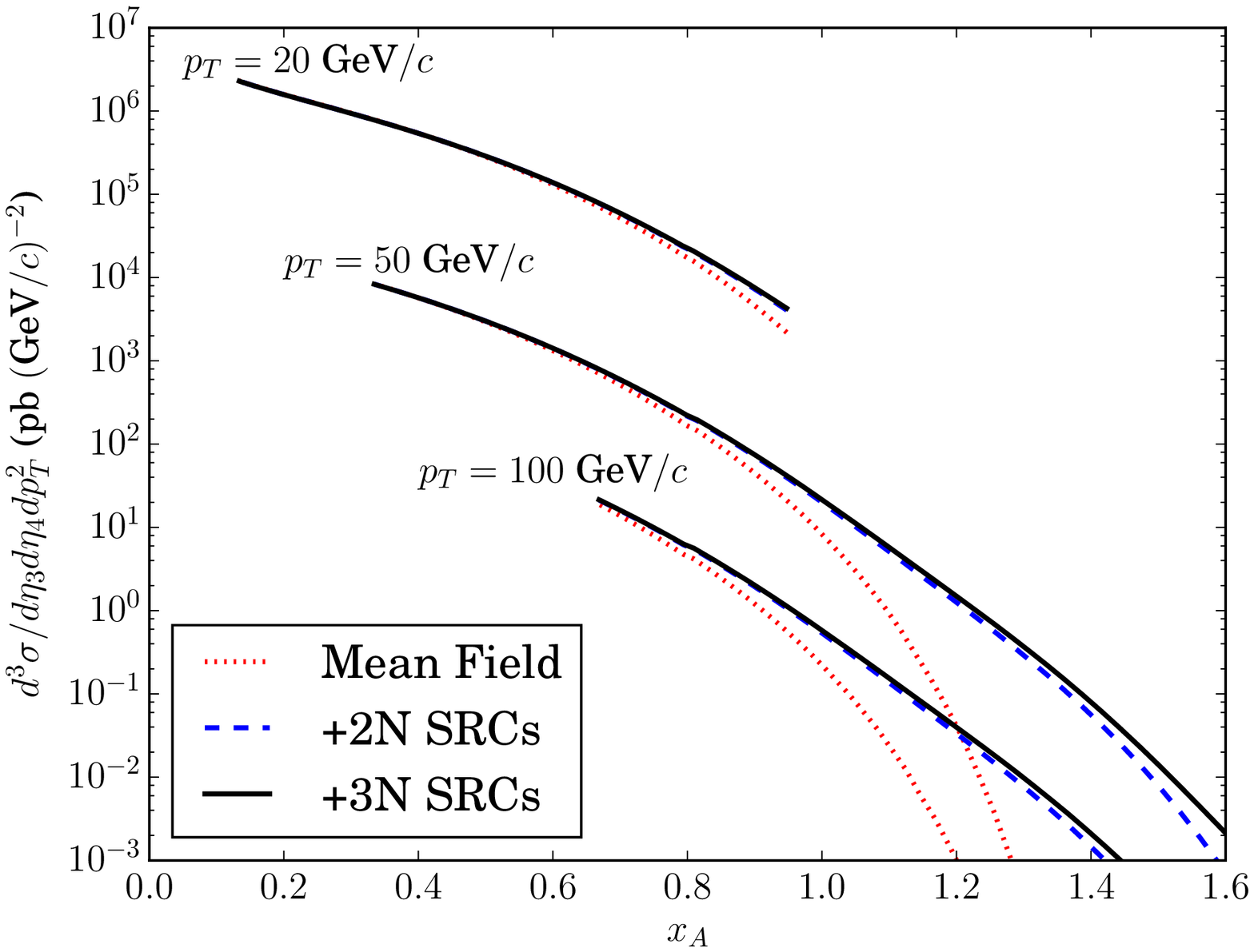}
    \caption{No medium modifications.}
    \label{fig:crx:unmod:log}
  \end{subfigure}
  \begin{subfigure}[t]{.45\textwidth}
    \includegraphics[width=\textwidth]{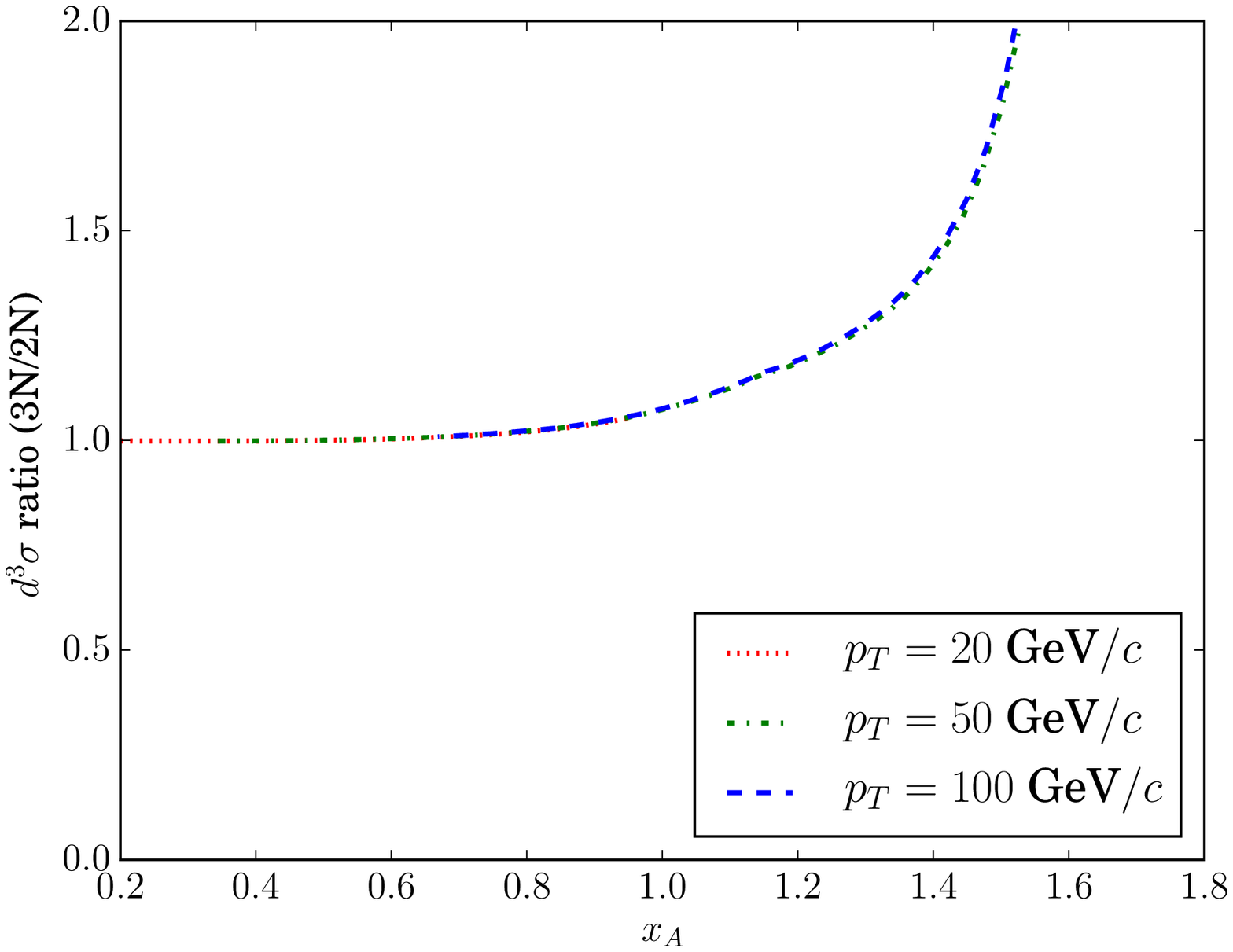}
    \caption{No medium modifications.}
    \label{fig:crx:unmod:ratio}
  \end{subfigure}
  \\
  \begin{subfigure}[t]{.45\textwidth}
    \includegraphics[width=\textwidth]{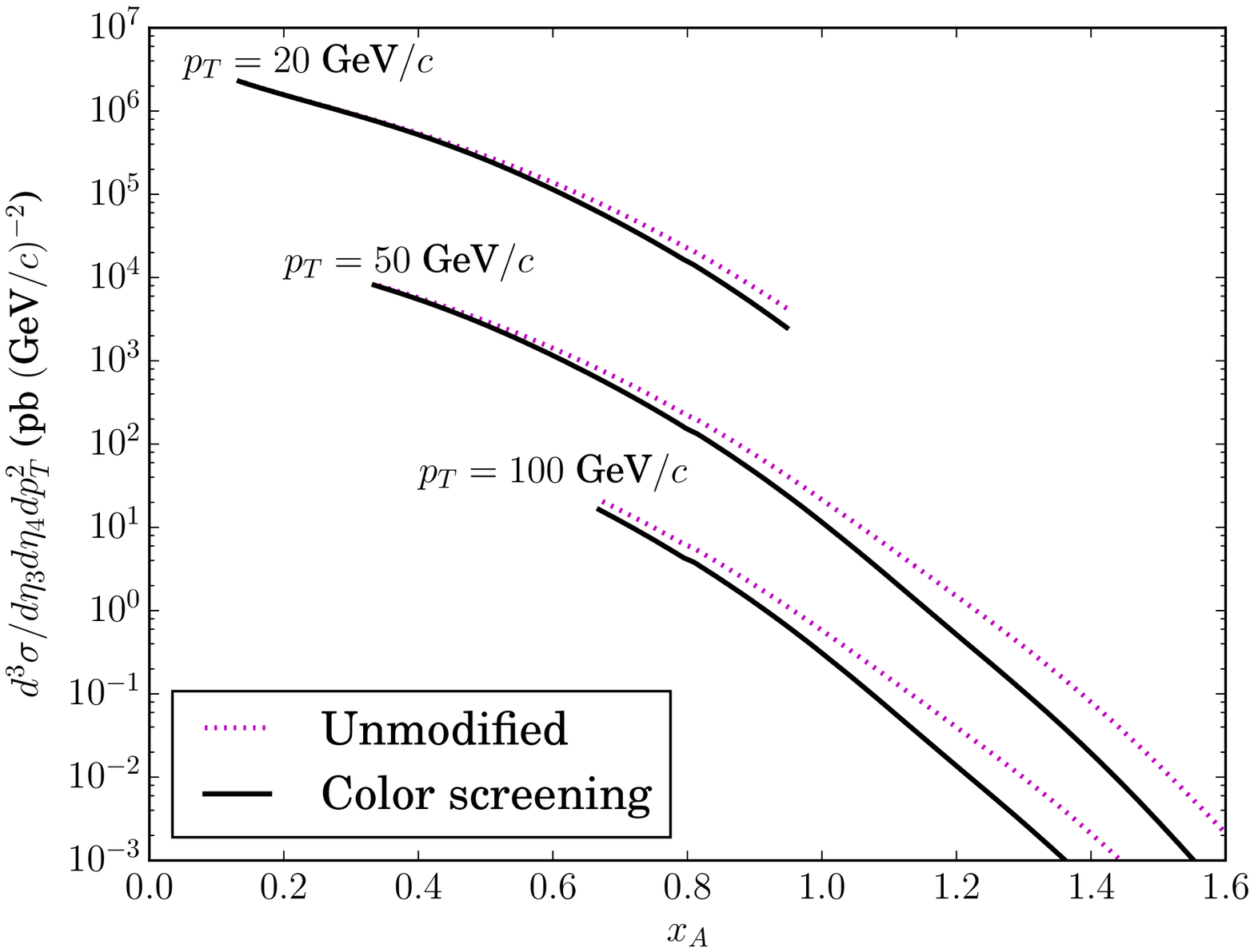}
    \caption{All SRCs contribute.}
    \label{fig:crx:clrscn:log}
  \end{subfigure}
  \begin{subfigure}[t]{.45\textwidth}
    \includegraphics[width=\textwidth]{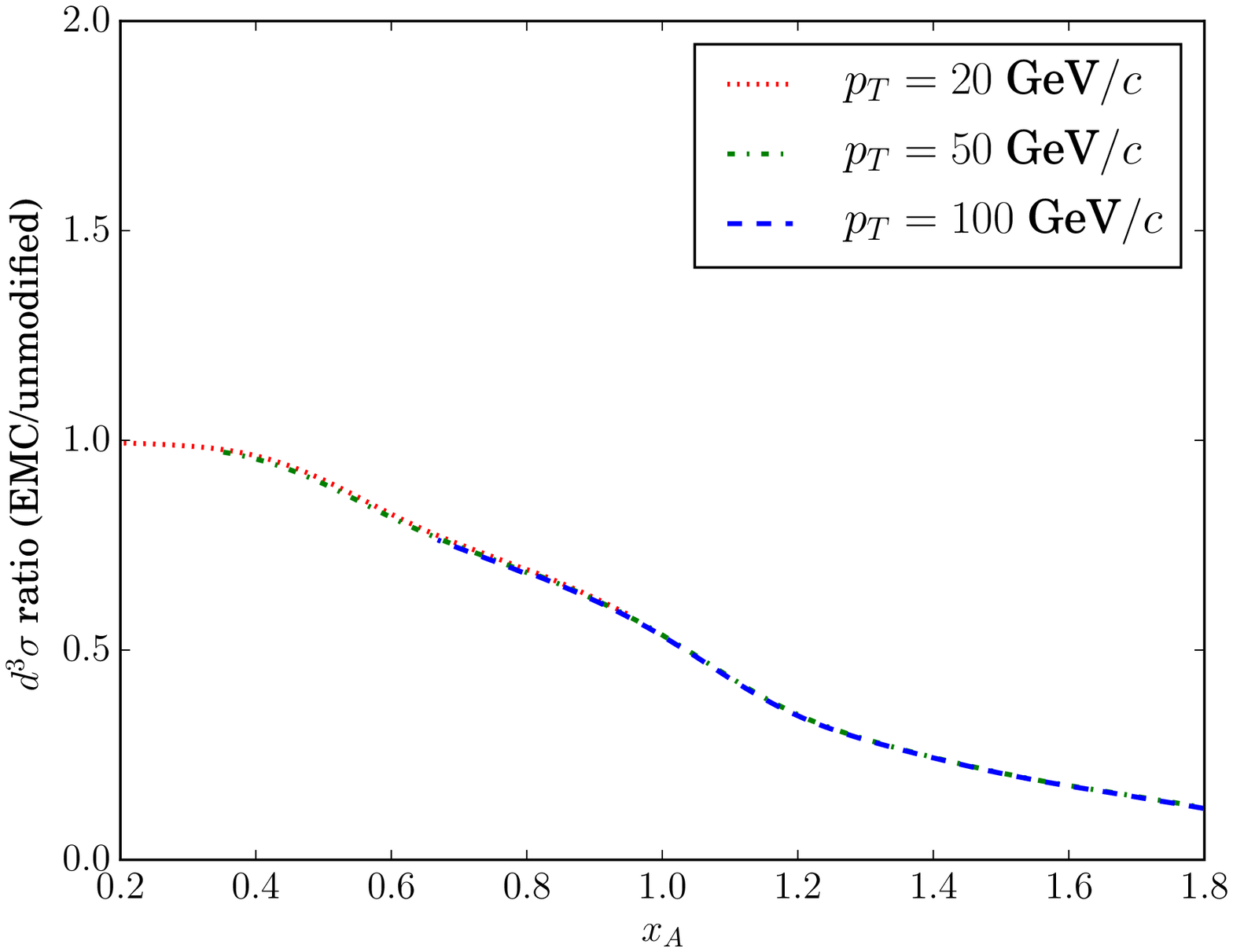}
    \caption{All SRCs contribute.}
    \label{fig:crx:clrscn:ratio}
  \end{subfigure}
  \caption{
    (Color online.)
    Three-fold differential cross section for $p+^{208}$Pb$\rightarrow\mathrm{dijet}+X$
    according to Eq.~(\ref{eqn:diffcrx}).
    (a) and (c) are absolute cross sections.
    (b) and (d) are ratios.
    (b) is the ratio of the three-fold differential cross section
    when 2N and 3N SRCs are both considered to when only 2N SRCs are considered.
    (b) is the ratio of the three-fold differential cross section
    with medium modifications to without.
    $\eta_4$ is constrained to the range $(-5,-3)$, and $\eta_3=0$.
    Proton PDF used is CT10\cite{Gao:2013xoa}.
  }
  \label{fig:crx:three}
\end{figure}

In addition to comparisons of the effects of SRCs and medium modifications within the model
for the nuclear light cone density presented in this work,
we compare in Fig.~\ref{fig:fs81} the prediction for the three-fold differential cross section using our model
to the one made using the FS81 model\cite{Frankfurt:1981mk}.
The FS81 model uses exponential decay to model the large $\alpha$ part of the nuclear light cone density,
and likewise the nuclear PDF calculated using this model exhibits exponential decay for large $x_A$.
It is for this reason the curve for the FS81 model appears as a straight line in the log plot of Fig.~\ref{fig:fs81}.
Qualitatively, the model of this work can be distinguished experimentally from the FS81 model
by determining whether the three-fold differential cross section of Eq.~(\ref{eqn:diffcrx}) falls exponentially
or faster than a simple exponential.

\begin{figure}
  \centering
  \includegraphics[scale=0.5]{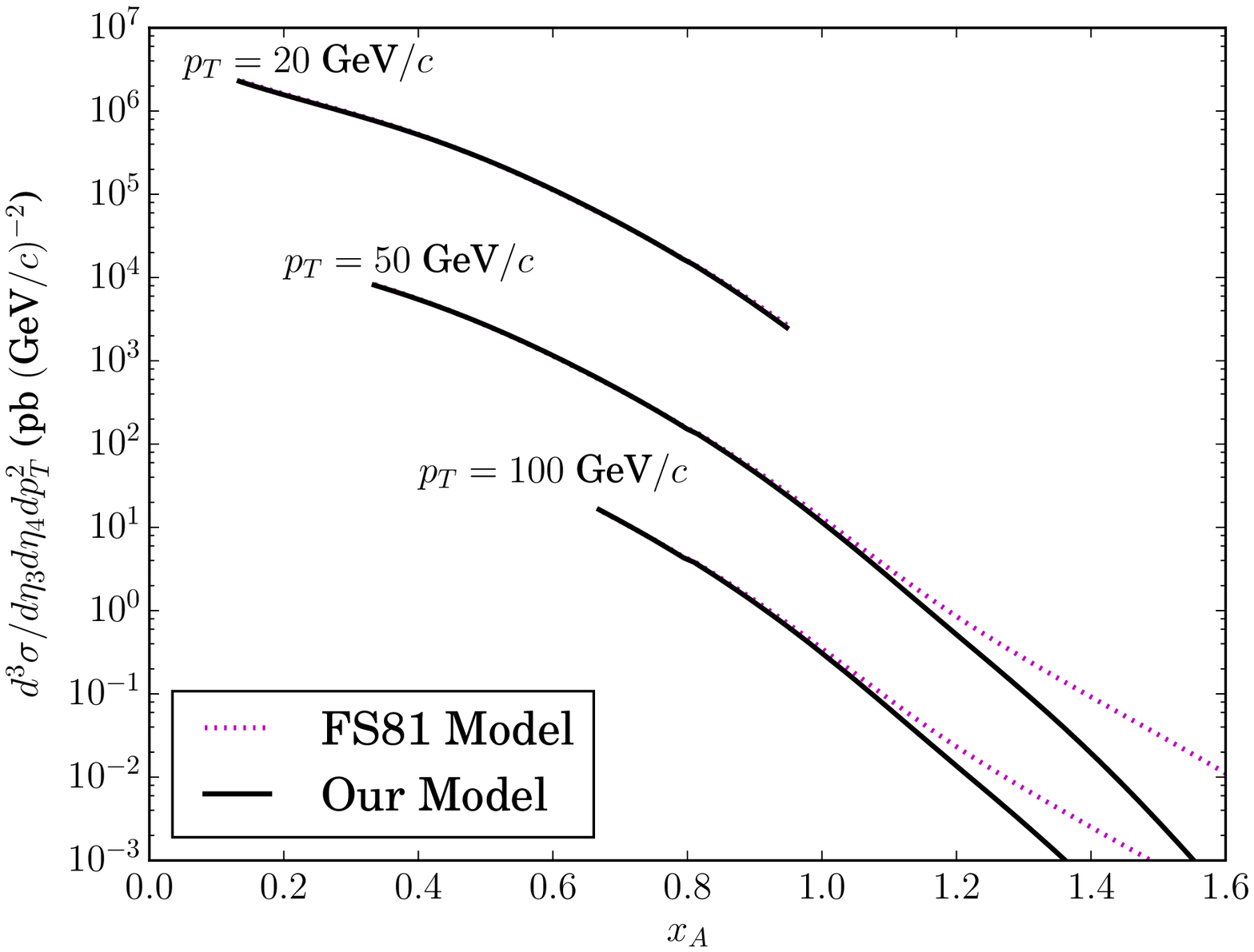}
  \caption{The color screening model accounts for medium modifications for both models.
    Calculation within our model considers 2N and 3N SRCs to contribute.}
  \label{fig:fs81}
\end{figure}

\subsection{Partially integrated cross section}

Next we study the sensitivity to SRCs and medium modification effects of the one-fold differential cross section
$\frac{d\sigma}{dp_T}$, which is obtained by integrating Eq.~(\ref{eqn:diffcrx}) over the pseudo-rapidities of the jets.
The main goal in doing so is to maximize the absolute cross section while retaining highest possible sensitivity to 
short-range nuclear phenomena, namely the $x>1$ region.
For this we choose the ranges of rapidity integrations in order to maximally preserve the sensitivity
from the $x>1$ domain.
The ranges of $\eta_3$ and $\eta_4$ that are considered have been discussed above,
namely the jet originating from the proton should be in the central region $-2.5 < \eta_3 < 2.5$
and the jet originating from the nucleus should be forward in the nucleus beam region,
{\sl i.e.}~$-5 < \eta_4 < -3$.
Thus, the one-fold differential cross section is:
\begin{equation}
  \frac{d\sigma}{dp_T}        = \int_{-2.5}^{2.5}d\eta_3 \int_{-5}^{-3}d\eta_4
    \frac{2p_T d^3\sigma}{d\eta_3 d\eta_4 dp_T^2}
    \label{eqn:crxpT:full}
  .
\end{equation}
In order to verify the sensitivity of Eq.~(\ref{eqn:crxpT:full}) to the kinematic domain of superfast quarks ($x>1$),
we compare it with the one-fold differential cross section over the same range of $\eta_3$ and $\eta_4$,
but with $x>1$ events selected for only, {\sl viz.}
\begin{equation}
  \frac{d\sigma(x_A>1)}{dp_T} = \int_{-2.5}^{2.5}d\eta_3 \int_{-5}^{-3}d\eta_4
    \frac{2p_T d^3\sigma}{d\eta_3 d\eta_4 dp_T^2} \Theta(x_A-1)
    \label{eqn:crxpT:fast}
  .
\end{equation}
By comparing the differential cross sections of Eqs.~(\ref{eqn:crxpT:full}) and Eqs.~(\ref{eqn:crxpT:fast}),
we can determine the range of $p_T$ in which the results are close, and accordingly identify the $p_T$ region
that is sensitive to short-range nuclear phenomena.

\begin{figure}
  \centering
  \begin{subfigure}[t]{.45\textwidth}
    \centering
    \includegraphics[width=\textwidth]{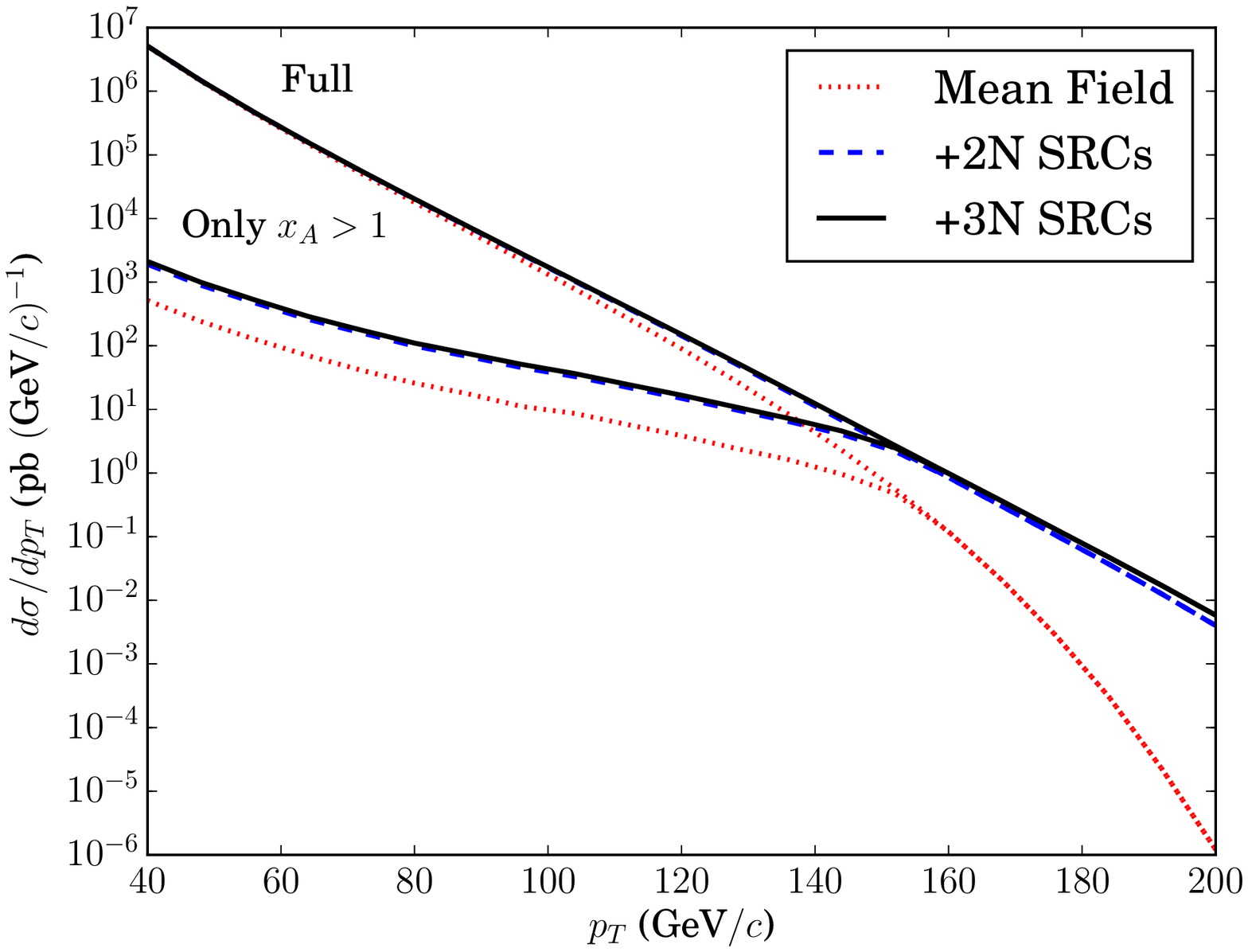}
    \caption{Without medium modifications.}
    \label{fig:pT:unmod}
  \end{subfigure}
  \begin{subfigure}[t]{.45\textwidth}
    \centering
    \includegraphics[width=\textwidth]{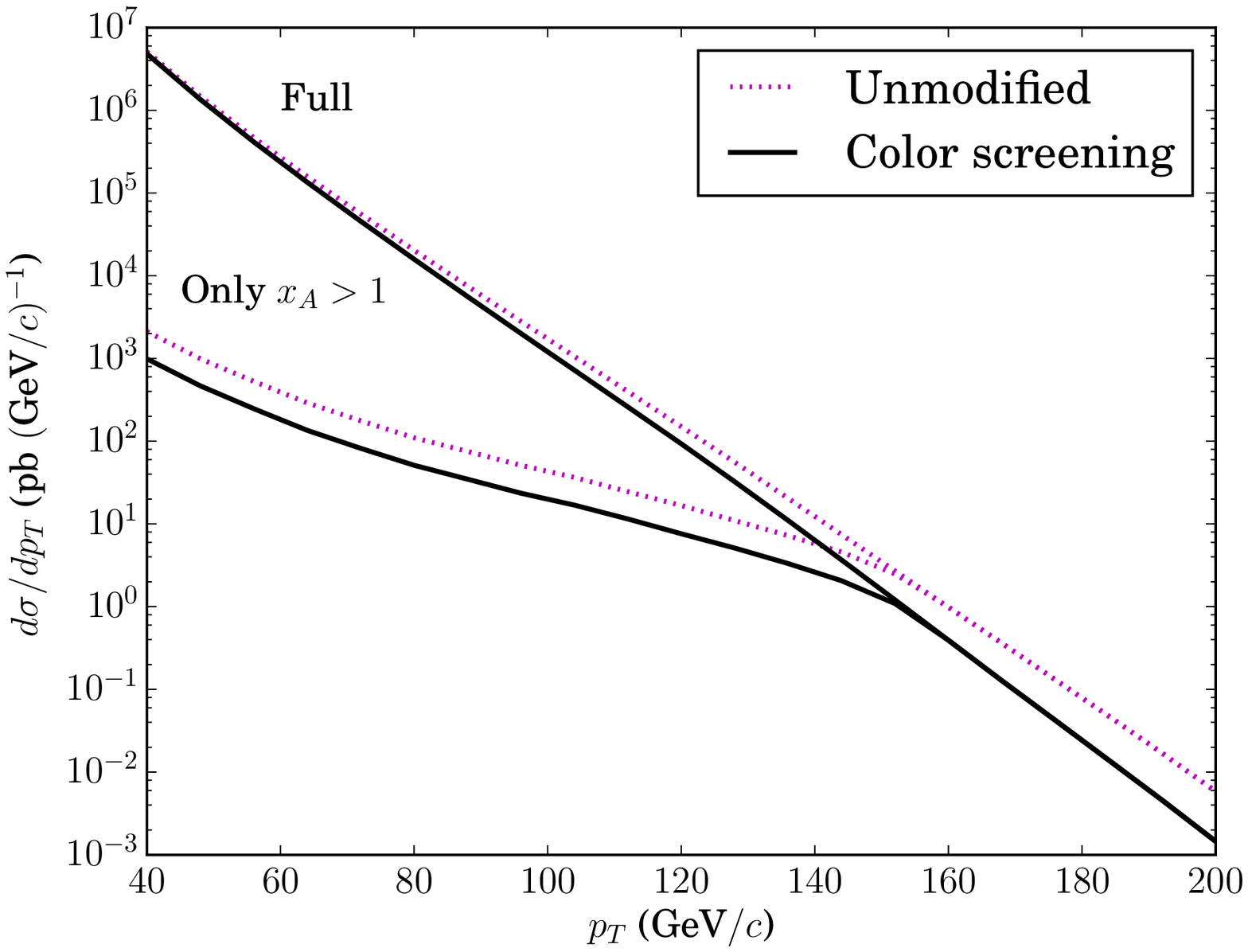}
    \caption{With and without color screening.}
    \label{fig:pT:clrscn}
  \end{subfigure}
  \caption{Differential cross section for $p+^{208}$Pb$\rightarrow\mathrm{dijet}+X$,
    according to Eqs.~(\ref{eqn:crxpT:full},\ref{eqn:crxpT:fast}).}
    Computed using CT10 for proton PDF\cite{Gao:2013xoa}.
  \label{fig:pT}
\end{figure}

In Fig.~\ref{fig:pT} we present estimates of the differential cross sections of
Eqs.~(\ref{eqn:crxpT:full},\ref{eqn:crxpT:fast}).
Fig.~\ref{fig:pT:unmod} compares the differential cross section, in the absence of medium modifications,
without and with short-range correlations accounted for.
Most dijets with transverse momentum up to about $100$~GeV$/c$ can be attributed to nucleons in the mean field.
By contrast, dijets with high $p_T \gtrsim 150$~GeV$/c$ can be attributed predominantly to short range correlations.
In addition, selecting only dijets with kinematics corresponding to an initial state nuclear parton with $x_A>1$
produces a differential cross section that is at least an order of magnitude larger when SRCs are considered
than for the mean field by itself.
This confirms that $x_A>1$ events can be attributed predominantly to the presence of SRCs in the nucleus,
and also indicates that finding a large yield for $x_A > 1$ events can demonstrate the presence of SRCs.

On the other hand, since SRCs are highly modified in the nuclear medium, $x_A > 1$ events in particular
are suppressed by the EMC effect, as can be seen in Fig.~\ref{fig:pT:clrscn}.
This means the value of the one-fold differential cross section for $x_A>1$ dijet events
experiences two effects that compete with one-another:
it is increased by the presence of SRCs, but it is at the same time suppressed by medium modifications,
which most strongly affect SRCs.

In addition to one-fold differential cross sections, we also present estimates for the total integrated cross section
in order to show that $x_A > 1$ events should produce a measurable yield.
Since jet identification at low $p_T$ is difficult,
we consider the total cross section integrated from a minimum $p_T$ value of 50~GeV$/c$. 
In Table~\ref{table:crx} we present numerical estimates for the total cross section
for different bins of $x$ as indicated in the table.
In addition to comparing the integrated cross section with and without medium modifications,
we have compared the model of this work to the FS81 model\cite{Frankfurt:1988nt}
(with medium modifications accounted for by the color screening model).
The FS81 model predicts a larger cross section than our model does at $x_A > 1$,
owing to the different form of 3N correletions,
as well as the implicit inclusion of 4, 5, {\sl etc.}~nucleon correlations in their light cone density.

\bgroup
\def\arraystretch{1}
\begin{table}
  \centering
  \begin{tabular}{| c | c | c | c | c |}
    \hline
    ~ &
    Unmodified &
    Modified (no SRCs) &
    Modified (SRCs) &
    FS81 (modified)
    \\   \hline
    All $x_A$       &    7.8~$\mu$b   &    6.8~$\mu$b   &    6.9~$\mu$b   &    6.9~$\mu$b   \\   \hline
    $0.6<x_A<0.7$   &   0.94~$\mu$b   &   0.73~$\mu$b   &   0.72~$\mu$b   &   0.76~$\mu$b   \\   \hline
    $0.7<x_A<0.8$   &   0.38~$\mu$b   &   0.25~$\mu$b   &   0.27~$\mu$b   &   0.28~$\mu$b   \\   \hline
    $0.8<x_A<0.9$   &   0.17~$\mu$b   &   0.07~$\mu$b   &   0.08~$\mu$b   &   0.09~$\mu$b   \\   \hline
    $0.9<x_A<1$     &     36~nb       &     14~nb       &     21~nb       &     23~nb       \\   \hline
    $1<x_A$         &     12~nb       &    2.3~nb       &    5.7~nb       &    7.3~nb       \\   \hline
  \end{tabular}
  \caption{Estimates of integrated cross sections,
    for different bins of $x_A$.
    Calculations include two- and three-nucleon SRCs.
  }
  \label{table:crx}
\end{table}

As can be seen in Table~\ref{table:crx}, medium modifications suppress
the integrated cross section, especially at high $x_A$.
This is consistent with the expectation that medium modifications affect SRCs especially strongly.
However, despite the suppression by medium modification, the cross section remains large enough to be observable.
In particular, with an integrated luminosity of $35.5$~nb$^{-1}$ (achieved in the 2013 $pA$ run of the LHC),
the expected yield for $x_A>1$ events (with suppression from the color screening model considered) is about 200 events.
Additionally, the expected yield for $x_A > 0.6$---the classical kinematical region of the EMC effect---is about 39000 events,
suggesting dijet production from $pA$ collisions could be used to study the EMC effect at high $Q^2$ where higher-twist
effects are negligible,
but where existing experimental data have large error bars.

\subsection{Jet resolution}
\label{sec:jetres}

To use the leading order kinematic relations Eqs.~(\ref{eqn:xp:eta},\ref{eqn:xA:eta}) requires unambiguous identification
of two jets with equal $p_T$, and neglects effects of jet fragmentation.
There could be a number of issues related to jet production and reconstruction which may spoil these relations.

Experimental measurement of the cross section of jet (or dijet) production involves using a particular algorithm
for extracting information about single (di-) jet production kinematics.
Currently, the anti-$k_t$ jet clustering algorithm\cite{Cacciari:2008gp} is widely used. 
For jets with $p_T \ge 150$~GeV$/c$, which we consider here, a cone with $R=0.2$-$0.4$ is usually used.
Numerical studies for the $p+A\rightarrow \mathrm{jet} +X$ process
with the jet produced in the proton fragmentation region,
performed in \cite{Perepelitsa:pc2015} using anti-$k_t$ clustering algorithm and PYTHIA,
find that in the kinematics studied by ATLAS\cite{ATLAS:2014cpa},
the value of $y= (E_{\mathrm{jet}} +p_{z,\mathrm{jet}})/2 E_p$ 
determined using anti-$k_t$ clustering algorithm with $R=0.4$
is very close to the $x_p$ of the colliding proton for $x_p\ge 0.2$ up to $x_p\sim 0.7$
(higher $x_p$ were not studied in Ref.~\cite{Perepelitsa:pc2015} due to the lack of statistics).
This allows good tracking of $x$ for the large $x$ parton.

In principle, fluctuations in the energy deposited in the calorimeter could have affect such a determination
for large $x$ where distribution drops very steeply,
leading to a large value of $\frac{\partial \ln F_{2}^{(A)}(x,Q^2)}{ \partial x }$.
However, it is easy to check that for $F_{2}^{(A)}(x,Q^2) \propto \exp (-bx)$ with $b \sim 8$,
variation with $x$ for $x > 0.6$ is weaker in the nucleus case.
Moreover, detection of the central jet we envision should help to suppress the effects of large fluctuations.
The observation of the central jets would additionally help to suppress contribution of multiparton interactions (MPI),
in which production of two jets with {\sl e.g.}\ $x_A\sim 0.5$ could mimic production of one $x_A\sim 1$ jet.
   
Another potential problem is the subtraction of the contributions of the underlying events for $x_A > 0.5$. 
On one hand, the underlying activity is higher in this case than in $pp$ collisions.
On the other hand, for $\eta \sim -4$ it is significantly smaller than for the central rapidities.
The anti-$k_t$ algorithm is well suited for dealing with such problems.
  
Obviously further simulations are necessary,
which would be based on more detailed experimental information about
the underlying event fluctuations in the backward kinematics than are available now.
Simulations should also account for energy and angular resolutions and
the acceptances of the detectors, as well as treating NLO effects like $p_T$ inbalance between the two jets.
Such a simulation would require a joint effort of experimentalists and theorists, and is beyond the scope of this paper.

\section{Conclusions}
\label{sec:theend}

In this work, we have shown that superfast partons inside a heavy nucleus with $x > 1$
are experimentally accessible at the LHC,
and can be probed in dijet production reactions from proton-nucleus collisions.
The value of $x$ can be related to jet kinematics in the LHC rest frame in a straightforward way,
so it is possible to select events that correspond to $x > 1$.
Partons with $x > 1$ can predominantly be attributed to the presence of short range correlations
between nucleons within the nucleus.
Expected medium modifications suppress the cross section appreciably,
but the total $x > 1$ cross section remains on the order of a few nanobarns,
making measurement feasible.
Additionally, it is possible to to perform novel studies of the EMC effect in the $x < 1$ region.

The observability of superfast partons in the nucleus owes itself to the presence of short range correlations.
This fact tells us that it is inadequate to treat the nucleus as a collection of free nucleons,
moving forward with equal light cone momenta.
The internal structure of the nucleus, and the relative momenta of its nucleons, are necessary to account for. 
We observe that even in the one-fold differential cross section for inclusive dijet production,
looking at large transverse jet momenta ($p_T\ge 150$~GeV$/c$)
allows one to select events originating from short range correlations in nuclei. 
Overall, our conclusion is that even at such high energies as those at the LHC,
short range nuclear structure can be explored,
and it will dominate the rate of the events with large transverse momenta
and negative pseudo-rapidities in nuclear fragmentation region.  

We expect the formalism to have further applications beyond dijet production in proton-nucleus collisions.
In particular,  short range nuclear structure and medium modification can be checked in 
Drell-Yan processes both in proton-nucleus and nucleus-nucleus collisions.
Additionally, it will prove necessary to account for multi-nucleon short range correlations
to correctly predict the high-$x$ cross section in inelastic nuclear DIS experiments at high energy.
The Electron Ion Collider\cite{Accardi:2012qut} in particular will be able to carry out such experiments,
where the formalism developed in this work will prove necessary.
In particular, the nuclear PDF at high $Q^2$ will need to be obtained through QCD evolution,
since the phenomenological models that exist to account for nuclear modifications are constructed using lower-$Q^2$
experimental data.

\section*{Acknowledgments}

We are thankful to Alberto Accardi, Brian Cole, Wim Cosyn, Leonid Frankfurt, Shunzo Kumano,
and Gerald Miller for many useful discussions.
Our special thanks are to Matteo Cacciari, Dennis Perepelitsa, Gavin Salam
for discussion of the problems of extraction of nuclear quark distributions at large $x_A$ from $pA$ jet production data.
This work is supported by US Department of Energy grants under contracts DE-FG02-01ER-41172 and DE-FG02-93ER40771.

%%%%%%%%%%%%%%%%%%%%%%%%%%%%%%%%%%%%%%%%%%%%%%%%%%%%%%%%%%%%%%%%%%%%%%%%%%%%%%%%%%%%%%%%%%%%%%%%%%%%%%%%%%%%%%%%%%%%%%%%
%  Derivation of impulse approximation
%%%%%%%%%%%%%%%%%%%%%%%%%%%%%%%%%%%%%%%%%%%%%%%%%%%%%%%%%%%%%%%%%%%%%%%%%%%%%%%%%%%%%%%%%%%%%%%%%%%%%%%%%%%%%%%%%%%%%%%%

\appendix

\section{Derivation of impulse approximation}
\label{appendix:factorize}

Here, we derive the impulse approximation for scattering of a hard probe $H$ from a nucleus,
which is assumed in the convolution approach taken in Eq.~(\ref{eqn:pdf:convolution}).
We utilize the assumption that the PDFs of the nucleus factorize
as in Eq.~(\ref{eqn:factorize}), 
We then proceed to derive the impulse approximation under a model where the nucleus consists of a system
of bound nucleons on the light cone. 
Corrections to the impulse approximation from nuclear shadowing become important
at $x_A \lesssim 0.15$\cite{Frankfurt:1988nt},
so the derivation to follow assumes $x_A > 0.15$.

\begin{figure}
  \centering
  \includegraphics[scale=0.75]{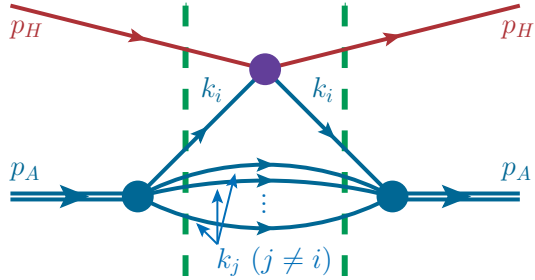}
  \caption{(Color online) The impulse approximation diagram for zero-degree $HA$ scattering.}
  \label{fig:lcnucleus}
\end{figure}

The derivation is performed using light cone perturbation theory.
The approach taken here is to calculate the forward $HA$ scattering amplitude
using the diagram depicted in Fig.~\ref{fig:lcnucleus},
and to use the optical theorem to relate this to the total inelastic $HA$ cross section.
In principle, $HA$ scattering can be accomplished in the impulse approximation by scattering
from any of the bound nucleons, and
for the full amplitude, a sum over $i$ is necessary.

To begin, the diagrammatic rules of light cone perturbation theory\cite{Brodsky:1997de} give
\begin{align}
  \mathcal{M}_{HA\rightarrow HA} &=
  \sum_{i=1}^A
  \int \prod_{j\neq i} \frac{dk_j^- d^2\mathbf{k}_{jT}}{2(2\pi)^3}
  \bigg\{
  \bar{u}^{\lambda_A}(p_A)
  \bar{u}^{\lambda_h}(p_h)
  \Gamma_{N/A}
  \left[\prod_{j\neq i} \frac{\sum_{\lambda_j}u^{\lambda_j}(k_j)\bar{u}^{\lambda_j}(k_j)\Theta(k_j^-)}{k_j^-}\right]
  \notag \\ & \qquad \times
  \frac{\sum_{\lambda_i}u^{\lambda_i}(k_i)\bar{u}^{\lambda_i}(k_i)\Theta(k_i^-)}{k_i^-}
  \frac{1}{\sum_{\mathrm{init.}}k^+ - \sum_{\mathrm{inter.}}k^+}
  \Gamma_{HN^*\rightarrow HN^*}
  \frac{\sum_{\lambda_i}u^{\lambda_i}(k_i)\bar{u}^{\lambda_i}(k_i)\Theta(k_i^-)}{k_i^-}
  \notag \\ & \qquad \times
  \frac{1}{\sum_{\mathrm{init.}}k^+ - \sum_{\mathrm{inter.}}k^+}
  \Gamma_{N/A}
  u^{\lambda_h}(p_h)
  u^{\lambda_A}(p_A) 
  \bigg\}
  \label{eqn:deriv:full}
  .
\end{align}
Here $\Gamma_{N/A}$ represents the effective vertex of transition of the nucleus 
with mass number $A$ to $A$ nucleons.
Since the nucleus is traveling rapidly in the $-z$ direction,
it is $x^-$ that is taken as the ``time'' coordinate, and thus $k^+$ is taken as the ``energy.''
Note that we have used $u^\lambda(p)$ to denote the part of a particle's wave function
that transforms under a representation of the Lorentz group, even if the particle in question
({\sl e.g.}, a spin-zero nucleus) is not a fermion.

The factors of $(\sum_{\mathrm{init.}}k^+ - \sum_{\mathrm{inter.}}k^+)^{-1}$ in Eq.~(\ref{eqn:deriv:full}) represent
intermediate states in the diagram, which are cut in Fig.~\ref{fig:lcnucleus} by dashed (green) lines.
All light cone energies are defined to be on shell, so each $k^+$ is determined by the on-shell relation
$k^+k^- = m^2 + \mathbf{k}_T^2$.
For the initial and intermediate states, this means:
\begin{align}
  \sum_{\mathrm{init.}}k^+ &= p_A^+ + p_h^+ \\
  \sum_{\mathrm{inter.}}k^+ &= p_h^+ + \sum_{j=1}^A k_j^+ \\
  \sum_{\mathrm{init.}}k^+ - \sum_{\mathrm{inter.}}k^+ &= p_A^+ - \sum_{j=1}^A k_j^+
  = \frac{1}{p_A^-}\left(M_A^2 - \sum_{j=1}^A \frac{m_N^2+\mathbf{k}_{jT}^2}{\alpha_j/A}\right)
  .
\end{align}
Accordingly, the intermediate state factors can be rewritten as
\begin{equation}
  \frac{1}{\sum_{\mathrm{init.}}k^+ - \sum_{\mathrm{inter.}}k^+}
  =
  \frac{p_A^-}{M_A^2 - \sum_{j=1}^A \frac{m_N^2+\mathbf{k}_{jT}^2}{\alpha_j/A}}
  .
\end{equation}
This, together with helicity conservation at the $HN^*$ vertex and some reshuffling of terms,
allows the full amplitude to be rewritten as
\begin{align}
  \mathcal{M}_{HA\rightarrow HA} &=
  \sum_{i=1}^A
  \sum_{\{\lambda_j\}}
  \int \prod_{j\neq i} \frac{dk_j^- d^2\mathbf{k}_{jT}}{2k_j^-(2\pi)^3}
  \bigg\{
  \left(\frac{p_A^-}{k_i^-}\right)^2
  \frac{\bar{u}^{\lambda_A}(p_A)\Gamma_{N/A}\left[\prod_{j=1}^A u^{\lambda_j}(k_j)\right]}
       {M_A^2 - \sum_{j=1}^A \frac{m_N^2+\mathbf{k}_{jT}^2}{\alpha_j/A}}
  \notag \\ & \qquad \times
  \left[\bar{u}^{\lambda_h}(p_h) \bar{u}^{\lambda_i}(k_i)
    \Gamma_{HN^*\rightarrow HN^*}
    u^{\lambda_i}(k_i) u^{\lambda_h}(p_h)\right]
  \frac{\left[\prod_{j=1}^A\bar{u}^{\lambda_j}(k_j)\right]\Gamma_{N/A}u^{\lambda_A}(p_A)}
       {M_A^2 - \sum_{j=1}^A \frac{m_N^2+\mathbf{k}_{jT}^2}{\alpha_j/A}}
  \bigg\}
  \label{eqn:deriv:step2}
  .
\end{align}
This can be simplified by using the subprocess amplitude
\begin{equation}
  \mathcal{M}_{HN^*\rightarrow HN^*}(\alpha_i,\mathbf{k}_{iT}) =
  \left[\bar{u}^{\lambda_h}(p_h) \bar{u}^{\lambda_i}(k_i)
    \Gamma_{HN^*\rightarrow HN^*}
    u^{\lambda_i}(k_i) u^{\lambda_h}(p_h)\right]  
\end{equation}
which depends on the momentum of the nucleon that was struck.
Using this amplitude and the $A$-body light cone nuclear wave function, defined by
\begin{equation}
  \psi_{N/A}^{\{\lambda_j\}}(\{\alpha_j\},\{\mathbf{k}_{jT}\})
  =
  - \frac{1}{\sqrt{A}}
  \left(\sqrt{\frac{1}{2(2\pi)}}\right)^{(A-1)}
  \frac{\left[\prod_{j=1}^A\bar{u}^{\lambda_j}(k_j)\right]\Gamma_{N/A}u^{\lambda_A}(p_A)}
       {M_A^2 - \sum_{j=1}^A \frac{m_N^2+\mathbf{k}_{jT}^2}{\alpha_j/A}}
  \label{eqn:lcwf:A}
       ,
\end{equation}
for the $HA$ forward amplitude one obtains
\begin{align}
  \mathcal{M}_{HA\rightarrow HA} &=
  \sum_{i=1}^A
  \int \prod_{j\neq i} \frac{d\alpha_j d^2\mathbf{k}_{jT}}{\alpha_j}
  \bigg\{
  \frac{A}{\alpha_i^2}
  \overline{\left|\psi_{N/A}^{\{\lambda_j\}}(\{\alpha_j\},\{\mathbf{k}_{jT}\})\right|^2}
  \mathcal{M}_{HN^*\rightarrow HN^*}(\alpha_i,\mathbf{k}_{iT})
  \bigg\}
  \label{eqn:deriv:step3}
  .
\end{align}
The line over the nuclear light cone wave function squared
signifies a sum over nucleon spins and an average over nucleus spin.

To proceed further, a completeness (containing momentum conservation) relation is used:
\begin{equation}
  1 = \int_0^A d\alpha_i \int d^2\mathbf{k}_{iT}
  \delta^{(1)}\left(A-\sum_{j=1}^A\alpha_j\right) \delta^{(2)}\left(\sum_{j=1}^A\mathbf{k}_{jT}\right)
  ,
\end{equation}
giving
\begin{align}
  \mathcal{M}_{HA\rightarrow HA} &=
  \sum_{i=1}^A
  \int \prod_{j=1}^A \frac{d\alpha_j d^2\mathbf{k}_{jT}}{\alpha_j}
  \bigg\{
  \delta^{(1)}\left(A-\sum_{j=1}^A\alpha_j\right) \delta^{(2)}\left(\sum_{j=1}^A\mathbf{k}_{jT}\right)
  \notag \\ & \qquad \times
  \frac{A}{\alpha_i}
  \overline{\left|\psi_{N/A}^{\{\lambda_j\}}(\{\alpha_j\},\{\mathbf{k}_{jT}\})\right|^2}
  \mathcal{M}_{HN^*\rightarrow HN^*}(\alpha_i,\mathbf{k}_{iT})
  \bigg\}
  \label{eqn:deriv:step4}
  .
\end{align}
Using optical theorem, applied the $HA$ and $HN$ scattering amplitudes in Eq.~(\ref{eqn:deriv:step4}):
\begin{align}
  \mathrm{Im}(\mathcal{M}_{HA\rightarrow HA}) &= s_{HA} \sigma_{HA} \\
  \mathrm{Im}\left(\mathcal{M}_{HN^*\rightarrow HN^*}(\alpha_i,\mathbf{k}_{iT})\right)
  &= s_{HN^*}(\alpha_i) \sigma_{HN^*}(\alpha_i,\mathbf{k}_{iT})
  = \frac{\alpha_i}{A} s_{HA} \sigma_{HN^*}(\alpha_i,\mathbf{k}_{iT})
  ,
\end{align}
one obtains for the $HA$ inelastic cross section
\begin{align}
  \sigma_{HA} &=
  \sum_{i=1}^A
  \int \prod_{j=1}^A \frac{d\alpha_j d^2\mathbf{k}_{jT}}{\alpha_j}
  \bigg\{
  \delta^{(1)}\left(A-\sum_{j=1}^A\alpha_j\right) \delta^{(2)}\left(\sum_{j=1}^A\mathbf{k}_{jT}\right)
  \notag \\ & \qquad \times
  \overline{\left|\psi_{N/A}^{\{\lambda_j\}}(\{\alpha_j\},\{\mathbf{k}_{jT}\})\right|^2}
  \sigma_{HN^*}(\alpha_i,\mathbf{k}_{iT})
  \bigg\}
  \label{eqn:deriv:step5}
  .
\end{align}
Furthermore, inserting the relation relation
\begin{equation}
  1 = \int_0^A \frac{d\alpha}{\alpha} \int d^2\mathbf{p}_T
  \alpha_i \delta^{(1)}(\alpha-\alpha_i) \delta^{(2)}(\mathbf{p}_T-\mathbf{k}_{iT})
  ,
\end{equation}
into Eq.~(\ref{eqn:deriv:step5}) gives
\begin{align}
  \sigma_{HA} &=
  \int_0^A \frac{d\alpha}{\alpha} \int d^2\mathbf{p}_T
  \Bigg[
    \int \prod_{j=1}^A \frac{d\alpha_j d^2\mathbf{k}_{jT}}{\alpha_j}
    \bigg\{
    \delta^{(1)}\left(A-\sum_{j=1}^A\alpha_j\right) \delta^{(2)}\left(\sum_{j=1}^A\mathbf{k}_{jT}\right)
    \notag \\ & \qquad \times
    \left(\sum_{i=1}^A \alpha_i \delta^{(1)}(\alpha-\alpha_i) \delta^{(2)}(\mathbf{p}_T-\mathbf{k}_{iT})\right)
    \overline{\left|\psi_{N/A}^{\{\lambda_j\}}(\{\alpha_j\},\{\mathbf{k}_{jT}\})\right|^2}
    \bigg\}
    \sigma_{HN^*}(\alpha,\mathbf{p}_{T})
  \Bigg]
  \label{eqn:deriv:step5}
  ,
\end{align}
which, using the definition of the light cone density matrix $\rho_{N/A}(\alpha,\mathbf{p}_T)$
({\sl cf.}~Eq.~(\ref{eqn:fNA}) and note that
$f_{N/A}(\alpha,\mathbf{p}_T) = \frac{1}{\alpha}\rho_{N/A}(\alpha,\mathbf{p}_T)$), gives
\begin{align}
  \sigma_{HA} &=
  \int_0^A \frac{d\alpha}{\alpha} \int d^2\mathbf{p}_T
  \rho_{N/A}(\alpha,\mathbf{p}_T)
  \sigma_{HN^*}(\alpha,\mathbf{p}_{T})
  \label{eqn:factorize:sigs}
  .
\end{align}
From here, we make use of the assumption that the $HN^*$ inclusive scattering cross section factorizes
in terms of universal PDFs, namely
\begin{equation}
  \sigma_{HN^*} = \sum_j \int_0^1 dx_{N^*} f_{j/N^*}(x_{N^*},Q^2,\alpha,\mathbf{p}_T)
  \sigma_{Hj}
  \label{eqn:factorize:HN}
  ,
\end{equation}
where $\sigma_{Hj}$ is the hard cross section for inclusive scattering of the hard probe $H$
from a parton with flavor $j$,
and where $x_{N^*}$ is the light cone momentum fraction of the parton in the bound nucleon.
In general, the the bound nucleon is off its light cone energy shell, so {\sl a priori},
its PDF $f_{j/N^*}(x_{N^*},Q^2,\alpha,\mathbf{p}_T)$ depends on the degree of off-shellness.
This fact is signified in the dependence of the bound nucleon PDF on $\alpha$ and $\mathbf{p}_T$.

Using now Eq.~(\ref{eqn:factorize:HN}) one arrives at:
\begin{align}
  \sigma_{HA} &=
  \sum_{j}
  \int_0^A \frac{d\alpha}{\alpha} \int d^2\mathbf{p}_T
  \rho_{N/A}(\alpha,\mathbf{p}_T)
  \int_0^1 dx_{N^*} f_{j/N^*}(x_{N^*},Q^2,\alpha,\mathbf{p}_T)
  \sigma_{Hj}
  \label{eqn:deriv:step6}
  .
\end{align}
The integration over $x_{N^*}$ can be converted into integration over $x_A$, with limits of $0$ and $A$,
provided 
$x_A < \alpha$, which can be satisfied by setting the lower integration limit of $\alpha$ to $x_A$.
This results in the expression:
\begin{align}
  \sigma_{HA} &=
  \sum_{j}
  \int_0^A dx_A
  \left[
    \int_{x_A}^A \frac{d\alpha}{\alpha^2} \int d^2\mathbf{p}_T
    \rho_{N/A}(\alpha,\mathbf{p}_T) f_{j/N^*}(x_{N^*},Q^2,\alpha,\mathbf{p}_T)
     \right]
    \sigma_{Hj}
 \label{eqn:deriv:step7}
 .
\end{align}
Under the assumption that the $HA$ inclusive scattering cross section factorizes,
the quantity in the square brackets of Eq.~(\ref{eqn:deriv:step7}) is equal to the nuclear PDF,
thus giving Eq.~(\ref{eqn:pdf:convolution}).

%%%%%%%%%%%%%%%%%%%%%%%%%%%%%%%%%%%%%%%%%%%%%%%%%%%%%%%%%%%%%%%%%%%%%%%%%%%%%%%%%%%%%%%%%%%%%%%%%%%%%%%%%%%%%%%%%%%%%%%%
%  Derivation of 2N and 3N models
%%%%%%%%%%%%%%%%%%%%%%%%%%%%%%%%%%%%%%%%%%%%%%%%%%%%%%%%%%%%%%%%%%%%%%%%%%%%%%%%%%%%%%%%%%%%%%%%%%%%%%%%%%%%%%%%%%%%%%%%

\section{Light cone densities for short range correlations}
\label{appendix:lcd}

In this Appendix, we will derive the functional forms of the light cone density matrices
for two-nucleon and three-nucleon short range correlations.
These can be related to the light cone fraction distributions through
$f_{N/A}(\alpha,\mathbf{p}_T) = \frac{1}{\alpha}\rho_{N/A}(\alpha,\mathbf{p}_T)$.
In the following calculations,
the direction of $z$ axis is defined by the nuclear momentum.
Therefore the ``$+$'' component of the nucleus momentum is a large parameter. 

%%%%%%%%%%%%%%%%%%%%%%%%%%%%%%%%%%%%%%%%%%%%%%%%%%%%%%%%%%%%%%%%%%%%%%%%%%%%%%%%%%%%%%%%%%%%%%%%%%%%%%%%%%%%%%%%%%%%%%%%
%  Derivation of 2N model
%%%%%%%%%%%%%%%%%%%%%%%%%%%%%%%%%%%%%%%%%%%%%%%%%%%%%%%%%%%%%%%%%%%%%%%%%%%%%%%%%%%%%%%%%%%%%%%%%%%%%%%%%%%%%%%%%%%%%%%%

\subsection{2N correlations}
\label{appendix:lcd:2N}

The light cone density matrix for two-nucleon correlations is calculated using a cut diagram, Fig.~\ref{fig:diagram:2N}.
The rules for calculating a cut diagram are similar to the Feynman rules for perturbation theory,
but contain additional terms, namely a vertex $\hat{V}_{2N}(\alpha,\mathbf{p}_T)$ for the probed nucleon in the SRC,
and an on-mass-shell factor $(2\pi)\delta^{(1)}(p_2^2-m_N^2)$ for the spectator nucleon.
This approach can be seen as a relativistic generalization of the method of Ref.~\cite{CiofidegliAtti:1995qe}
(see also Ref.~\cite{Frankfurt:1981mk} for application to the light cone).
In this case, for the unpolarized density matrix, we have
\begin{equation}
  \hat{V}_{2N}(\alpha,\mathbf{p}_T) = \sum_\lambda
  a_N^\dagger(p,\lambda)
  \alpha^2
  \delta^{(1)}(\alpha-\alpha_1) \delta^{(2)}(\mathbf{p}_T-\mathbf{p}_{1T})
  a_N(p,\lambda)
  \label{eqn:V2N}
  ,
\end{equation}
where $a^\dagger_N(p,\lambda)$ and $a_N(p,\lambda)$ are creation and annihilation operators defined so that
\begin{align}
  &a_N(p,\lambda)
  u^{\lambda_1^\prime}(p_1)
  =
  \delta_{\lambda \lambda_1^\prime}
  ~~~
  &\bar{u}^{\lambda_1}(p_1)
  a^\dagger_N(p,\lambda)
  =
  \delta_{\lambda \lambda_1}
  \label{eqn:new:aat}
  .
\end{align}
The operator $\hat{V}_{2N}(\alpha,\mathbf{p}_T)$ essentially picks a nucleon with light cone fraction $\alpha$
and transverse momentum $\mathbf{p}_T$ out of the 2N SRC. 
Additionally, the vertices $\Gamma_{dpn}$ are accompanied by a factor of $\frac{1}{\sqrt{2}}$ in this derivation,
since we are deriving the distribution of a single nucleon within a 2N SRC.
This requires each $NN$ vertex be accompanied by division by the square root of the number of baryons involved.

\begin{figure}
  \centering
  \includegraphics[scale=0.75]{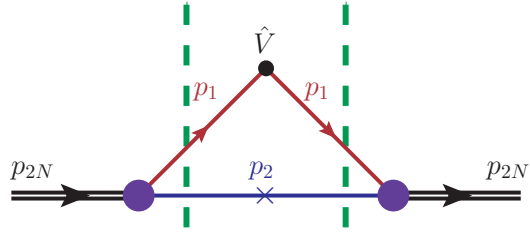}
  \caption{(Color online) Cut diagram for calculating 2N SRCs.}
  \label{fig:diagram:2N}
\end{figure}

First, the application of Feynman rules to the diagram of Fig.~\ref{fig:diagram:2N} gives
\begin{align}
  \rho_{2N}(\alpha,\mathbf{p}_T) &=
  \frac{1}{3} \sum_{\lambda_d}
  \int \frac{dp_2^+ dp_2^- d^2\mathbf{p}_{2T}}{2(2\pi)^4}
  \bigg[
    \chi^{\dagger(\lambda_d)}_d(p_{2N})
    \frac{ \Gamma_{dpn}(\alpha_1,\mathbf{p}_{1T}) }{\sqrt{2}}
    \left(\sum_{\lambda_2}u^{\lambda_2}(p_2)\bar{u}^{\lambda_2}(p_2)\Theta(p_2^+)\right)
    \notag \\ & \qquad \times
    \left((2\pi)\delta^{(1)}(p_2^2-m_N^2)\right)  
    \frac{\sum_{\lambda_1}u^{\lambda_1}(p_1)\bar{u}^{\lambda_1}(p_1)\Theta(p_1^+)}{p_1^+\mathcal{D}_{2N}}
    \hat{V}_{2N}(\alpha,\mathbf{p}_T)
    \notag \\ & \qquad \times
    \frac{\sum_{\lambda_1^\prime}u^{\lambda_1^\prime}(p_1)\bar{u}^{\lambda_1^\prime}(p_1)\Theta(p_1^+)}{p_1^+\mathcal{D}_{2N}}
    \frac{ \Gamma_{dpn}(\alpha_1,\mathbf{p}_{1T}) }{\sqrt{2}}
    \chi^{(\lambda_d)}_d(p_{2N})
    \bigg]
  \label{eqn:2Nderiv:step1}
  ,
\end{align}
where $\chi_d^{(\lambda_d)}(p_{2N})$ is the spin wave function of the deuteron,
the factor of $\frac{1}{3}$ comes from assuming the short-range 2N configuration is in a spin-one state, 
$\Gamma_{dpn}(\alpha_1,\mathbf{p}_{1T})$ is a $d\rightarrow pn$ transition vertex, 
and $\mathcal{D}_{2N}$ is an intermediate state factor, given by
\begin{equation}
  \mathcal{D}_{2N} = \sum_{\mathrm{init.}}p^- - \sum_{\mathrm{inter.}}p^-
  .
\end{equation}
The next step is to rewrite the intermediate state factors:
\begin{align}
  \sum_{\mathrm{init.}}p^- &= p_{2N}^- = \frac{m_{2N}^2 + \mathbf{p}_{2N,T}^2}{p_{2N}^+} \\
  \sum_{\mathrm{inter.}}p^- &= p_1^{-,\mathrm{on}} + p_2^{-,\mathrm{on}} = \frac{m_N^2+\mathbf{p}_{1T}^2}{p_1^+}
  + \frac{m_N^2 + (\mathbf{p}_{2N,T} - \mathbf{p}_{1T})^2}{p_2^+} \\
  \sum_{\mathrm{init.}}p^- - \sum_{\mathrm{inter.}}p^- &=
  \frac{1}{p_{2N}^+}\left[
    m_{2N}^2 - 4\frac{m_N^2 + \left(\mathbf{p}_{1T}-\frac{\alpha_1}{2}\mathbf{p}_{2N,T}\right)^2}{\alpha_1\alpha_2}
  \right]
  \equiv \mathcal{D}_{2N}
  \label{eqn:D2N}
  ,
\end{align}
where $\alpha_1$ and $\alpha_2$ are the light cone momentum fractions of the probed and spectator nucleons respectively.
In the case that $\mathbf{p}_{2N,T} = \mathbf{0}$,
\begin{equation}
  \mathcal{D}_{2N} = \frac{1}{p_{2N}^+}\left(m_{2N}^2 - 4\frac{m_N^2+\mathbf{p}_{1T}^2}{\alpha_1\alpha_2}\right)
  .
\end{equation}
Using Eq.~(\ref{eqn:D2N}) we can, in analogy with Eq.~(\ref{eqn:lcwf:A}),
introduce the light-cone wave function of an NN correlation in the form: 
\begin{equation}
  \psi^{(\lambda_1,\lambda_2;\lambda_d)}_{2N}(\alpha_1,\alpha_2,\mathbf{p}_{1T},\mathbf{p}_{2T},p_{2N}) =
  - \frac{1}{\sqrt{2}\sqrt{2(2\pi)^3}}
  \frac{\bar{u}^{\lambda_1}(p_1) \bar{u}^{\lambda_2}(p_2) \Gamma_{dpn}(\alpha_1,\mathbf{p}_{1T})\chi_d^{(\lambda_d)}(p_{2N})}
       {\frac{1}{2}\left[m_{2N}^2 - 4\frac{m_N^2 + \left(\mathbf{p}_{1T}-\frac{\alpha_1}{2}\mathbf{p}_{2N,T}\right)^2}{\alpha_1\alpha_2}\right]}
       \label{eqn:lcwf:2N}
       ,
\end{equation}
giving (where we assume the 2N correlation has negligible center-of-mass motion and note $p_2=p_{2N}-p_1$, making
$\alpha_1$ and $\mathbf{p}_{1T}$ the only independent parameters in the wave function),
\begin{align}
  \rho_{2N}(\alpha,\mathbf{p}_T) &=
  \frac{1}{3} \sum_{\lambda_d,\lambda_1,\lambda_1^\prime,\lambda_2}
  \int \frac{dp_2^+ dp_2^- d^2\mathbf{p}_{2T}}{2(2\pi)^4}
  \bigg[
    \frac{2(2\pi)^3}{\alpha_1^2}
    \psi^{\dagger(\lambda_1,\lambda_2;\lambda_d)}_{2N}(\alpha_1,\mathbf{p}_{1T})
    \bar{u}^{\lambda_1}(p_1)
    \hat{V}_{2N}(\alpha,\mathbf{p}_T)
    \notag \\ & \qquad \times
    u^{\lambda_1^\prime}(p_1)
    \psi^{(\lambda_1,\lambda_2;\lambda_d)}_{2N}(\alpha_1,\mathbf{p}_{1T})
    \left((2\pi)\delta^{(1)}(p_2^2-m_N^2)\right)      
  \bigg]
  \label{eqn:2Nderiv:step2}
  ,
\end{align}
where the $\Theta(p_{1/2}^+)$ are now implicit.
Next, the on-mass-shell condition $(2\pi)\delta^{(1)}(p_2^2-m_N^2)$ is eliminated through the integration over $p_2^-$.
In particular, since $p_2^+ > 0$,
\begin{align*}
  (2\pi)\delta^{(1)}(p_2^2-m_N^2) \frac{dp_2^+ dp_2^- d^2\mathbf{p}_{2T}}{2(2\pi)^4}
  =
  \frac{dp_2^+ d^2\mathbf{p}_{2T}}{2p_2^+(2\pi)^3}
  =
  \frac{d\alpha_2 d^2\mathbf{p}_{2T}}{2\alpha_2(2\pi)^3}
  .
\end{align*}
Now, we replace the integration variables $(\alpha_2,\mathbf{p}_{2T})$ with $(\alpha_1,\mathbf{p}_{1T})$, obtaining
\begin{align*}
  \delta^{(1)}(\alpha-\alpha_1) \delta^{(2)}(\mathbf{p}_T-\mathbf{p}_{1T})
  \frac{d\alpha_2 d^2\mathbf{p}_{2T}}{2\alpha_2(2\pi)^3}
  =
  \delta^{(1)}(\alpha-\alpha_1) \delta^{(2)}(\mathbf{p}_T-\mathbf{p}_{1T})  
  \frac{d\alpha_1 d^2\mathbf{p}_{1T}}{2\alpha_2(2\pi)^3}
  =
  \frac{1}{2(2\pi)^3\alpha_2}
  ,
\end{align*}
with $\alpha_1=\alpha$ and $\mathbf{p}_{1T}=\mathbf{p}_T$ by virtue of the delta functions.
With the integrations eliminated, and by using Eq.~(\ref{eqn:new:aat}), one obtains
\begin{align}
  \rho_{2N}(\alpha,\mathbf{p}_T) &=
  \frac{1}{2-\alpha}
  \frac{1}{3} \sum_{\lambda_d,\lambda_1,\lambda_2}
  \left|\psi^{(\lambda_1,\lambda_2;\lambda_d)}_{2N}(\alpha,\mathbf{p}_T)\right|^2
  \label{eqn:2Nderiv:step3}
  .
\end{align}
We proceed by further introducing the light cone relative momentum $k$ through the relation
\begin{equation}
  \alpha = 2 \frac{E_k + k_z}{2E_k}
  ,
\end{equation}
which allows one to represent the denominator $\mathcal{D}_{2N}$ in Eq.~(\ref{eqn:D2N}) in the one-parameter form:
\begin{equation}
  \frac{m^2+\mathbf{p}_T^2}{\alpha(2-\alpha)} = m^2 + \frac{(\alpha-1)^2m^2+\mathbf{p}_T^2}{\alpha(2-\alpha)}
  \equiv m^2 + k^2
  .
\end{equation}
Based on the requirement that the light cone equation for the NN interaction leads to the rotationally invariant
on-shell NN scattering amplitude, one can demonstrate that
$\Gamma_{dpn}(\alpha,\mathbf{p}_T) = \Gamma_{dpn}(k^2(\alpha,\mathbf{p_T}))$,
which allows one to write the NN wave function in terms of a single parameter $k$ with the normalization
condition\cite{Frankfurt:1988nt}:
\begin{equation}
  \int d^3\mathbf{k} \frac{\left|\psi_{2N}(k)\right|^2}{E_k} = 1
  .
\end{equation}
It is straightforward to show that such a wave function will satisfy also the momentum sum rule:
\begin{equation}
  \int_0^A \frac{d\alpha}{\alpha} \int d^2\mathbf{p}_T \alpha \rho_{2N}(\alpha,\mathbf{p}_T)
  =
  \int d^3\mathbf{k} \frac{\left|\psi_{2N}(k)\right|^2}{E_k}
  +
  \int d^3\mathbf{k} \frac{k_z}{E_k} \frac{\left|\psi_{2N}(k)\right|^2}{E_k}
  =
  1 + 0
  .
\end{equation}
Another advantage of assuming rotational invariance is that with it,
one can show\cite{Frankfurt:1977vc,Frankfurt:1981mk}
that a Weinberg-type equation for the 2N bound system reduces to the non-relativistic Lippmann-Schwinger equation
with the light-cone relative momentum $k$.
This results in the relation:
\begin{equation}
  \left|\psi_{2N}(k)\right|^2 = \left|\psi_{NR}(k)\right|^2 E_k
  ,
\end{equation}
where $\psi_{NR}(k)$ represents the non-relativistic wave function of NN system.

In further discussions, we introduce the light-cone momentum fraction distribution
\begin{equation}
  f_{N/2N}(\alpha,\mathbf{p}_T) = \frac{1}{\alpha}\rho_{2N}(\alpha,\mathbf{p}_T)
  ,
\end{equation}
which is analogous to a parton distribution for a two-parton system,
but for a nucleon in a two-nucleon SRC instead.
It's worth noting that $f_{N/2N}(\alpha,\mathbf{p}_T)$ is symmetric
when $\alpha$ is substituted with $(2-\alpha)$,
since it represents a two ``parton'' composite system.

%%%%%%%%%%%%%%%%%%%%%%%%%%%%%%%%%%%%%%%%%%%%%%%%%%%%%%%%%%%%%%%%%%%%%%%%%%%%%%%%%%%%%%%%%%%%%%%%%%%%%%%%%%%%%%%%%%%%%%%%
%  Derivation of 3N model
%%%%%%%%%%%%%%%%%%%%%%%%%%%%%%%%%%%%%%%%%%%%%%%%%%%%%%%%%%%%%%%%%%%%%%%%%%%%%%%%%%%%%%%%%%%%%%%%%%%%%%%%%%%%%%%%%%%%%%%%

\subsection{3N correlations}
\label{appendix:lcd:3N}

In this section, we derive the functional form of the three-nucleon SRC density,
namely Eq.~(\ref{eqn:rho3}).

A similar approach is taken here as in Appendix~\ref{appendix:lcd:2N}.
However, unlike with 2N SRCs, the 3N SRC does not correspond to a scaled version of an $A=3$ nucleus.
In particular, the spin and isospin state of a 3N SRC are not {\sl a priori} known,
and at most behave only like a particular, short-range configuration that an $A=3$ nucleus
(which also has a mean field and a 2N SRC part) can take.
However, the approach of calculating $\rho_{3N}(\alpha,\mathbf{p}_T)$ through cut diagrams will still
be taken, with the caveat that each diagram should be multiplied by an unknown constant.

As described in Sec.~\ref{sec:pdf:lcd:3N}, we use a model in which a 3N SRC is generated by a sequence of
two short-range 2N interactions.
Since we consider 2N SRCs to be dominated by $pn$ pairs, this schematic calculation for 3N SRCs takes into
account only $ppn$ and $nnp$ configurations, which can be mediation through two consecutive $pn$ interactions.
A typical diagram of such a process is given in Fig.~\ref{fig:diagram:3N}.
While there are several other topologies for this generation mechanism (depicted in Fig.~\ref{fig:diagram:pnn}),
it can be shown, within an approximation where the three nucleons are collinear before the SRC generation mechanism,
that the results of all diagrams contain the same functional dependence on $\alpha$ and $\mathbf{p}_T$,
so can at most differ in unknwon isospin factors multiplying them.
Accordingly, these constants factor out and can be absorbed into an overall normalization factor.
In particular, we define $\rho_{3N}(\alpha,\mathbf{p}_T)$ to satisfy the normalization condition:
\begin{equation}
  \int \frac{d\alpha}{\alpha} d^2 \mathbf{p}_T \rho_{3N}(\alpha,\mathbf{p}_T) = 1
  \label{eqn:new:norm}
  .
\end{equation}

Another difference between 3N SRCs and 2N SRCs will present itself in the momentum threshold for 2N SRCs to occur. 
For two-nucleon SRCs, it is sufficient to multiply $\rho_{2N}(\alpha,\mathbf{p}_T)$ by $\Theta(k-k_F)$.
However, for three-nucleon SRCs,
one requires that the relative light-cone momentum entering into each $NN$ vertex exceeds $k_F$.

With this in mind, $\rho_{3N}(\alpha,\mathbf{p}_T)$ should be related to $\rho_{N/A}^{(3)}(\alpha,\mathbf{p}_T)$ by
\begin{equation}
  \rho^{(3)}_{N/A}(\alpha,\mathbf{p}_T) = \left\{a_2(A)\right\}^2
  \rho_{3N}(\alpha,\mathbf{p}_T) \Theta_{3N}
  \label{eqn:new:rho3}
  ,
\end{equation}
where $\Theta_{3N}$ indicates the constraints mentioned above on the relative momenta at the $NN$ vertices.
For large nuclei, within the considered model of 3N SRCs, the overall scaling factor is approximately proportional
to the square of the probability for finding a 2N SRC in the nucleus.
Due to surface effects, isospin asymmetry, and the combinatorics of selecting several nucleons,
Eq.~(\ref{eqn:new:rho3}) is not exact,
but should have corrections on the order of $\frac{1}{A}$ that can be neglected for large nuclei.

\begin{figure}
  \centering
  \includegraphics[scale=0.8]{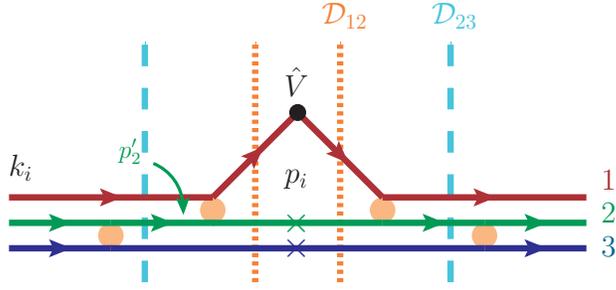}
  \caption{(Color online) Characteristic cut diagram for three-nucleon SRC.}
  \label{fig:diagram:3N}
\end{figure}

Now, we proceed to calculate the cut diagram Fig.~\ref{fig:diagram:3N}.
We have $\hat{V}_{3N} = \hat{V}_{2N}$, with $\hat{V}_{2N}$ given in Eq.~(\ref{eqn:V2N}).
Besides the on-shell specators,
Fig.~\ref{fig:diagram:3N} has two (off-shell) intermediate states on each side of $\hat{V}_{3N}$.
The energy denominators of these states are defined by $\mathcal{D}_{12}$ and $\mathcal{D}_{23}$,
with the notational details given in Fig.~\ref{fig:diagram:3N}. 
The nucleons' initial and final momenta are denoted $k_i$,
and light cone fractions are given by $\beta_i = 3\frac{k_i^+}{p_{3N}^+}$.
We will work in an approximation where the initial nucleons are collinear,
so $\beta_i=1$ and $\mathbf{k}_{iT}=\mathbf{0}$.
The innermost momenta
are denoted $p_i$ and their fractions $\alpha_i = 3\frac{p_i^+}{p_{3N}^+}$.
The intermediate momentum of nucleon ``2'' between $k_2$ and $p_2$ is denoted $p_2^\prime$,
and its light cone fraction is $\alpha_2^\prime = 3\frac{{p_2^\prime}^+}{p_{3N}^+}$.
$\sigma_i$ and $\lambda_i$ are helicities of nucleons with momenta $k_i$ and $p_i$,
while $\lambda_i^\prime$ and $\lambda_i^{\prime\prime}$ are helicities at intermediate states.

In the calculation that follows, we use an approximation where the off-shell NN to of-shell NN transition
depends only on the larger off-shellness,
which in principle requires the smaller off-shellness to be much smaller than the larger one.

\begin{figure}
  \centering
  \begin{subfigure}[t]{.45\textwidth}
    \centering
    \includegraphics[width=\textwidth]{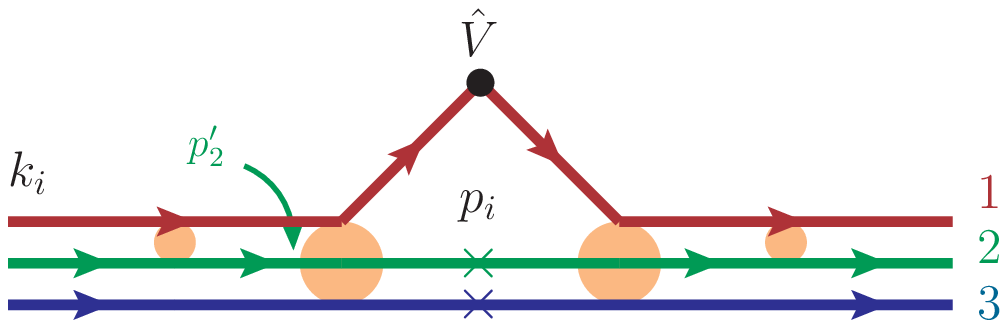}
    \caption{~}
    \label{fig:diagram:pnn:a}
  \end{subfigure}
  \begin{subfigure}[t]{.45\textwidth}
    \centering
    \includegraphics[width=\textwidth]{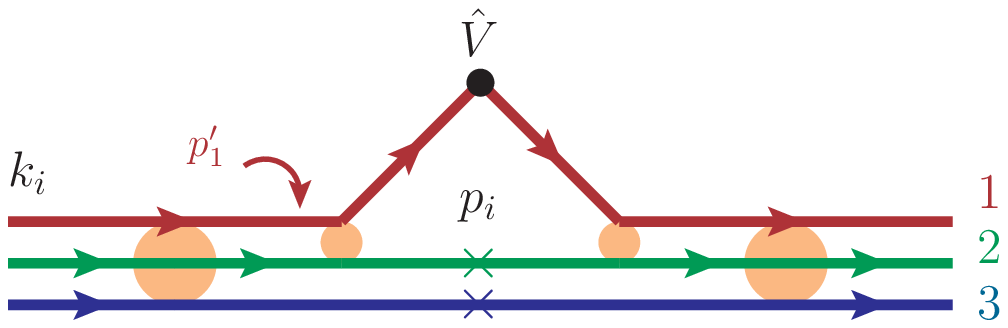}
    \caption{~}
    \label{fig:diagram:pnn:b}
  \end{subfigure}
  \caption{(Color online) Alternate topologies for cut diagrams for 3N SRCs.}
  \label{fig:diagram:pnn}
\end{figure}

Applying Feynman rules for light-cone perturbation theory, 
one obtains:
\begin{align}
  \rho_{3N}(\alpha,\mathbf{p}_T)
  &=
  \frac{1}{8}\sum_{\sigma_1,\sigma_2,\sigma_3}
  \int
  \frac{dp_2^+ dp_2^- d^2\mathbf{p}_{2T}}{2(2\pi)^4}
  \frac{dp_3^+ dp_3^- d^2\mathbf{p}_{3T}}{2(2\pi)^4}
  \bigg[
    \bar{u}^{\sigma_1}(k_1) \bar{u}^{\sigma_2}(k_2) \bar{u}^{\sigma_3}(k_3)
    \frac{ \Gamma_{pn}^{(2,3)} }{\sqrt{2}}
    \notag \\ & \qquad \times
    \left(\sum_{\lambda_3}u^{\lambda_3}(p_3)\bar{u}^{\lambda_3}(p_3)\Theta(p_3^+)\right) \left((2\pi)\delta^{(1)}(p_3^2-m_N^2)\right)
    \frac{\sum_{\lambda_2^\prime}u^{\lambda_2^\prime}(p_2^\prime)\bar{u}^{\lambda_2^\prime}(p_2^\prime)
      \Theta({p_2^\prime}^+)}{{p_2^\prime}^+\mathcal{D}_{23}}
    \notag \\ & \qquad \times
    \frac{ \Gamma^{(1,2)}_{pn} }{\sqrt{2}}
    \frac{1}{\mathcal{D}_{12}}
    \left(\sum_{\lambda_2}u^{\lambda_2}(p_2)\bar{u}^{\lambda_2}(p_2)\Theta(p_2^+)\right) \left((2\pi)\delta^{(1)}(p_2^2-m_N^2)\right)
    \notag \\ & \qquad \times
    \frac{\sum_{\lambda_1}u^{\lambda_1}(p_1)\bar{u}^{\lambda_1}(p_1)\Theta(p_1^+)}{p_1^+}
    \hat{V}_{3N}(\alpha,\mathbf{p}_T)
    \frac{\sum_{\lambda_1^\prime}u^{\lambda_1^\prime}(p_1)\bar{u}^{\lambda_1^\prime}(p_1)\Theta(p_1^+)}{p_1^+}
    \frac{1}{\mathcal{D}_{12}}
    \frac{ \Gamma^{(1,2)}_{pn} }{\sqrt{2}}
    \notag \\ & \qquad \times
    \frac{\sum_{\lambda_2^{\prime\prime}}u^{\lambda_2^{\prime\prime}}(p_2^\prime)\bar{u}^{\lambda_2^{\prime\prime}}(p_2^\prime)
      \Theta({p_2^\prime}^+)}{{p_2^\prime}^+\mathcal{D}_{23}}
    \frac{ \Gamma_{pn}^{(2,3)} } {\sqrt{2}}  
    u^{\sigma_1}(k_1) u^{\sigma_2}(k_2) u^{\sigma_3}(k_3)
    \bigg]
  \label{eqn:deriv3N:step1}
  .
\end{align}
First, we examine the intermediate state denominators. For $\mathcal{D}_{23}$ one has:
\begin{align}
  \mathcal{D}_{23} &=
  \left(k_1^-+k_2^-+k_3^-\right) - \left(k_1^-+{p_2^\prime}^-+p_3^-\right)
  =
  k_2^- + k_3^- - {p_2^\prime}^- - p_3^-
  \notag \\ &=
  \frac{m_N^2}{k_2^+} + \frac{m_N^2}{k_3^+}
  - \frac{m_N^2+\mathbf{p}_{3T}^2}{{p_2^\prime}^+} - \frac{m_N^2+\mathbf{p}_{3T}^2}{p_3^+}
  ,
\end{align}
where we used the fact that $\mathbf{p}_{2T}^\prime = -\mathbf{p}_{3T}$.
We define the total momentum of the two-nucleon pair $(2,3)$
\begin{equation}
  p_{23}^+ = {p_2^\prime}^+ + p_3^+ = k_2^+ + k_3^+
  ,
\end{equation}
and two relative light cone fractions
\begin{align}
  \gamma_2^\prime &= 2\frac{{p_2^\prime}^+}{p_{23}^+} = 2 - \alpha_3 \\
  \gamma_3 &= 2\frac{p_3^+}{p_{23}^+} = \alpha_3
  .
\end{align}
Using the collinear approximation, in which case $\frac{k_2^+}{p_{23}^+} = \frac{k_3^+}{p_{23}^+} = \frac{1}{2}$,
one obtains
\begin{align}
  \mathcal{D}_{23} &=
  \frac{1}{p_{23}^+}\left(
  \frac{m_N^2}{1/2} + \frac{m_N^2}{1/2}
  - \frac{m_N^2+\mathbf{p}_{3T}^2}{\gamma_2^\prime/2} - \frac{m_N^2+\mathbf{p}_{3T}^2}{\gamma_3/2}
  \right)
  =
  \frac{1}{p_{23}^+}\left(4m_N^2 - 4\left[\frac{m_N^2+\mathbf{p}_{3T}^2}{\gamma_2^\prime\gamma_3}\right]\right)
  .
\end{align}
Except for the factor of $\frac{1}{p_{23}^+}$, this is part of the denominator of the two-nucleon wave function, and
in analogy to the two-nucleon SRC we introduce the light-cone relative momentum of the ``23'' nucleon pair as
\begin{equation}
  k_{23}^2 \equiv \frac{(\gamma_2^\prime-1)^2m_N^2 + \mathbf{p}_{3T}^2}{\gamma_2^\prime\gamma_3}
  = \frac{(1+\alpha_3)^2 m_N^2 + \mathbf{p}_{3T}^2}{\alpha_3(2-\alpha_3)}
  \label{eqn:k23}
  .
\end{equation}
For the energy denominator of ``12'' state one has:
\begin{align}
  \mathcal{D}_{12} &=
  \left(k_1^- + k_2^- + k_3^-\right) - \left(p_1^- + p_2^- + p_3^-\right)
  \notag \\ &=
  m_N^2\left(\frac{1}{k_1^+} + \frac{1}{k_2^+} + \frac{1}{k_3^+}\right)
  - \left(\frac{m_N^2+\mathbf{p}_{1T}^2}{p_1^+} + \frac{m_N^2+\left[\mathbf{p}_{1T}+\mathbf{p}_{3T}\right]^2}{p_2^+}
  + \frac{m_N^2+\mathbf{p}_{3T}^2}{p_3^+}\right)
  .
\end{align}
Introducing the total momentum of the pair $(1,2)$
\begin{equation}
  p_{12}^+ = p_1^+ + p_2^+ = k_1^+ + {p_2^\prime}^+
  ,
\end{equation}
and relative light cone fractions
\begin{align}
  \gamma_1 &= 2\frac{p_1^+}{p_{12}^+} = 2\frac{\alpha_1}{3-\alpha_3} \\
  \gamma_2 &= 2\frac{p_2^+}{p_{12}^+} = 2\frac{3-\alpha_1-\alpha_3}{3-\alpha_3}
  ,
\end{align}
a similar derivation to that for the ``23'' energy denominator result in:
\begin{equation}
  \mathcal{D}_{12} = \frac{1}{p_{12}^+}\left(
  m_{12}^2 - 4\left[\frac{m_N^2+\left[\mathbf{p}_{1T}
        +\frac{\gamma_1}{2}\mathbf{p}_{3T}\right]^2}{\gamma_1\gamma_2}\right]
  \right)
  \label{eqn:new:D12}
  ,
\end{equation}
where $m_{12}^2$ is the invariant mass squared of the 2N system, given by:
\begin{align}
  m_{12}^2 &=
  m_N^2\left(\frac{9p_{12}^+}{p_{3N}^+} - \frac{p_{12}^+}{p_3^+}\right)
  - \mathbf{p}_{3T}^2\left(\frac{p_{12}^+}{p_3^+} + 1\right)
  =
  m_N^2\left(9\frac{3-\alpha_3}{3} - \frac{3-\alpha_3}{\alpha_3}\right)
  - \mathbf{p}_{3T}^2 \frac{p_{3N}^+}{p_3^+}
  \notag \\ &=
  \left[10 - 3\left(\alpha_3+\frac{1}{\alpha_3}\right)\right]m_N^2 - \frac{3}{\alpha_3}\mathbf{p}_{3T}^2
  =
  4m_N^2 - \Delta
  ,
\end{align}
where
\begin{equation}
  \Delta = 3\left[\alpha_3 + \frac{1}{\alpha_3}-2\right]m_N^2 + \frac{3}{\alpha_3}\mathbf{p}_{3T}^2 \geq 0
  .
\end{equation}
The term $\Delta$ accounts for the off-shellness of the 2N pair in the intermediate state
and decreases its invariant mass.
If $m_{12}^2 \approx 4m_N^2$, then Eq.~(\ref{eqn:new:D12}) can be interpreted as the denominator of a $2N$ wave function
({\sl cf.} Eq.~(\ref{eqn:lcwf:2N})), with $-\mathbf{p}_{3T}$ as the total transverse momentum of the $2N$ pair.
Its momentum argument can be defined as
\begin{align}
  k_{12}^2
  &=
  \frac{(\gamma_1-1)^2m_N^2 +
    \left[\mathbf{p}_{1T}+\frac{\gamma_1}{2}\mathbf{p}_{3T}\right]^2}{\gamma_1\gamma_2}
  \notag \\ &=
  \frac{(3-\alpha_3)^2}{4}\left[
    \frac{\left(\frac{2\alpha_1}{3-\alpha_3}-1\right)^2 m_N^2
      + \left(\mathbf{p}_{1T} + \frac{\alpha_1}{3-\alpha_3}\mathbf{p}_{3T}\right)^2}{\alpha_1(3-\alpha_1-\alpha_3)}
  \right]
  \label{eqn:k12}
  .
\end{align}
The ability to interpret $\mathcal{D}_{12}$ as the denominator of a two-nucleon wave function is dependent on
$k_{12}^2 \gg \Delta$, so that the difference between $4m_N^2 - 4(k_{12}^2+m_N^2)$ and $m_{12}^2 - 4(k_{12}^2+m_N^2)$,
which is $\Delta$, is negligible compared to the scales considered.
This approximation is valid for large $k_{12}^2$, which is the domain of relevance of three-nucleon SRCs. 

Using the intermediate state denominators, and Eq.~(\ref{eqn:lcwf:2N}) for the two-nucleon wave function, we have
\begin{align}
  \rho_{3N}(\alpha,\mathbf{p}_T)
  &=
  \frac{1}{8}\sum_{\sigma_1,\sigma_2,\sigma_3}
  \sum_{\substack{\lambda_1,\lambda_1^\prime,\lambda_2 \\ \lambda_2^\prime,\lambda_2^{\prime\prime},\lambda_3}}
  \int
  \frac{dp_2^+ dp_2^- d^2\mathbf{p}_{2T}}{2(2\pi)^4}
  \frac{dp_3^+ dp_3^- d^2\mathbf{p}_{3T}}{2(2\pi)^4}
  \bigg[
    \frac{(2\pi)^6(p_{12}^+ p_{23}^+)^2}{4(p_1^+ {p_2^\prime}^+)^2}
    \psi^{\dagger(\lambda_2^\prime,\lambda_3;\sigma_2,\sigma_3)}(\gamma_2^\prime,-\mathbf{p}_{3T})
    \notag \\ & \qquad \times
    \psi^{\dagger(\lambda_1,\lambda_2;\sigma_2,\lambda_2^\prime)}\left(\gamma_1,\mathbf{p}_{1T}+\frac{\gamma_1}{2}\mathbf{p}_{3T}\right)
    \bar{u}^{\lambda_1}(p_1)
    \hat{V}_{3N}(\alpha,\mathbf{p}_T)
    u^{\lambda_1^\prime}(p_1)
    \notag \\ & \qquad \times
    \psi^{(\lambda_1^\prime,\lambda_2;\sigma_2,\lambda_2^{\prime\prime})}\left(\gamma_1,\mathbf{p}_{1T}+\frac{\gamma_1}{2}\mathbf{p}_{3T}\right)
    \psi^{(\lambda_2^{\prime\prime},\lambda_3;\sigma_2,\sigma_3)}(\gamma_2^\prime,-\mathbf{p}_{3T})
    \notag \\ & \qquad \times
    \left((2\pi)\delta^{(1)}(p_2^2-m_N^2)\right)
    \left((2\pi)\delta^{(1)}(p_3^2-m_N^2)\right)
    \bigg]
  \label{eqn:deriv3N:step2}
  .
\end{align}
Next, the delta functions available are used to eliminate several of the integrations.
In particular,
\begin{align*}
  \left((2\pi)\delta^{(1)}(p_2^2-m_N^2)\right)
  \frac{dp_2^+ dp_2^- d^2\mathbf{p}_{2T}}{2(2\pi)^4}
  =
  \frac{dp_2^+ d^2\mathbf{p}_{2T}}{2p_2^+(2\pi)^3}
  =
  \frac{d\alpha_2 d^2\mathbf{p}_{2T}}{2\alpha_2(2\pi)^3}
  ,
\end{align*}
and likewise for nucleon ``3.'' Additionally,
\begin{align*}
  \delta^{(1)}(\alpha-\alpha_1)\delta^{(2)}(\mathbf{p}_T-\mathbf{p}_{1T})
  \frac{d\alpha_2 d^2\mathbf{p}_{2T}}{2\alpha_2(2\pi)^3}
  =
  \delta^{(1)}(\alpha-\alpha_1)\delta^{(2)}(\mathbf{p}_T-\mathbf{p}_{1T})
  \frac{d\alpha_1 d^2\mathbf{p}_{1T}}{2\alpha_2(2\pi)^3}
  =
  \frac{1}{2\alpha_2(2\pi)^3}
  .
\end{align*}
Applying also the rules for creation and annihilation operators according to Eq.~(\ref{eqn:new:aat}),
and noticing that integrands with $\lambda_2^\prime\neq\lambda_2^{\prime\prime}$ diminish, one arrives at
\begin{align}
  \rho_{3N}(\alpha,\mathbf{p}_T)
  &=
  \int
  \frac{d\alpha_3 d^2\mathbf{p}_{3T}}{\alpha_2\alpha_3}
  \bigg[
    \frac{\alpha^2(p_{12}^+ p_{23}^+)^2}{16(p^+ {p_2^\prime}^+)^2}
    \left|\psi_{2N}(k_{12})\right|^2
    \left|\psi_{2N}(k_{23})\right|^2
    \bigg]
  \label{eqn:deriv3N:step6}
  ,
\end{align}
where initial spins have been averaged over and final spins have been summed.
Finally, using the relation
\begin{align*}
    \frac{\alpha^2(p_{12}^+ p_{23}^+)^2}{16(p^+ {p_2^\prime}^+)^2}
    =
    \left\{\frac{3-\alpha_3}{2(2-\alpha_3)}\right\}^2
    ,
\end{align*}
one obtains
\begin{align}
  \rho_{3N}(\alpha,\mathbf{p}_T)
  &=
  \int
  \frac{d\alpha_3 d^2\mathbf{p}_{3T}}{\alpha_2\alpha_3}
  \bigg[
    \left\{\frac{3-\alpha_3}{2(2-\alpha_3)}\right\}^2
    \left|\psi_{2N}(k_{12})\right|^2
    \left|\psi_{2N}(k_{23})\right|^2
    \bigg]
  \label{eqn:deriv3N:step7}
  .
\end{align}
We find numerically that $\rho_{3N}(\alpha,\mathbf{p}_T)$ satisfies the required normalization condition
of Eq.~(\ref{eqn:new:norm}). % if $\mathcal{C}=\frac{1}{2}$.
Thus, through Eq.~(\ref{eqn:new:rho3}), and the threshold conditions $k_{12} > k_F$ and $k_{23} > k_F$, we have
\begin{align}
  \rho^{(3)}_{N/A}(\alpha,\mathbf{p}_T) &= 
  \left\{a_2(A)\right\}^2
  \int \frac{d\alpha_3 d^2\mathbf{p}_{3T}}{\alpha_2\alpha_3}
  \left\{\frac{3-\alpha_3}{2(2-\alpha_3)}\right\}^2
  ~\overline{\left|\psi_{d}(k_{12})\right|^2}~
  \Theta(k_{12}-k_F)
  \notag \\ & \qquad \times
  ~\overline{\left|\psi_{d}(k_{23})\right|^2}~
  \Theta(k_{23}-k_F)
  \label{eqn:deriv3N:step8}
  .
\end{align}
Eq.~(\ref{eqn:rho3}) now follows by using the relation
$f_{N/A}^{(3)}(\alpha,\mathbf{p}_T) = \frac{1}{\alpha}\rho_{N/A}^{(3)}(\alpha,\mathbf{p}_T)$.

A more detailed account of the derivation of the 3N SRC density matrix can be found
in an upcoming work\cite{Artilles:pc2014}.

%%%%%%%%%%%%%%%%%%%%%%%%%%%%%%%%%%%%%%%%%%%%%%%%%%%%%%%%%%%%%%%%%%%%%%%%%%%%%%%%%%%%%%%%%%%%%%%%%%%%%%%%%%%%%%%%%%%%%%%%
%  Differential cross section, two-fold
%%%%%%%%%%%%%%%%%%%%%%%%%%%%%%%%%%%%%%%%%%%%%%%%%%%%%%%%%%%%%%%%%%%%%%%%%%%%%%%%%%%%%%%%%%%%%%%%%%%%%%%%%%%%%%%%%%%%%%%%

\section{Relationship between two- and three-fold differential cross sections}
\label{appendix:dsig2}

Here, we present a brief derivation of the relationship between
$d^2\sigma/dm_{JJ}dy^*$ used in Ref.~\cite{Aad:2011fc} and $d^3\sigma/dp_T^2d\eta_3d\eta_4$ used here.

Ref.~\cite{Aad:2011fc} distinguishes between rapidity $y$ and pseudo-rapidity $\eta$,
but calculations in this work are done at leading order,
for which $y=\eta$.
Using this equality, we denote rapidity using $\eta$ instead of $y$.
With jet rapidities $\eta_3$ and $\eta_4$,
$\bar{\eta}$ is the rapidity of the dijet as a whole,
and $\eta^*$ is the rapidity of an individual jet in the dijet center-of-mass frame, where
\begin{align}
  \bar{\eta} &= \frac{\eta_3+\eta_4}{2} \\
  \bar{\eta}^* &= \frac{\eta_3-\eta_4}{2}
  .
\end{align}
The Jacobian for the transformation from $(\eta_3,\eta_4)$ to $(\bar{\eta},\eta^*)$ gives us
$d\eta_3d\eta_4=2d\bar{\eta}d\eta^*$.
In addition to the rapidity in the center-of-mass frame $\eta^*$,
Ref.~\cite{Aad:2011fc} uses the dijet mass $m_{JJ}$, which is given by
\begin{equation}
  m_{JJ}^2 = (p_3+p_4)^2 = 4p_T^2\cosh^2(\eta^*)
  ,
\end{equation}
or $m_{JJ} = 2p_T\cosh(\eta^*)$.
We note that the overall Jacobian is given by
$dp_T^2 d\eta_3 d\eta_4 = \frac{2 p_T}{\cosh(\eta^*)} dm_{JJ} d\bar{\eta} d\eta^*$,
and thus the two-fold differential cross section given in Ref.~\cite{Aad:2011fc} can be written as
\begin{equation}
  \frac{2d^2\sigma}{dm_{JJ}d\eta^*} = \frac{4 p_T}{\cosh(\eta^*)}\int d\bar{\eta}
    \frac{d^3\sigma}{d\eta_3 d\eta_4 dp_T^2}
  \label{eqn:dsig2}
  .
\end{equation}

%  --------------------------------------------------------------------------------------

\bibliographystyle{apsrev4-1}
\bibliography{references}

\end{document}